\documentclass{aa}   

\usepackage{subcaption}
\usepackage{graphicx}
\usepackage{txfonts}

\usepackage{amstext}

\begin{document}

   \title{Kinematic signatures of reverberation mapping of close binaries of supermassive black holes in active galactic nuclei}

   \subtitle{III. The case of  elliptical orbits}

   \author{Andjelka B. Kova{\v c}evi{\'c}  \inst{1},
                   Jian-Min Wang \inst{2}, 
                    Luka  {\v C}. Popovi{\'c}  \inst{3}  
          }

   \institute{Department of astronomy, Faculty of mathematics, University of Belgrade \\
                  Studentski trg 16, Belgrade, 11000, Serbia
              \email{andjelka@matf.bg.ac.rs}
             \and
             Key Laboratory for Particle Astrophysics, Institute of High Energy Physics, CAS, \\
             19B Yuquan Road, Beijing 100049, China\\
             \email{wangjm@ihep.ac.cn}
             \and
             Astronomical observatory Belgrade \\
Volgina 7, P.O.Box 74 11060, Belgrade,  11060, Serbia
              \email{lpopovic@aob.rs}
             }

   \date{Received , ; accepted ,  }

 
  \abstract
   {An  unresolved  region in the relative vicinity  of the event horizon of a  supermassive black holes (SMBH) in active galactic nuclei (AGN) radiates  strongly variable optical   continuum  and broad-line emission flux.
These fluxes  can be processed  into two-dimensional transfer functions (2DTF) of material flows that encrypt various   information about these unresolved structures. An intense search for kinematic signatures  of reverberation mapping of close binary SMBH (SMBBH)  is currently
ongoing.}
   {Elliptical  SMBBH systems (i.e. both orbits and disc-like broad-line regions  (BLR) are elliptic)   have not been assessed  in 2DTF  studies. We aim to numerically infer such a 2DTF because the geometry of    the unresolved region  is imprinted on their optical emission. Through this, we determine their  specific kinematical signatures. }
   {We simulated the geometry and kinematics of  SMBBH whose components are on elliptical orbits. 
Each SMBH had a  disc-like  elliptical BLR. The SMBHs were active and orbited each other  tightly  at a subparsec distance.}
   {Here we calculate for the first time 2DTF, as defined in the velocity-time delay plane, for  several  elliptical configurations of  SMBBH orbits and their BLRs.  
We find that these very complex configurations are clearly resolved in maps. These results are  distinct from those obtained from  circular   and disc-wind geometry.  We calculate the expected line variability for all SMBBH configurations. We show that the line shapes  are influenced by the orbital phase of the SMBBH. Some line profiles resemble observed profiles, but they can also be  much deformed to look like those  from the disc-wind model.}
   {First, our results imply that using our 2DTF, we can detect and quantify kinematic signatures of elliptical SMBBH. Second,  the calculated expected  line profiles   share some intriguing similarities with observed profiles, but also with  some  profiles that are synthesised in disc-wind models. To overcome the non-uniqueness of the spectral line shapes as markers of SMBBH, they must be accompanied with  2DTF.}


\titlerunning{Reverberation mapping of close binaries of supermassive black holes in active galactic nuclei.III}

\authorrunning{A. B. Kova{\v c}evi{\'c} et al.}

 \maketitle
%

\section{Introduction}

 Active galactic nuclei (AGN) are among the most distant objects that can be observed that also play a fundamental role in the evolution   of galaxies and the Universe because their activity  is tied to galaxy growth. It is well known that galaxy growth by merging of two or several galaxies \citep{ji12,bu13,ca15} is very effective.  The last
phase of a galaxy merger is the collision of the central parts of the merging galaxies. Each of
the galaxies probably hosted a supermassive black hole and eventually formed a
bound supermassive binary black hole (SMBBH). Therefore, it 
 is believed that   SMBBH assemble in galaxy mergers and reside in galactic nuclei with high and poorly constrained concentrations of gas and stars. These systems are expected to emit nanohertz gravitational waves (GW) that will be detectable by pulsar timing arrays in ten years \citep{2017NatAs...1..886M}. Moreover, if a substantial population of binary or recoiling SMBHs were characterised, it would better delineate the rate of coalescence events because they may produce strong millihertz-frequency GWs that could be detected by space-based laser interferometers \citep{2003ApJ...595..614W}.

An SMBH is located in the centre of an AGN. The SMBH  is surrounded by an accretion disc that emits in the X-ray and UV. This high-energy radiation is able to photoionise the gas located in the so-called  broad-line region (BLR). Information about the motion and  size of the BLR  that lies close to the SMBH is crucial for   probing the physics under extreme conditions and for measuring SMBH masses. Through recombination, the BLR emits broad emission lines that often show a high variability (as does the central continuum source).  
So far, the size of the BLR has been inferred  mainly by a method called reverberation mapping \citep{doi: 10.1086/151300, 1982ApJ...255..419B, 1993PASP..105..247P}. 
The goal of this method is to use the observable  variability of the central AGN power source and the response of the gas in the BLR to solve the transfer function, which depends on the BLR geometry, kinematics, and reprocessing physics, and thus to obtain the  BLR geometry and  measure a time delay \citep{1997ASSL..218...85N}. Reverberation mapping has   successfully been employed for several dozen  AGN \citep[see some recent campaigns, e.g.][]{doi: 10.1088/0004-637X/806/1/22, doi: 10.1086/679601, doi: 10.3847/0004-637X/818/1/30, 2017MNRAS.466.4759S, 2018ApJ...866..133D, 2019MNRAS.485.4790S}.
These results, derived from timing information, can now be compared with spatially resolved information  
\citep{2018Natur.563..657G}. The spatial  observation  of rapidly moving gas around the central black hole of 3C 273 supports the fundamental assumptions of reverberation mapping and confirms results of SMBH masses derived by the reverberation method.

Ever-increasing   spectroscopic samples  of AGNs  offer a unique opportunity to search for SMBBH candidates based on their spectral characteristics
 \citep[see e.g.][]{2009ApJ...705L..76W,2009ApJ...705L..20X, 2012NewAR..56...74P}.
 The majority of  SMBH binary candidates has been identified by peculiar features  of their optical and near-infrared spectra.  For example, the broad lines emitted by gas in the BLR  of each SMBH can be blue- or redshifted through the Keplerian motion of the binary \citep{1980Natur.287..307B}.  Gravitational interaction between the components in an SMBBH system can also  perturb and even remove a broad-line emitting region so that some non-typical  flux ratios between broad lines can appear \citep[see][]{2011MNRAS.412...26M, 2012MNRAS.425.1633M}. As many studies showed, the spectroscopic approach is not bijective, that is to say, the spectroscopic signatures of  binary systems  can be interpreted by  some other single SMBH models \citep[see review of][for more detail]{2012NewAR..56...74P}.

 \cite{1996LNP...471..165G} suggested that broad double-peaked emission lines could
come from binary SMBHs. \cite{1994ApJS...90....1E} showed that
the binary black hole interpretation of double-peaked lines was
unlikely. Furthermore, the double-line spectroscopic binary case has been
tested and rejected for some quasars with double-peaked broad emission line profiles 
by several follow-up tests \citep[see e.g.][]{1997ApJ...490..216E, 2016ApJ...817...42L, Doan19}.

 The double-peaked broad-line profile does not mean that there is an SMBBH system 
\citep[see][for a review]{2012NewAR..56...74P}, and it seems that a double-peaked profile alone  is not an indicator for SMBBHs. However,  single-peaked broad lines might still indicate SMBBHs
 \citep[see e.g.][]{
 2011ApJ...738...20T, doi.org/10.1088/0067-0049/201/2/23,  2013MNRAS.433.1492D, 
 J2013ApJ...777...44J,  2013ApJ...775...49S,2014ApJ...789..140L,2017MNRAS.468.1683R, 2017ApJ...834..129W,  
2019MNRAS.482.3288G}. It seems that the
probability that  only one  SMBH in a system is active  is much higher than that both
SMBHs are simultaneously active. Single broad lines  also allow investigating a  larger  binary parameter space  than double-peaked broad lines \citep{2014ApJ...789..140L}.
 Continuing the point made above, we describe recent work by several
groups who tried to find SMBBH with only one active black hole. The observational signature used in these studies is a single
displaced peak in the broad H$\beta$ line.
The most effective  way to improve  the broad-line diagnostics
to identify SMBBHs is multi-epoch spectroscopy. Long-term spectroscopic
monitoring of the broad lines can confirm or  refute  the SMBBH hypothesis
based on the variability  in the broad-line radial velocity if
the monitoring time baseline is long enough to detect the expected binary motion
and if the quality of spectra is good enough \citep{2012ApJ...759..118B, 2016ApJ...822....4L, 
2016ApJS..222...25S, 2019ApJS..241...33L}.  Several
systematic searches have been conducted so far through the Sloan Digital Sky Survey (SDSS) database for
single-peaked broad emission lines with large velocity offsets from the
quasar rest frame. The studies have been focused on  
quasars with broad lines positioned at their systemic velocities (for the SMBBH hypothesis, this would mean that  binaries appear at conjuction; see \cite{J2013ApJ...777...44J,2013ApJ...775...49S, 2017ApJ...834..129W}) and
those that the broad emission lines have offset from the rest frame by
thousands of km \, s$^{-1}$ \citep{2011ApJ...738...20T,doi.org/10.1088/0067-0049/201/2/23,
2013MNRAS.433.1492D,2014ApJ...789..140L, 2017MNRAS.468.1683R}. 
\cite{2011ApJ...738...20T} found five new candidates with large velocity offsets in the Sloan Digital Sky Survey (SDSS)
DR 7. Recently, \cite{doi.org/10.1088/0067-0049/201/2/23} carried out the first
systematic spectroscopic follow-up study of quasars with offset broad
H$\beta$ lines. This study detected  88 quasars with broad-line offset
velocities and also conducted second-epoch spectroscopy of 68 objects.
Significant (at 99$\% $confidence) velocity shifts in 14
objects were found. \cite{2013MNRAS.433.1492D} also obtained second-epoch spectra for 32
SDSS quasars selected with peculiar
broad-line profiles \citep{2011ApJ...738...20T},  large velocity offsets,  and double-peaked or asymmetric line profiles. However, the conclusions
from \cite{2013MNRAS.433.1492D} are slightly less decisive because  the authors obtained  velocity
shifts using model fits to the emission-line profiles rather than the more
robust cross-correlation approach adopted by \cite{doi.org/10.1088/0067-0049/201/2/23},
\cite{2013ApJ...775...49S} and \cite{2014ApJ...789..140L}.
\cite{2013ApJ...775...49S} performed systematic search for sub-parsec binary SMBH
in normal broad-line quasars at z $<$ 0.8, using multi-epoch SDSS spectroscopy of the broad H$\beta$ line. Using the set of
700 pairs of spectra, the authors detected 28 objects with significant
velocity shifts in the broad H$\beta$, of which 7 are  classified as the best candidates for
the  SMBBH, 4 as most likely due to broad-line variability in a single
SMBH, and the rest are inconclusive.
\cite{2014ApJ...789..140L} selected a sample of 399 quasars with kinematically offset
broad H$\beta$ lines from the SDSS  DR 7
and have conducted second-epoch optical spectroscopy for 50 of them. The
authors detected significant (99$\%$ confidence) radial accelerations in the
broad H$\beta$ lines in 24 of the 50 objects and found  that 9 of the 24
objects  are sub-parsec SMBBH candidates, showing consistent velocity
shifts independently measured from a second broad line (either H$\alpha$
or Mg II) without prominent  variability of  the broad-line profiles.
As a continuation of  the previous two works, \cite{2019MNRAS.482.3288G} presented further
third- and fourth-epoch spectroscopy for 12 of the 16 candidates for
continued radial velocity tests, spanning 5-15 yr in the quasar rest
frames. Cross-correlation analysis of the broad H$\beta$  suggested that
5 of the 12 quasars remain SMBBH candidates. These objects show broad
H$\beta$ radial velocity curves that are consistent with binary orbital
motion without notable variability in the broad-line profiles. Their broad
H$\alpha$ (or Mg II) lines display radial velocity shifts that are either
consistent with or smaller than those seen in broad H$\beta$.
Conversely,  \cite{J2013ApJ...777...44J} used a cross-correlation to search for temporal
velocity shifts in the Mg II broad emission lines of
0.36 $<$z$<$ 2 quasars  with multiple observations in the SDSS and found 7 candidate sub-parsec-scale binaries.  \cite{2017ApJ...834..129W} searched for binary SMBHs using time-variable velocity shifts in
broad Mg II emission lines of quasars with multi-epoch observations and
found only one object with an peculiar velocity. This  indicates that
$\lesssim 1 \% $ of the SMBHs are   binaries with 0.1 pc separations. \cite{2017MNRAS.468.1683R}  identified
3 objects (SDSS J093844, SDSS J095036, and SDSS J161911) in their sample that showed
systematic and monotonic velocity changes consistent with the binary
hypothesis. The authors  observed substantial profile shape variability in
at least one spectrum  of the SMBH candidates, indicating that variability
of the broad-line profile shape can mimic radial velocity variations, as was theoretically predicted
\citep[see e.g.][]{2016Ap&SS.361...59S}.

 An odd example of the non-uniqueness of line shapes as markers of geometry  is  SDSS J092712.65$+$294344.0, which was identified in the SDSS as a quasar, but with  the unusual
property of having two emission-line systems offset by  2650 km\, s$^{-1}$. One of these systems resembles the  usual mixture 
of broad and narrow lines, and the other system shows only narrow lines.
\cite{2008ApJ...678L..81K} interpreted  this as a galaxy  with a merged  pair of SMBHs, producing a recoil of several thousand  km\,s$^{-1}$  to the
new, larger SMBH. In two other papers by  \cite{2009ApJ...697..288B}
and \cite{2009MNRAS.398L..73D},  the unusual spectroscopic feature was explained as arising from an SMBH binary at
a separation of $\sim 0.1-0.3$ pc  whose mass ratio is  $q \sim 0.1-0.3$. Finally, a third interpretation was given  by  \cite{2009ApJ...695..363H}, who proposed  a random spatial superposition coincidence of two AGNs within the angular resolution of the spectrograph.

Even the periodicity detection method for distinguishing between the single and binary SMBH hypothesis based on  integrated light curves  is not bijective.  The most famous example is the subparsec binary candidate PG 1302-123 found by 
\cite{2015Natur.518...74G}. However, based on  newly added observations,  \cite{2018ApJ...859L..12L} reported a decrease in the periodicity significance, which may suggest that the binary model is less favourable. Simultaneously, \cite{2019ApJ...871...32K} confirmed by their hybrid method  the periodicity detection in the same data set.
 
Clearly, both spectra and time-domain light curves are just  a 1D projection of complex 3D physical objects. The mapping relation between them seems non-invertible.  To overcome this situation, both theoretical and observational improvements are needed.  
One promising modern  approach to investigating the physical processes in binary SMBH systems at scales that cannot be reached by  telescopes is reverberation mapping.
Using spectral monitoring and reverberation mapping, it  might be possible to  distinguish the binary scenario from its alternatives \citep[e.g. see][]{1997ApJ...490..216E, 2010Natur.463E...1G,2012NewAR..56...74P,2016ApJS..222...25S,2018MNRAS.475.2051K,2019ApJS..241...33L}. 
Clearly, in order to make progress in using the spectroscopic variability data for the above purposes, it is necessary to construct detailed reverberation maps.

To systematically study the reverberation signatures, it is imperative to exploit   those BLR models that can describe  line profile signatures without invoking axisymmetric assumptions.
Some recent simulations show  the formation of non-axisymmetric individual
accretion discs, or mini-discs, around each member of the binary
\citep[see e.g.][]{2014ApJ...783..134F, 2015MNRAS.447L..80F, 2015MNRAS.446L..36F, 2016MNRAS.459.2379D, 2017ApJ...838...42B, 2017ApJ...835..199R, 2018ApJ...865..140D, 2018ApJ...853L..17B, 2019ApJ...875...66M,  2019ApJ...879...76B}.
 It is well known that some broad spectral lines show a clear asymmetry, a   more prominent red peak   than blue peak, or their profiles vary
with  successive blue- and red-dominated peaks 
 \citep{1994ApJS...90....1E, 2000A&A...356...41S,
2003ApJ...599..886E, 2003AJ....126.1720S, 2010ApJS..187..416L, 2011A&A...528A.130P, 2017ApJ...835..236S} .

This asymmetry is induced by the asymmetry of the environment in which the spectral lines originate.
Thus circular disc
emission models must be replaced with  an asymmetrical  disc that can  reproduce
the observed line asymmetry. Among these models, an elliptical disc is generic because it  requires the lowest number of free parameters. Observations of line profiles
indicate that  accretion discs in  AGN are often non-axisymmetric 
\cite[in at least 60$\%$ of
the cases, see][]{2003AJ....126.1720S}.

\cite{2018ApJ...862..171W}  (hereafter Paper I) calculated detailed   reverberation maps of  spherical BLR of  SMBBHs on circular orbits, identifying changes in the maps due to the inclination to the observer as well as depending on whether the gas is orbiting, is outflowing-inflowing, or some combination of these processes. The novelty of this approach relies  in the conjugate  effects of the two SMBHs in order to obtain the composite 2D transfer function (2DTF) of such a system, which is clearly different from the simple 2DTF  of a single SMBH.
In contrast, elliptical  BLRs are very interesting for some particular AGNs, such as the very broad double-peaked AGNs monitored by \cite{1995ApJ...438..610E} and \cite{2003ApJ...598..956S}. 
Recently, an atlas of the 2DTFs has been compiled by  \citet[][submitted, hereafter Paper II]{Songsheng19}.

 No theoretical  reverberation maps of elliptical disc-like BLRs,  either of single or binary SMBH, are available so far for comparison with observations. 
Motivated by the lack of  systematic 2DTF for single  SMBH and  elliptical binary SMBH with  elliptical disc-like BLRs,  the main goal of this paper is to   calculate  such maps.
We aim to numerically infer the 2D reverberation maps of these systems because the geometry of the   unresolved region  is imprinted on their optical emission, in order to find their  specific kinematical signatures. 
We simulate the geometry and kinematics of  binary SMBHs whose components are on elliptical orbits. 
A separate disc-like  elliptical broad-line region is attached to each SMBH. Both components are active at a mutual subparsec distance. 
We aim to predict the expected  kinematical signatures from those sources. 
The constructed  model  produces reverberation signatures that are clearly distinct from those of a circular case.

The paper is organised as follows: in  Section \ref{sec:meth}   we present our geometrical and kinematical  model  as well as that transfer function that we used to calculate reverberation maps of elliptical binary SMBH systems. We give and discuss  the predicted reverberation signatures for different   elliptical configurations of a binary SMBH system to show how they are affected by different SMBH orbital and BLR parameters in  Section \ref{sec:discus}. Finally, brief conclusions are given in Section \ref{sec:con}.
 

\section{Formalism} \label{sec:meth}
 
 In this section we review and expand  the geometry  model of disc-like BLRs  in the elliptical disc-like case. We simultaneously assume that both components are on elliptical orbits. The simplest possible binary system consists of two  objects (see Paper I)  in a perfectly circular orbit.  We complicated the binary system
  by assuming that the total orbital energy is higher than the angular momentum of a circular orbit. This excess energy causes the orbital radius to oscillate in synchrony with the orbital period, which sends the two objects  into opposing elliptical orbits, defined by the orbital eccentricity.
  We briefly review  the general approach to solve for the orbit of an eccentric binary system, including some of the notation and formalism
of celestial mechanics.  We refer to \cite{BC61} for exhaustive details.
Regarding the notation, bold face letters refer to vectors (lower case).
State vectors  of the binary components and clouds in disc-like BLRs  are distinguished  by 
subscripts   in a left-right fashion. Leftmost indices in subscripts denote the binary component (b) or the cloud (c), the next index represents the primary or secondary SMBH (i.e. 1 or 2), and the rightmost index, if present, denotes the orbit of the cloud with respect to its innermost edge, where the innermost edge is always designed as 1.
For example, $\vec{r}_{{\mathrm{b}1}}$ is the position vector of the primary component in the SMBH system. Similarly, $\vec{r}_{{\mathrm{c}12}}$ represents the state vector of the cloud in a BLR around the primary component, and its orbit is the second with respect to the edge of its BLR.
In particular, the subscripts of  the SMBH orbital parameters account for the primary or secondary component  alone.  The orbital parameters of the clouds do not contain the rightmost index, that is, the designation with respect to the inner BLR edge.

  \subsection{Geometry of the  SMBH binary}\label{binsys}
  
  \begin{figure*}
  \begin{minipage}[b]{0.49\textwidth}
    \includegraphics[width=\textwidth]{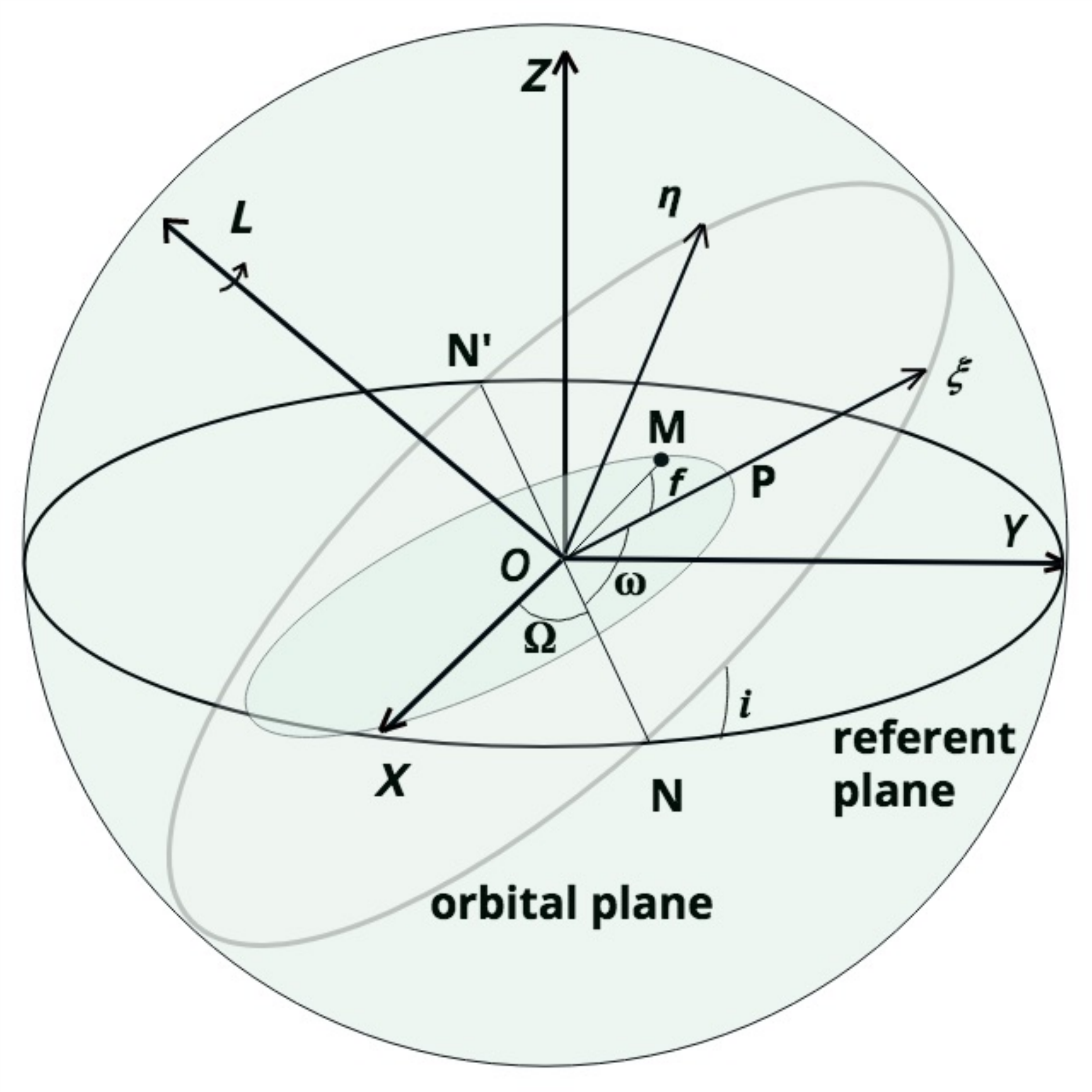}
   \hspace*{125pt} (a)
  \end{minipage}
 \hfill
  \begin{minipage}[b]{0.5\textwidth}
    \includegraphics[trim=800 300 800 500,clip,width=\textwidth]{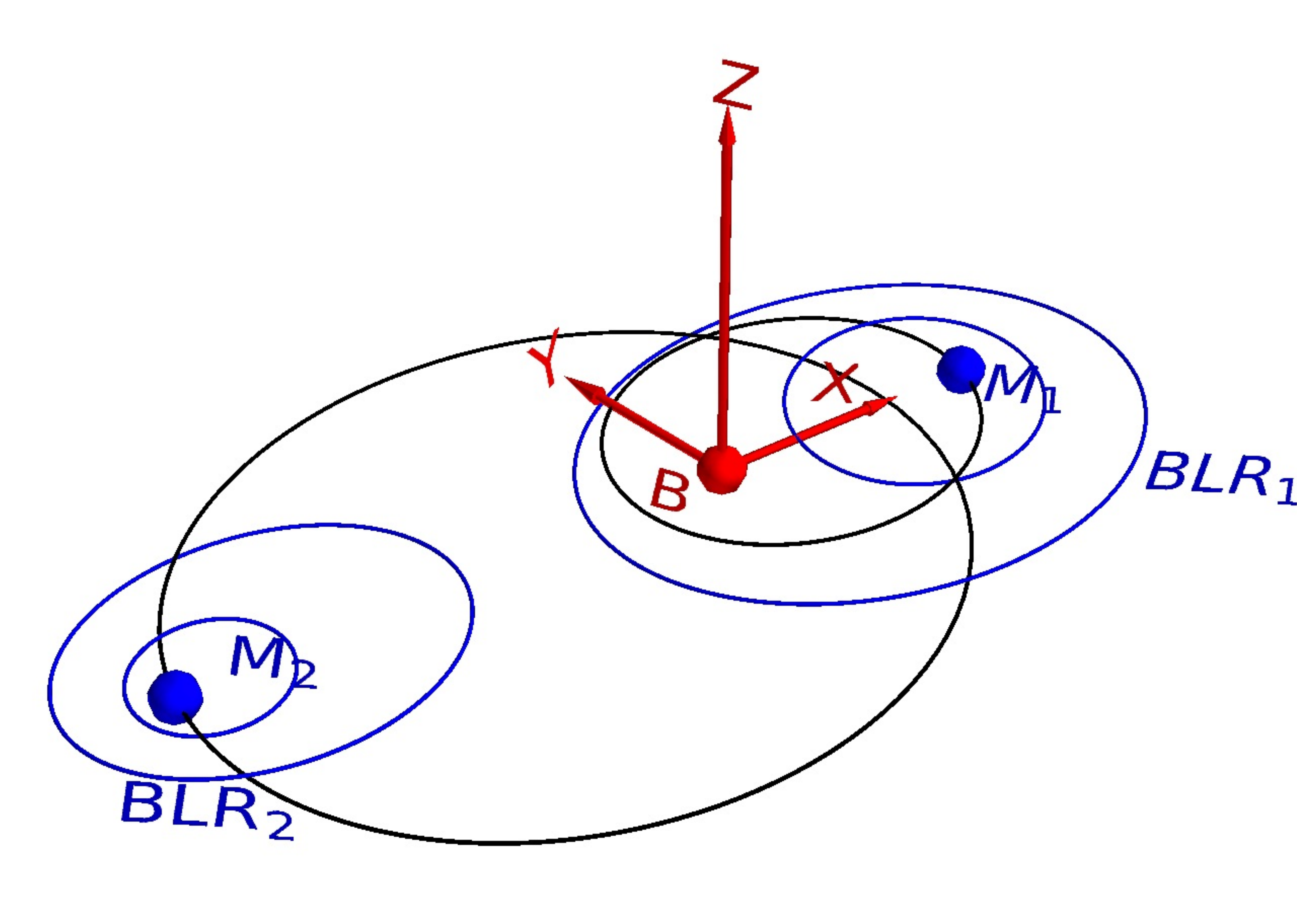} 
    \hspace*{125pt} (b)
  \end{minipage}
  \caption{(a) Orbit of body M described with the standard  astrometric set of orbital elements $(a, e, i, \omega, \Omega ], f)$, where $POM=f$ is the true anomaly and $\bf{L}$ is the angular momentum vector of the motion of the body. The inertial $OXYZ$ coordinate system can be chosen arbitrarily (see the detailed explanation in Appendix~\ref{app:geo} .) (b) A realisation of the coplanar elliptical SMBBH system  at  the half orbital period calculated from our binary system model. $M_{1} \text{ and }M_{2}$ are the locations of the SMBHs in a binary system. Blue ellipses are the inner and outer boundaries of disc-like BLRs. The reference plane ($BXY$) is the plane of the relative orbit of $M_2$ with respect to $M_{1}$, and the origin of the coordinate system is the barycenter  ($B$) of the system.  }
    \label{fig:cosis}
  \hfill
\end{figure*}

As we showed in Paper I,  to study the 2DTF of binary systems, it is practical to use a 3D orthogonal frame. In this frame,  the mutual relative motion of the SMBHs and the relative motion of the BLR to the SMBH can be expressed in the form of Keplerian orbits.  Here, we investigate elliptical orbits of the SMBH system and their elliptical  disc-like BLR, whose 2D orbital planes  are embedded in 3D space.
Using standard astrometric notation, the SMBH orbits and the cloud orbits in the BLR can be expressed  via Keplerian orbital elements. We can write an orbit directly in terms
of the semimajor axis ($a$), the inclination  of the orbit ($i$) to the reference plane,
the position angle of the ascending node ($\Omega$), the argument
of pericenter ($\omega$), and the true  ($f$) and eccentric ($E$ ) anomaly as shown in  Fig. \ref{fig:cosis}(a).

In addition to orbital elements, the positions and velocity vectors in 3D (i.e. orbital state vectors) describe the motion of a body. The Keplerian orbital elements  can be obtained from state vectors and vice versa (see Appendix \ref{app:geo}). In contrast to the state vectors, five of six orbital elements can be considered constant, which makes using them convenient for the characterization of the 2DTF maps.  
We note  that we restricted our analysis  to non-spinning objects. When spin interactions occur, the orbital plane is no longer fixed in space, and then it is not possible to define the longitude of ascending node $\Omega$.

Moreover, investigating the binary 2DTF map topology in relation to  the binary orbital phase relies on using the centre-of-mass (barycentric) frame.   In particular, the barycentric frame for binary SMBH motion tracking is a spatial realisation  of the general 
astrometric 3D coordinate frame (see Fig. \ref{fig:cosis}(a) and Appendix \ref{app:geo}). Although the explicit forms of the velocity and position of the clouds depend on
the coordinate frame, their projections on the  observer's line of sight do not because the scalar product does not change  its values when the basis changes.

We assumed   two SMBH on elliptical orbits  with a common barycenter, and each of which had elliptical disc-like BLRs.  In addition,  each SMBH was in the focus of its own elliptical   disc-like BLR.
Both BLRs consisted of  nested elliptical annuli  whose eccentricity might be different.  Line profile modelling implies  that the BLR consists  of  numerous  clouds  of negligible  individual sizes relative  to the BLR size \citep{1998MNRAS.297..990A}. According to the unification model \citep[see][and references therein]{2015ARA&A..53..365N}, most BLR clouds are seen directly by the observer if they are not masked by the dusty torus. 

Let $M_1$ and $M_2$ be the masses of the two SMBH components, while the mass of each cloud is considered negligible in comparison  to the mass of the corresponding  SMBH.
The spatial motion of  $M_1$ and $M_2$  around the system's barycenter ($B$) lies  in the relative orbital
plane  of binary (see Fig. \ref{fig:cosis}(b)). This is called coplanar binary SMBH system.
The common binary orbital plane is perpendicular to  the vector of the binary orbital angular momentum relative to the barycenter. We set this vector  as the $Z$ -axis of the barycentric frame and the barycenter $B$ as  the origin of the frame.
 The common orbital plane is set as the reference plane of the barycentric frame. 
Moreover, the reference plane  is  spanned 
by the $X$ -axis (aligned with the semimajor axis of the binary relative orbit and pointing from the barycenter to the pericenter of the binary relative orbit) and the $Y$ -axis (perpendicular to both the $X$ - and $Z$ -axis, making a right-handed triad). Thus, the  barycentric reference frame $(B,X,Y,Z)$ for tracking the relative motion of  components in binary system is completed.
Both SMBHs are assumed initially to lie on the $X$-axis.  If not otherwise stated,  the more massive SMBH  is  initially set  at  $(X, Y,Z)=(a\cdot(1-e),0,0)$ and the secondary SMBH  at  $(X,Y,Z)=(a\cdot(e-1),0 ,0)$, where $ a$ is the semimajor axis and $e$ is the eccentricity of their common barycentric orbit.

We are interested in  the motion of the clouds relative to the binary components in the barycentric frame,  therefore it is more convenient to follow   particles in the local moving frame of the corresponding binary component  and then  to transform their state vectors  into the barycentric frame (see Appendix \ref{app:expandtauv}). The $z$ -axis of the moving frame coincides with the orbital angular momentum direction of the SMBH. Its $x$ -axis is colinear with the barycenter-SMBH line, pointing outward. The right-hand triad is completed with the $y$ -axis, which is orthogonal to the previous two axes. We note that the reference planes of the local moving frames  and  the binary barycentric frame coincide in a coplanar case. For a  non-coplanar case, the local and general  reference planes coincide for  the  SMBH whose orbital plane is taken as reference.
The transformation between the barycentric and the local frame is given in Appendix \ref{app:expandtauv}.
The cloud orbital elements   are also defined with respect to the reference plane in the local moving coordinate system.  

Then 3D  barycentric orbits of  $M_1$ and $M_2$ are given as follows: 

\begin{eqnarray}
\vec{r}_{\mathrm b1}&=&\frac{M_{2}}{M_{1}+M_{2}} \vec{r}^{{\prime}{\prime}}, \\
\vec{v}_{\mathrm b1}&=& \frac{M_{2}}{M_{1}+M_{2}}\vec{v}^{{\prime}{\prime}},\\
\vec{r}_{\mathrm b2}&=&-\frac{M_{1}}{M_{1}+M_{2}}\vec{r}^{{\prime}{\prime}}, \\
\vec{v}_{\mathrm b2}&=& -\frac{M_{1}}{M_{1}+M_{2}}\vec{v}^{{\prime}{\prime}},\\
T_{1}&=&T_{2}=T_{\mathrm rel},\\
e_{1}&=&e_{2}
\label{baricentric}
\end{eqnarray}
where $\vec{r}_{\mathrm {b}i}$ and $\vec{v}_{\mathrm {b}{i}}$, $T_{i}, e_{i}, i=1,2$ 
are the barycentric radius  and velocity vectors, orbital period, and eccentricity of the primary (1) and secondary (2) SMBH,  
and   $T_{\mathrm rel}$ is the orbital period of   the barycenter.
The  state vectors $ (\vec{r}^{{\prime}{\prime}}, \vec{v}^{{\prime}{\prime}}) $  of the secondary motion relative to the primary component are given by  
Eq \ref{inplane0} and Eq. \ref{inplane} (see Appendix \ref{app:geo}).

 Then  the barycentric position of cloud $\vec{r}_{\mathrm {c}ij}$
 (see Fig. \ref{fig:sistem1})  is given  as

\begin{equation}
\vec{r}_{\mathrm {c}ij}=[\vec{\varrho}_{ ij}]_\mathrm{B}+\vec{r}_{\mathrm {b}i}
\label{rdisk}
 ,\end{equation}
  where $ \vec{r}_{\mathrm {b}i}$ is   the barycentric position of the SMBH  (defined by Eq. \ref{baricentric}), and 
 $[\vec{\varrho}]_\mathrm{B}=\mathbb{Q}^{-1}[\vec{\varrho}]_{\bullet}$
  is the relative  vector    as measured  from the SMBH to the cloud in the barycentric frame (see Eq.\ref{qrhotrans}  in Appendix  \ref  {app:expandtauv} and  Appendix \ref{app:relmotapp}).
The indexes  $i=1, 2$ and $ j=1, N$ refer to the primary and secondary  SMBH and the  elliptical rings in the disc-like BLR, respectively. 
From
the inner  $R_{\mathrm{in}}$ to the  outer boundary $R_\mathrm{out}$ of the disc-like  BLR, the pericenter
distance for the elliptical rings is evenly separated into N
bins,
\begin{equation}
 \left.  R_\mathrm{in}\leq {\varrho}_\mathrm{ ij} \right|_{(f=0)}  \leq R_\mathrm{out}
 \label{bbins}
,\end{equation} 
  with steps
of $(R_\mathrm{out}-R_\mathrm{in})/(N)$. For the simulations, we used $N=100$ because of computational memory requirements.
The orientation of each annulus (i.e. their $\Omega_\mathrm{c},  \omega_\mathrm{c}  $) in both discs was chosen so that the annulus apocenters faced each other. 
 This configuration was chosen for several reasons. Recent multidimensional numerical simulations \citep[see e.g.][]{2014ApJ...783..134F, 2016ApJ...827...43M} have clearly demonstrated that individual
mini discs form around each SMBH over many binary orbital periods.
\cite{1995ApJ...438..610E} suggested that  the tidal effects of the secondary
black hole on a disc around the primary could be analogous to the effects of
the secondary star in a cataclysmic variable on the accretion disc around
a white dwarf. In these studies it is generally found that for higher mass
ratios (>0.25), the disc is unstable to tidal perturbations. The same
instability in SMBBH could cause the disc to elongate in response to the
tidal field of the secondary.
\cite{2008ApJ...682.1134H} performed high-resolution smoothed particle
hydrodynamics (SPH ) simulations of  equal-mass SMBBH of moderate orbital
eccentricity (0.5) surrounded by a circumbinary disc. Their study showed that periodic mass transfer causes the mini discs
to become eccentric because the gas particles from the circumbinary disc originally
have elliptical orbits around the black holes. This enables
accretion directly onto the black holes during the binary orbit. Finally,
the discs are tidally deformed by the time-varying binary potential owing
to the orbital eccentricity. The apocenters of the discs face each other
while SMBHs sit in pericenter and apocenter of their orbital
configuration. Only in the beginning of accretion is the gas added to the
outer parts of discs (see their Fig. 2).
Recently, \cite{2017ApJ...838...42B, 2018ApJ...853L..17B} reported the first simulations of
mini-disc dynamics   for a binary consisting of  an equal-mass pair of
non-spinning SMBHs when their mutual separation is small enough, causing  the
mini discs to stretch toward the L1 point  of a binary system.

 For notation simplicity, from now on we drop the iterative indexes $ i$  and $j$.
To provide a more comprehensive practical illustration of our SMBBH geometrical model set-up, we applied it to the coplanar case of binary SMBHs whose elliptical disc-like BLRs  are aligned with the binary common orbital plane. Each disc-like BLR is defined by the inner and outer radius, and an SMBH is located in its focus. The material in the discs rotates around the SMBH on the elliptical  orbits with Keplerian rotation velocity (see Figure \ref{fig:cosis}(b)).
The SMBH  masses are
$M_{1}=10^{8} M_{\odot}$ and $ M_{2}=0.5 \times 10^{8} M_{\odot}$.   Their orbital eccentricities  are 0.5 and the semimajor axis of the relative orbit    is not larger than 30 ld.   The inner and outer edges for the smaller and larger BLR are (4, 10) and (7, 15) ld, respectively. The eccentricities of  the cloud orbits are 0.5.

In addition to a coplanar SMBBH, we also    consider a non-coplanar case, when the SMBH orbits are mutually inclined.
Such cases are well known in the multiple star population \citep[see e.g.][]{2016AJ....152..213S}.
In hierarchical formation scenarios, if  binary SMBH do not coalesce before the  merger with a third galaxy, the formation of  a triple SMBH system is possible. Most of such systems could be  long-lived ($\sim 10^9 $ year) \citep[see][and references therein]{doi.org/10.1111/j.1365-2966.2009.16104.x}. 
However, we  assumed a simple scenario, where mutually inclined SMBH orbits arise due  to perturbation during a close encounter with some unseen massive object (such as an SMBH). This perturber has a much  longer orbital period  than the two components in a subparsec binary.  The third object would be more distant and far away from the gravitational sphere of influence of  SMBHs. Thus, we can omit a long-period perturber from consideration. This seems  not an unrealistic  assumption because there is  an observed triple system of supermassive black holes, SDSS J150243.09+111557.3,  at redshift $z=0.39$, in which the closest pair is separated
by $\sim 140$ pc and  the third active nucleus is at 7.4 kpc \citep{DOI:10.1038/nature13454}.  In this case, the third SMBH is very far away from the gravitational influence of the tight pair because  the gravitational sphere of influence of a black hole with masses  of $10^6 M\odot$  (Sagitarius A) is about 3 to 5 pc
  \citep{http://www.aspbooks.org/publications/439/129.pdf} and  $10^9 M\odot$ is  about 100 pc  \citep{DOI:10.1038/nature13454}. In this case, the  reference plane is  the orbital plane of  one  SMBH in the binary system and the frame of origin is set  to the barycenter of the binary.

 \subsection{Kinematics of the disc-like BLR model} \label{kin}
 
\begin{figure*}[ht!]
\includegraphics[width=0.7\textwidth]{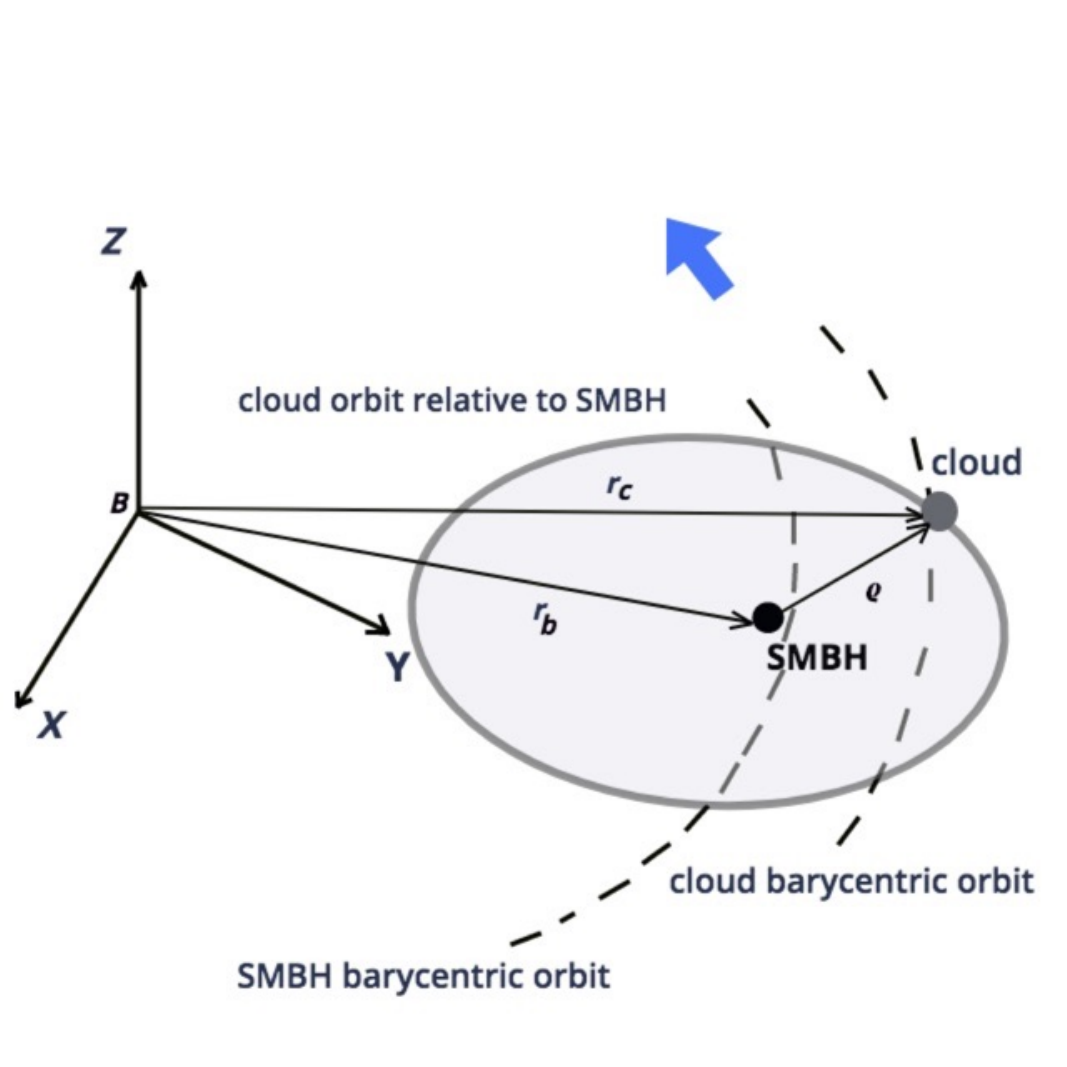}
\vspace{1pt}
\caption{Motion of a cloud in an elliptical disc-like BLR around an SMBH in the barycentric coordinate frame. The blue arrow denotes the direction of the SMBH motion around the barycenter of the SMBBH system.\label{fig:sistem1}}
\end{figure*}

 Now we derive the  velocity of a cloud  in the barycentric frame  BXYZ  (see Fig.
  \ref{fig:sistem1}). Here we list the formulas only for a cloud in a disc-like BLR of the more massive SMBH, and the derivation  for less massive SMBH  is similar.
   For the purpose of derivation, we rearrange  Eq. \ref{rdisk} so that the  vector  of the relative position of a cloud  $[\vec{\varrho}]_\mathrm{B}]$  in barycentric frame is given as
\begin{equation}
[\vec{\varrho}]_\mathrm{B}=\vec{r}_{\mathrm c}-\vec{r}_{\mathrm b}
\label{disk2}
 ,\end{equation}
where $\vec{r}_{\mathrm b}$ and $\vec{r}_{\mathrm c}$ are barycentric positions of the SMBH and the cloud, respectively.

Simply from the transport theorem, we have that the barycentric velocity of a cloud 
is given by 
\begin{equation}
[\dot{\vec{\varrho}}]_\mathrm{B}=\dot{\vec{\varrho}}_{\bullet}+\vec{\omega}_{\bullet/\mathrm{B}}\times \vec{\varrho}_{\bullet}
\label{barvel}
 ,\end{equation}
where  $\dot{\vec{\varrho}}_{\bullet}$ is the velocity  vector  of the cloud relative to the SMBH in the comoving frame of the SMBH (see Appendix \ref{app:expandtauv}), and $\vec{\omega}_{\bullet/\mathrm{B}}$ is equivalent to the angular velocity vector of the SMBH at each instant, which can be calculated from the known SMBH orbital angular momentum:
\begin{equation}
\vec{\omega}_{\bullet/\mathrm{B}}=\frac{\vec{r}_{\mathrm b}\times{\dot{\vec{r}}_{\mathrm b}}}{\vec{r}_{\mathrm b}\cdot \vec{r}_{\mathrm b}}
\label{bulet}
.\end{equation}
We note that $ \vec{\omega}_{\bullet/\mathrm{B}}$ can also be calculated using the barycentric orbital elements of the SMBH (see  Eq.\ref{L}).
We  can rephrase Eq. \ref{bulet}   in our set-up:

\begin{equation}
{\vec{V}}={\vec{V}}_{\bullet}+\vec{\omega}_{\bullet/\mathrm{B}}\times \vec{\varrho}_{\bullet}
\label{barvel1}
 ,\end{equation}
where   ${\vec{V}}_{\bullet}=\dot{\vec{\varrho}}_{\bullet}$. The relative state vectors ($\vec{\varrho}_{\bullet}$ and     ${\vec{V}}_{\bullet}$) of the cloud are given by Eqs. \ref{comorho} and \ref{comovel}, respectively.

Having in mind  Eq. \ref{disk2}, the time delay $\tau$  of each cloud is obtained  as 
\begin{equation}
\tau=\frac{|[\vec{\varrho}]_\mathrm{B}|+\vec{n}_{\mathrm obs}\cdot[\vec{\varrho}]_\mathrm{B}}{c},
\label{delay}
\end{equation}
 where $|[\vec{\varrho}]_\mathrm{B}|$ is the norm of the relative position vector of a cloud in the barycentric frame and $\vec{n}_{\mathrm obs}=(0, -\sin{i_{0}},-\cos{i_{0}})$   is the  barycentric  vector of the observer's line of sight defined by the angle $i_0$. The relative  barycentric position of the cloud can be obtained by transforming its relative position in the local comoving frame given by Eq. \ref{qrhotrans}.

The velocity of the cloud (see Eq. \ref{barvel1}) projected onto the direction $\vec{n}_{\mathrm obs}$ is 
\begin{equation}
V_\mathrm{z}={\vec{V}}\cdot \vec{n}_{\mathrm obs}
\label{radvel}
.\end{equation}
Expansions of equations for the projections of $\vec{\varrho} $ and velocity $\vec{V_\mathrm{z}}$ are given in Appendix \ref{app:expandtauv}.

\subsection{Simple and composite transfer function}

In the simple linear theory, the broad emission-line  radial velocity $ V_\mathrm{z}$ (given by Eq. \ref{radvel}) and time-dependent response $\mathcal {L}(V_\mathrm{z},t)$ is a convolution of prior time-delayed continuum variations $\mathcal{C}(t-\tau)$ with a transfer function $\Psi(V_\mathrm{z},\tau)$ such that \citep{1982ApJ...255..419B}

\begin{equation}
\mathcal{L}(V_\mathrm{z},\tau)=\int_{-\infty}^{+\infty}\mathcal{C}(t-\tau) \Psi(V_\mathrm{z},\tau) d\tau 
\label{bland}
.\end{equation}
The transfer function is a projection of 6D (three spatial and three kinematical) phase space distribution into 2D phase space  (defined by the radial velocity $V_\mathrm z$ and the time lag $\tau$).
The contribution of a particular cloud in a  single BLR  to  the overall response depends  on three parameters:  its distance from the continuum source (setting
the time delay of its response),  its radial velocity (i.e. the velocity at which its response is observed), and  the emissivity (the parameter describing the efficiency of the cloud in reprocessing the continuum into line photons in steady state).
Based on the above discussion, we can predict the response of an emission line to continuum variations for any
physical description  of the BLR  and probe  the modelled transfer functions by comparing them  to those that are inferred 
from observations. We aim  here  to predict these observational reverberation signatures for an elliptical disc
model of the BLR of an SMBBH on elliptical orbits.

Thus the transfer function for a single elliptical disc can be written as follows:
\begin{equation}
\Psi(v,\tau) =\int \epsilon({\varrho} )\delta(v-\vec{V} \vec{n}_{\mathrm {obs}}) \delta(ct-(|\vec{\varrho}|-\vec{\varrho}\cdot  \vec{n}_{\mathrm {obs}})) \mathrm{d} {\vec{\varrho}}\, \mathrm{d}\vec{V}
,\end{equation}
where $ \epsilon{\varrho}))$ is the (assumed isotropic) responding volume
emissivity of the emission region as a function of position,
and $\vec{\varrho}, \vec{V}$ are the barycentric state vectors of a cloud, but for clarity, the subscript $B$ is omitted.
We adopted the emissivity law $\epsilon(\varrho)=\epsilon_{0} \varrho^{-q}$ \citep[see][and references therein]{1995ApJ...438..610E} for calculating the 2DTF, where
$\varrho$ is the polar form of the trajectory of the cloud determined for a given time span starting from the solution of the Kepler equation. Because the trajectory is  an ellipse, it implies that
emissivity varies both with radial distance and the given true or eccentric  anomaly and pericenter position of  the cloud.
The parameter $q$ can take different values \citep[see e.g.][]{1995ApJ...438..610E,2004A&A...423..909P,af19}.  Here we used $q=$2.5  \citep{1995ApJ...438..610E}, which is expected in the case of moderate elongated annuli ($e<\sim 0.55$).

Because the  orbital plane  of a cloud is defined by  inclination ($i$) and longitude of the ascending node ($\Omega $), the transfer function of the elliptical disc can be given  as follows:

\begin{equation}
\Psi(v,\tau) =\epsilon_{0} \int_{R_{\mathrm in}}^{R_{\mathrm {out}}}  \varrho ^{-q} \mathrm{d}\varrho \int_{0}^{2\pi} \mathrm{d} \Omega \int_{-i_{\mathrm{ min}}}^{i_{\mathrm {max}}} \sin i \mathrm{d}i \int_{0}^{2\pi}\delta (X_1)\delta(X_2) \mathrm{d}E
,\end{equation}
where $X_{1}=v-{V}_\mathrm{z}, X_{2}=ct-c\tau, $ and $E$ is the eccentric anomaly  of the cloud on its orbital plane.
Limits of integration $ i_{\mathrm {min}},i_{\mathrm {max}}$ indicate the range of the orbit inclination of the cloud in a disc-like BLR, so that $\Theta=|i_{\mathrm {max}}-i_{\mathrm {min}}|$.

Based on prescription from Paper I, the composite transfer function for an SMBBH system is obtained by calculating  $\Psi_{1}(v,\tau)$ and  $\Psi_{2}(v,\tau)$ for each BLR and coupling them as follows:
 
\begin{equation}
\Psi(v,\tau)_{\mathrm coupled} =\frac{\Psi_{1}(v,\tau)}{1+\Gamma_0}+\frac{\Psi_{2}(v,\tau)}{1+\Gamma^{-1}_0}
,\end{equation}
where $\Gamma_0$ is the coupling factor that is obtained by normalisation of the continuum variation of one of  the SMBHs  with the continuum of the other SMBH. Here we used  the constant $\Gamma_{0} \sim 1$ as the simplest case when the binary black holes have the same properties as the continuum variations.

\section{Results and discussion: Reverberation signatures of elliptical BLR model } \label{sec:discus}

The aim of this study
is  the differentiation  of elliptical disc-like BLR models on the basis
of  2DTF maps, therefore we calculated these maps for  various adopted  orbital configurations of the clouds and binary components. The time -averaged line profiles are also calculated and compared to  those observed and  inferred from disc-wind models.

For the initial conditions, we  considered an   SMBH binary with the same parameters as given in Paper I. The masses of the components are
$M_{1}=1\times 10^{8} M{\odot} \text{ and } M_{2}=0.5\times10^{8} M{\odot}$.  We let  the pericenters of the cloud orbits  be uniformly distributed for the BLR around the primary  $(R_{\mathrm {in}}, R_{\mathrm {out}})_{\mathrm c1}=(7,15)$ ld   and  secondary component  $(R_{\mathrm in}, R_{\mathrm {out}})_{\mathrm c2}=(4,10)$ ld  (see Eq.  \ref{bbins}). We generated 100 cloud orbits with  a resolution of 1000 points each within the range of the given inner and outer boundaries of the disc-like BLRs.  The  inclination range for the cloud trajectories in  the two disc-like BLR  are $\Theta=5^ {\circ}$. This assumption  is in agreement with the hypothesis that a near-coplanar accretion disc  and BLR could be expected  because in gas-rich mergers, the evolution of the SMBBHs is due  to  interaction with the surrounding gas, so that the accretion onto
the black holes  align  their spins with
the angular momentum of the binary \citep{2007ApJ...661L.147B}. Our elliptical BLR models and binary geometry should be reduced to circular models when the eccentricity and  the orbital orientation parameters are set to zero. Thus, the parameters  presented in Paper I  were used as a consistency check for our model.  In this set-up, each mini-disc is illuminated by the continuum source at
its own centre and not by the continuum source in the other disc.

In particular we considered models (i) with non-randomised orbital elements of clouds  and  (ii) with  randomised eccentricities and/or an orbital orientation of the clouds. 
Specifically, for  type (ii) simulations, we assumed that at the beginning, 
the clouds have random eccentricities and/or angles of orbital orientation.
While eccentricities  (and/or angles of orbital orientation) of clouds
were chosen at random, their semimajor axes and inclinations were kept
at their values as in the non- randomised case, thus ensuring non-intersecting
orbital planes. In addition,   clouds are considered to  be dimensionless
points whose motion has been followed on timescales of
decades. 
 We note here that our type (ii) model represents an  idealised case where no cloud collision is taken into account.
However, randomised cloud  motion can be  more complex than we have
assumed in
this work, and collisions between clouds could be present.
Firstly, in a real situation the size of the clouds cannot be taken to be arbitrarily small.
The BLR covering factor (i.e. the
fraction of the sky that is covered by BLR clouds, as seen by the central source)
is proportional to  the space density of clouds \citep{1985aagq.conf..185M}. On the other hand, the averaged BLR  covering factor is
determined by estimating the fraction of ionising continuum photons
that are absorbed by BLR clouds and  reprocessed into broad emission lines, and it
is constrained to be of  $\sim$ 10$\%$ \citep[using  the  equivalent widths
of the broad emission lines, see e.g.][]{https://doi.org/10.1007/3-540-34621-X_3}.
 Combining these two facts means that clouds on randomised orbits collide
in about one
orbital period \citep{1985aagq.conf..185M}.  As a result, the system of clouds (or fluid elements) cannot remain on elliptical orbits  for more than a few dynamical times, and it will settle down in a different configuration where the orbits are circular. When the  elliptical orbits are to survive, they  must therefore not cross.
Second, as suggested by \cite{1985aagq.conf..185M} collisions may be reduced
if the  clouds resemble the motion of school of fish on a nearly circular orbit.
Having this in mind,   our model imputes a certain order  in the cloud motion
by constraints on their semimajor axes and inclinations, which can reduce the collision rates.
For example,  based on a geometrical approach  \citep{C1996A&A...308L..17C, DOI: 10.1051/0004-6361:20040527}, we can
roughly estimate  that  the collision rate of clouds is not so high (of about some hundred
collisions per year for $10^{10}$ clouds at a distance of $\sim 10$ ld from the SMBH),
and probably
that a small fraction of  clouds on randomised orbits (at the beginning of
our simulations)
may collide during one orbital period, and after this, the emitted clouds follow orbits with a low probability of colliding. Otherwise, if the semimajor axes and inclinations could be random as well, the probability would be higher of crossing (intersecting) orbits, and
collisions between clouds would be inevitable.
In any
case, the cloud collisions during randomised motion and implications on orbital stability will be considered in some further work, however.

A first glance reveals that  the resulting response functions (for single and binary systems) are complex and differ substantially from those estimated by spherical disc-like BLR  models in single  SMBH and those in  circular binaries.  The geometry and kinematics of
the elliptical BLR models determine the appearance of these maps. In the following  subsections, we present a detailed description and discussion of the inferred 2DTF of a single SMBH, a binary SMBH system,
and time-averaged line profiles for each of the given maps.

\subsection{Simple 2DTF}
\begin{figure*}
\centering
\begin{subfigure}{6cm}
    \centering
    \includegraphics[width=5cm]{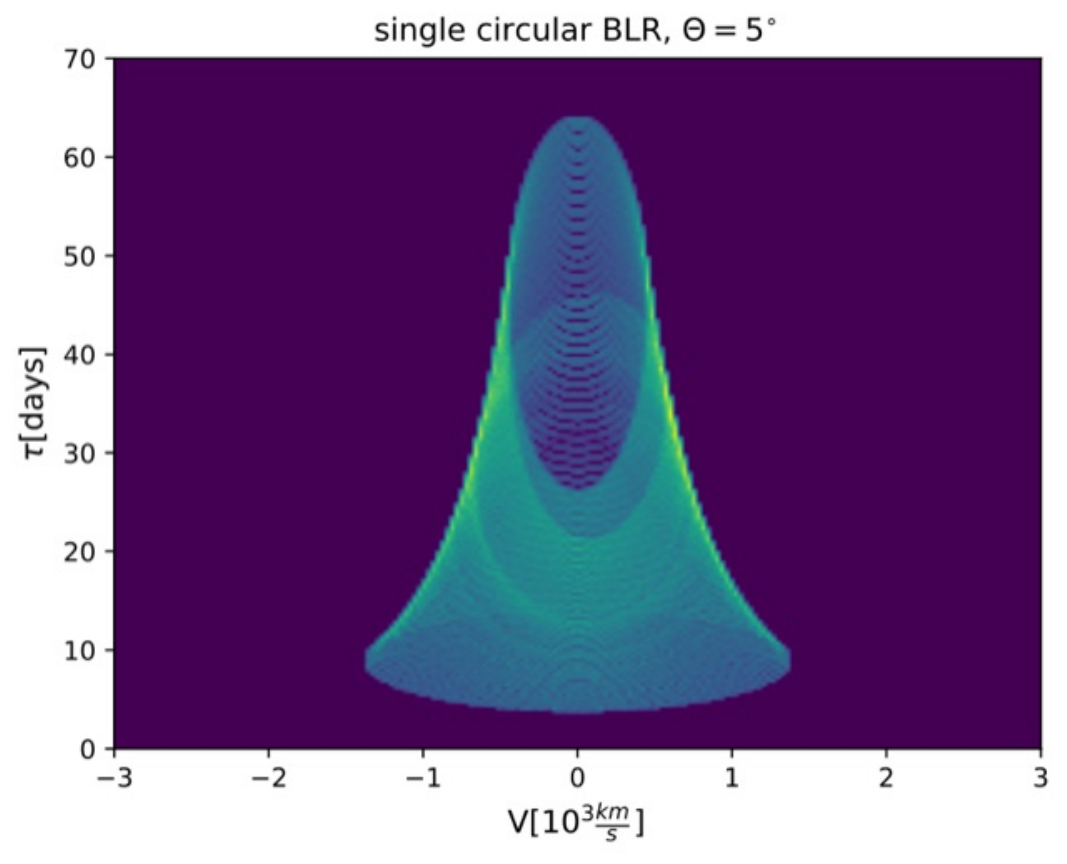}
    \caption{}
\end{subfigure}%
\begin{subfigure}{6cm}
    \centering
    \includegraphics[width=5cm]{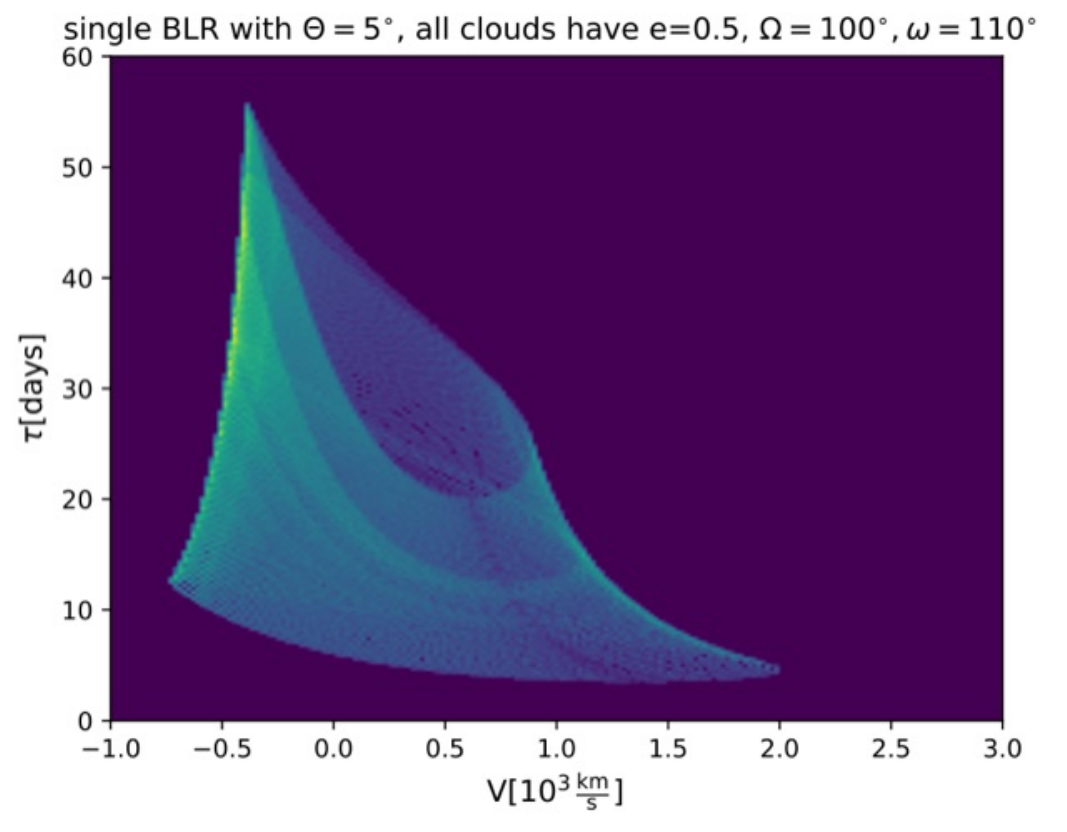}
    \caption{}
\end{subfigure}
\begin{subfigure}{6cm}
    \centering
    \includegraphics[width=5cm]{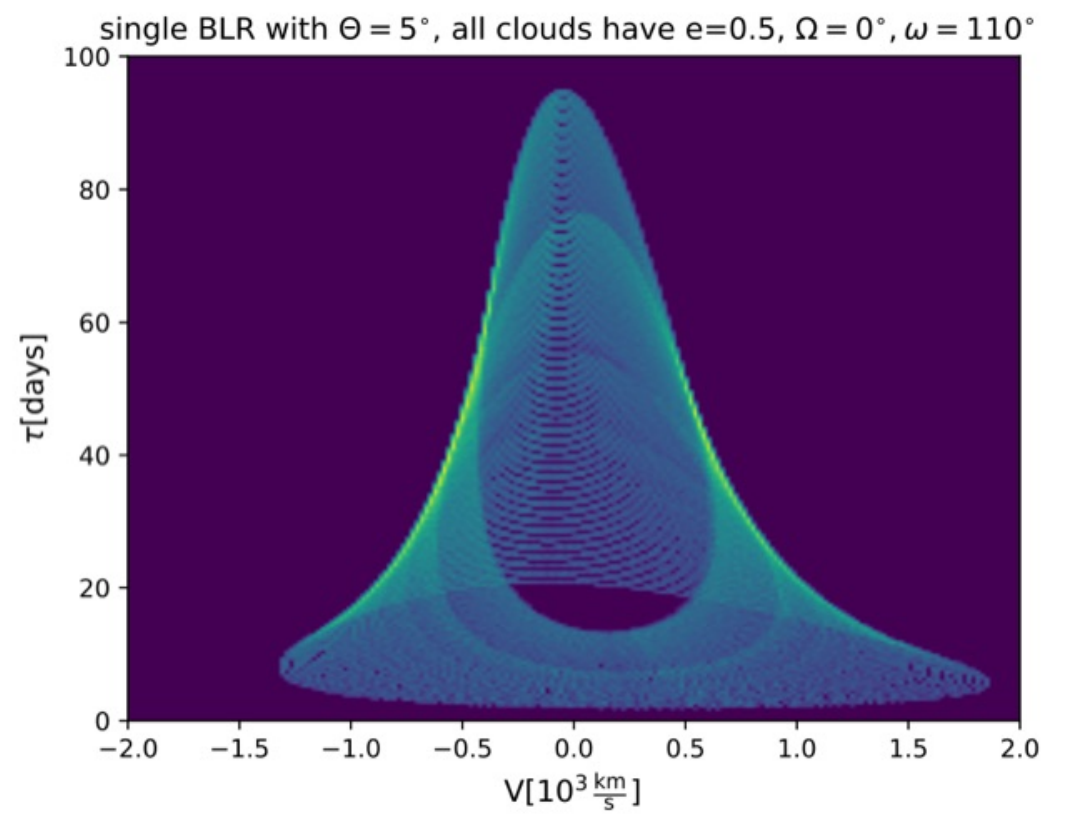}
    \caption{}
\end{subfigure}
\begin{subfigure}{6cm}
    \centering
    \includegraphics[width=5cm]{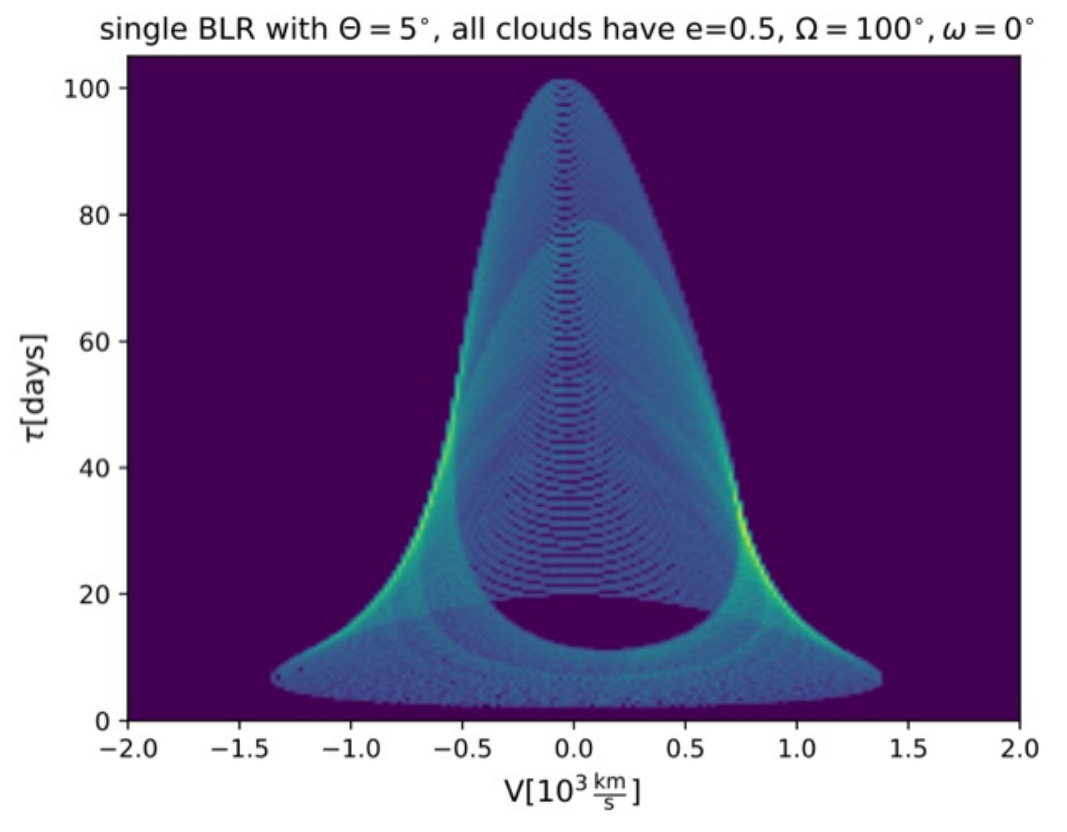}
    \caption{}
\end{subfigure}
\begin{subfigure}{6cm}
    \centering
    \includegraphics[width=5cm]{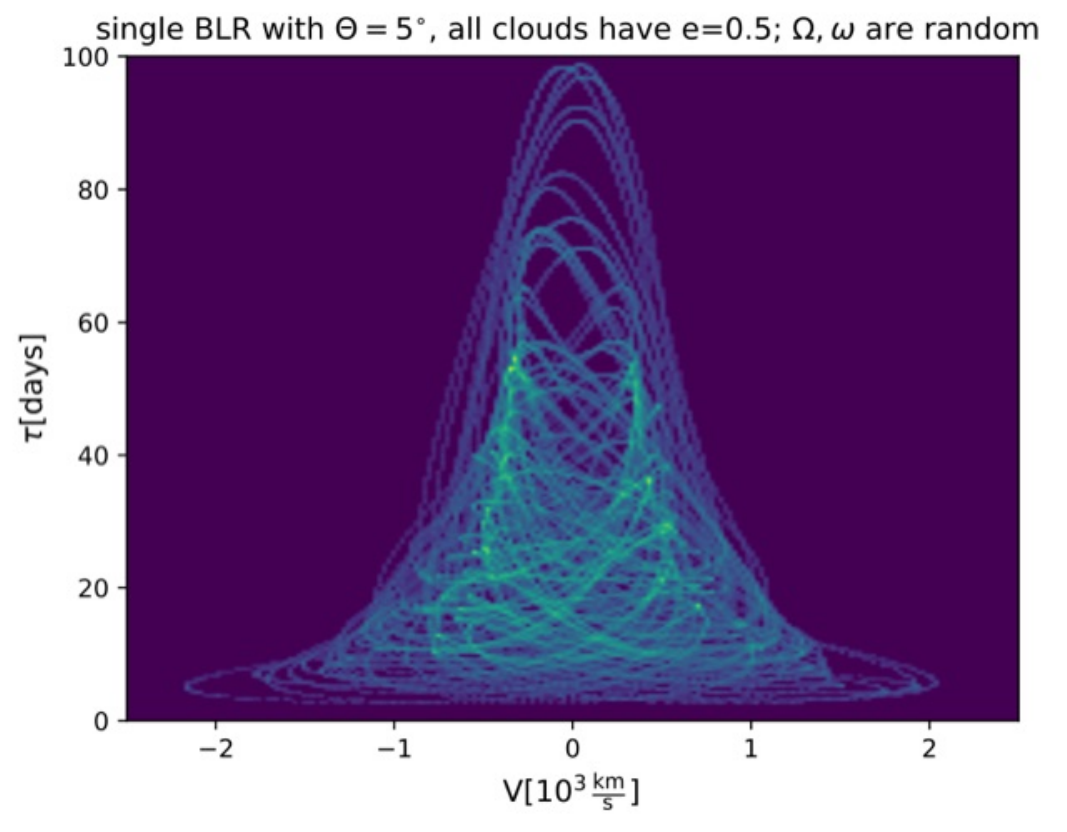}
    \caption{}
\end{subfigure}
\begin{subfigure}{6cm}
    \centering
    \includegraphics[width=5cm]{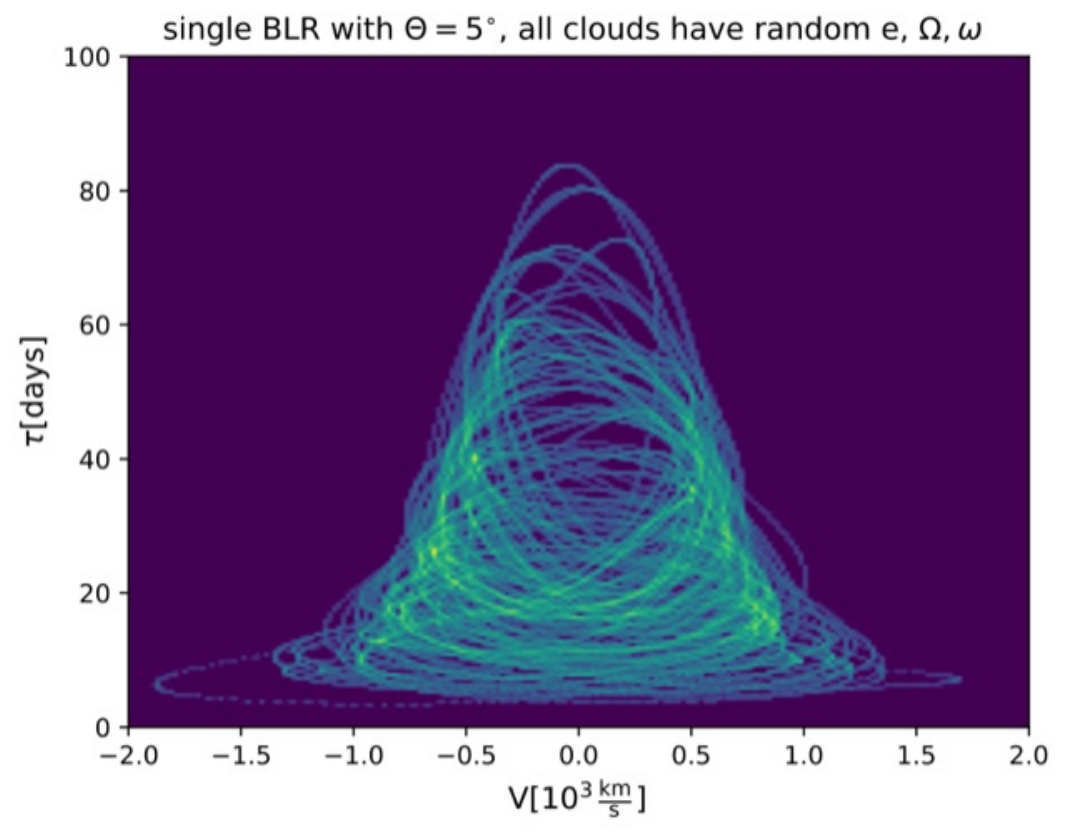}
    \caption{}
\end{subfigure}
\begin{subfigure}{6cm}
    \centering
    \includegraphics[width=5cm]{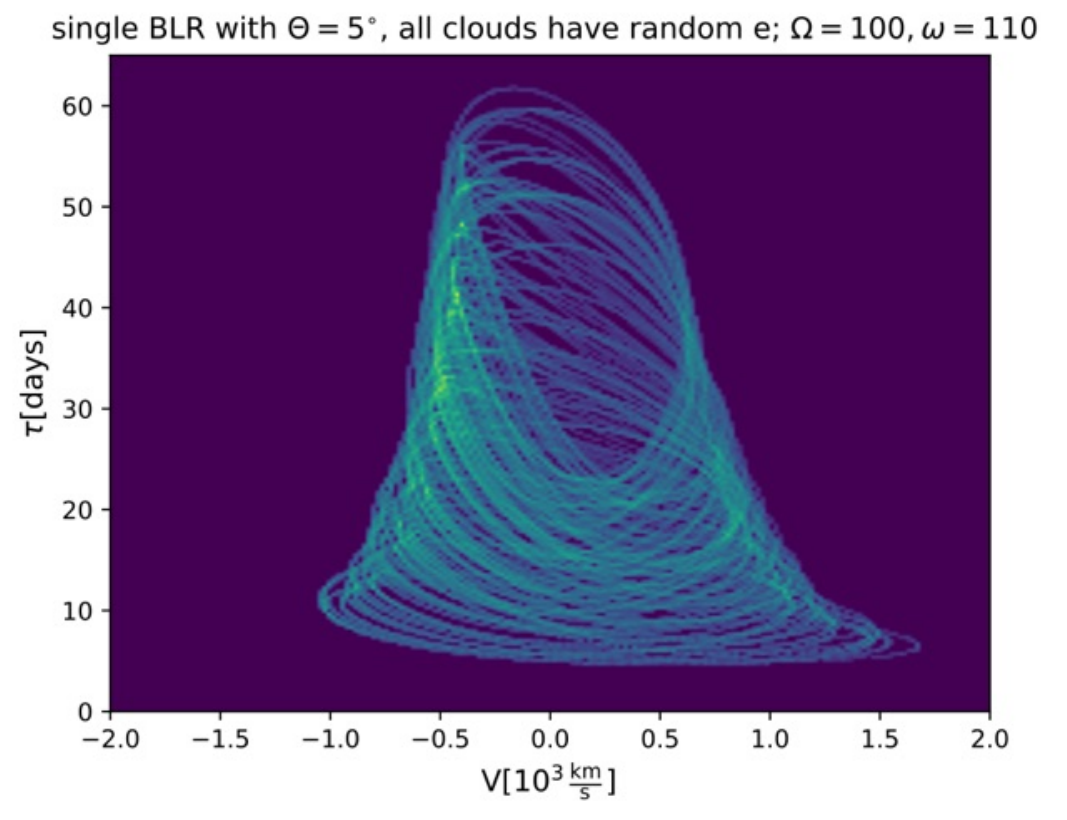}
    \caption{}
\end{subfigure}
\begin{subfigure}{6cm}
\centering
    \includegraphics[width=5cm]{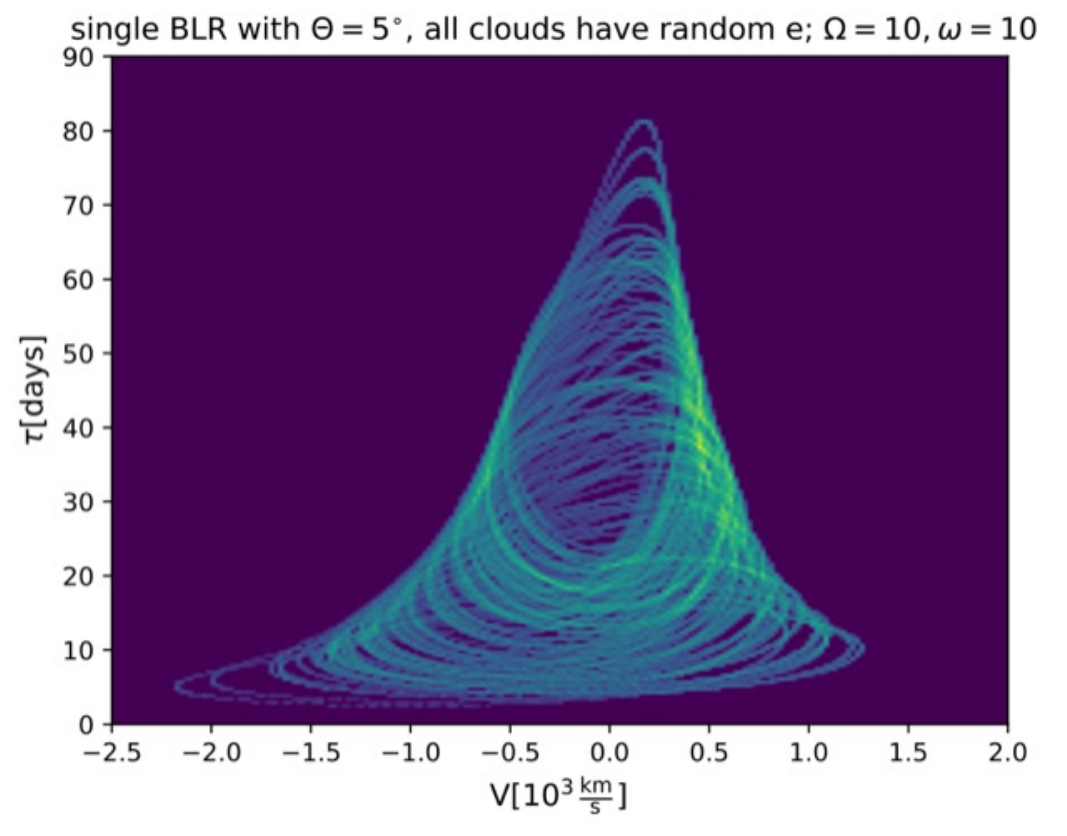}
    \caption{}
\end{subfigure}
\begin{subfigure}{6cm}
  \hspace*{\fill}%
  \end{subfigure}
  \caption{2DTF  maps obtained for different geometries of a single disc-like BLR. The panels show signatures of (a) circular  BLR as obtained using the parameters  from  Paper I   (see text).  Compare this with Fig. 2  in Paper I.   (b) An elliptical BLR with { clouds} orbital parameters $e=0.5, \Omega=100^{\circ},\text{and }\omega=110^{\circ}$. (c) The same orbital { configurations} of the { clouds} as given in  (b), but with  $\Omega=0^{\circ}$; (d) the same orbital parameters as given in  (b), but with $\omega=0^{\circ}$; (e) the same as panel (b), but the  angles of the orbital orientation of { clouds} are random; (f) all three orbital parameters are random; (g) the orbital eccentricities of the { clouds} are random, but  the orientation is the same as in panel (b);   (h) the orbital eccentricities of the { clouds} are random, but $\Omega=10^{\circ} \text{ and } \omega=10^{\circ}$.  We did not consider the stability of the elliptical orbits in the randomised motion. }
\label{fig:tf1}
\end{figure*}

The 2DTF maps  of a  single   elliptical BLR  that we presented in  Fig. \ref{fig:tf1}  have previously not been discussed, to our knowledge.
Each map consists of two parts, a central region (2DTF core) and low flat wings on either side. A closer look reveals important nuances, however.

Panel (a) shows the 2DTF that was obtained  from our model using  parameters for a single circular BLR ($\Omega=\omega=0^{\circ}$, and $e=0$)  as given in Paper I  (see their Fig 2 (a)).
 The effects of eccentricity and orientation of the cloud  elliptical orbits are illustrated in panels (b)$-$(h). 
 The   shape of the maps is clearly asymmetric, where the well-known  bell-like structure  of the circular case is deformed. 
 The orientation of the trajectories of the clouds visibly change  the orientation  of the  map cores. It is interesting that both the longitude of ascending node and the angle of periastron are important for the map orientation. The effects of randomisation of eccentricities and orientation of orbits are shown in  panels (e)-(h) (as we discussed above,  in our model of randomised motion we did not
 consider the cloud collisions, which should be taken  into
 account, and we postpone this consideration to a future study). The filaments  of the 2DTF  displayed a more chaotic appearance  when  the orbital orientations  of clouds were randomised (e)  in comparison with  the regular and robust  maps when only orbital  eccentricities are random  ((g), (h)).
The elongation of the red and blue wings depends on the angle of the pericenter.
The red wing is more prominent when the angle of  the pericenter increases and vice versa. The same holds when we consider that only eccentricities are random. Additionally, there is asymmetry in  the monotonic appearance of  the  left  and right sides of the 2DTF core. As shown in panels (b), (c), (d), and (g), the slopes of the left wings are steeper then the right  ones, and this is controlled  by the orbital orientation angles. For example, less steeply sloping  gradients  of the left  wings occur when both orientation angles have low values  (see panel (h)). 
The asymmetric  2DTF inferred from asymmetric models for the Seyfert 1 galaxy Mrk 50 \citep{10.1088/0004-637X/754/1/49}  is similar to the map in panel (h).

\subsection{Composite 2DTF}

\begin{figure*}
\centering
\begin{subfigure}{6cm}
    \centering
    \includegraphics[width=5cm]{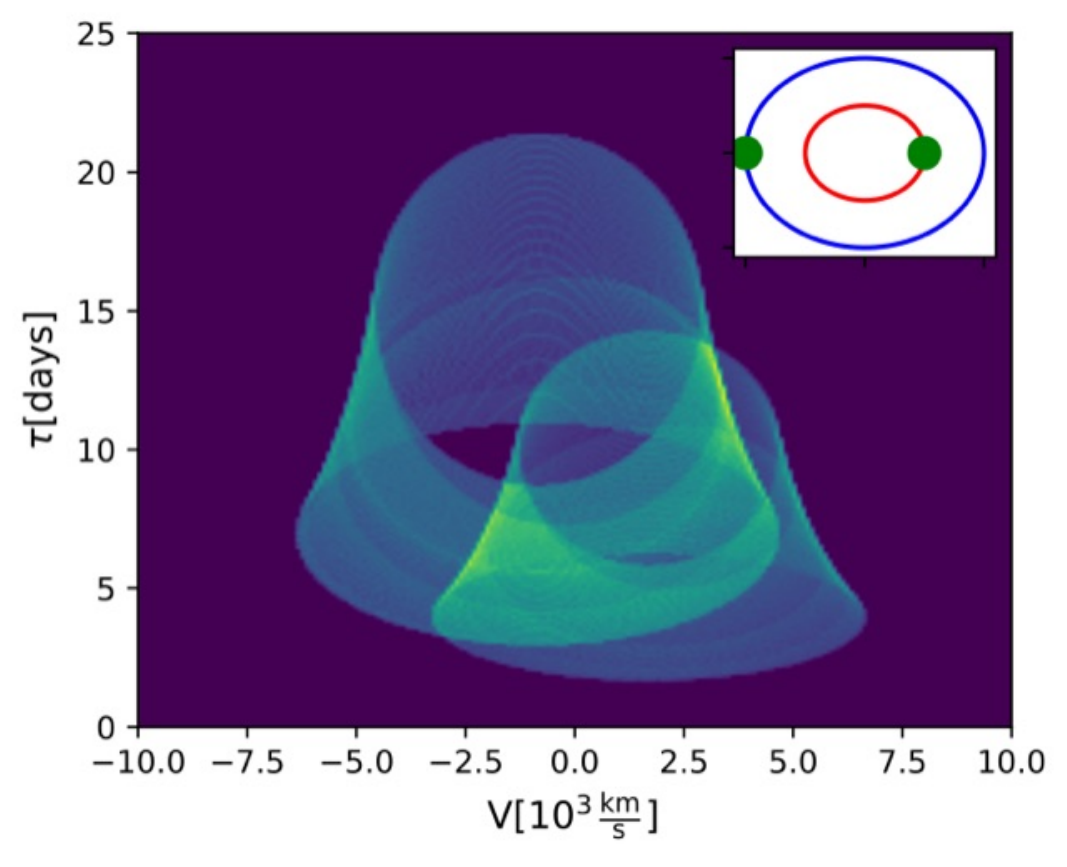}
    \caption{}
\end{subfigure}%
\begin{subfigure}{6cm}
    \centering
    \includegraphics[width=5cm]{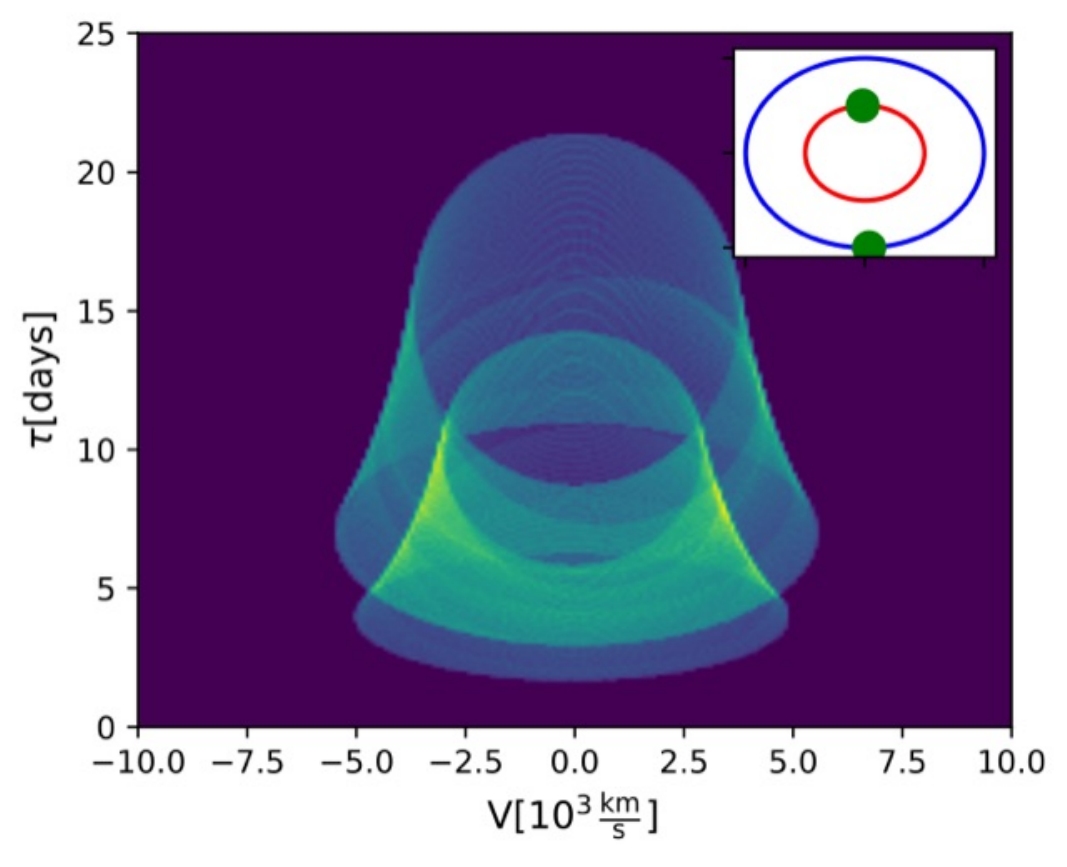}
    \caption{}
\end{subfigure}
\begin{subfigure}{6cm}
    \centering
    \includegraphics[width=5cm]{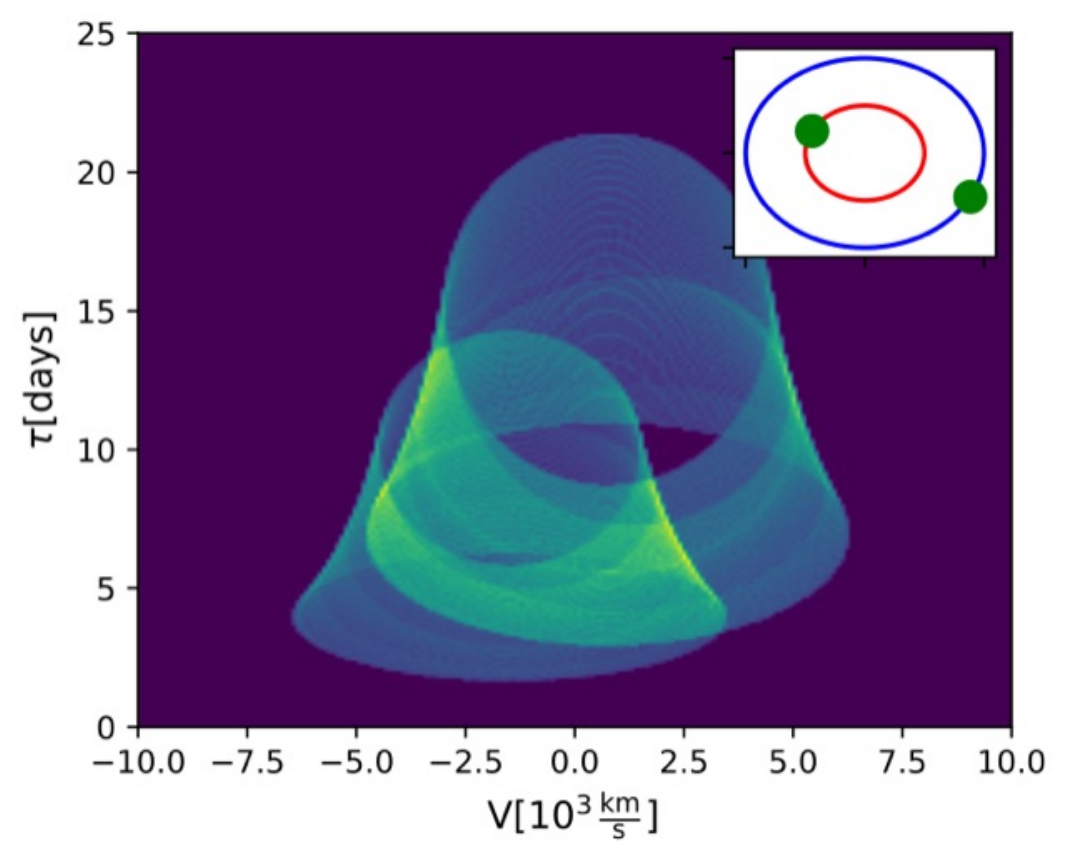}
    \caption{}
\end{subfigure}
\begin{subfigure}{6cm}
    \centering
    \includegraphics[width=5cm]{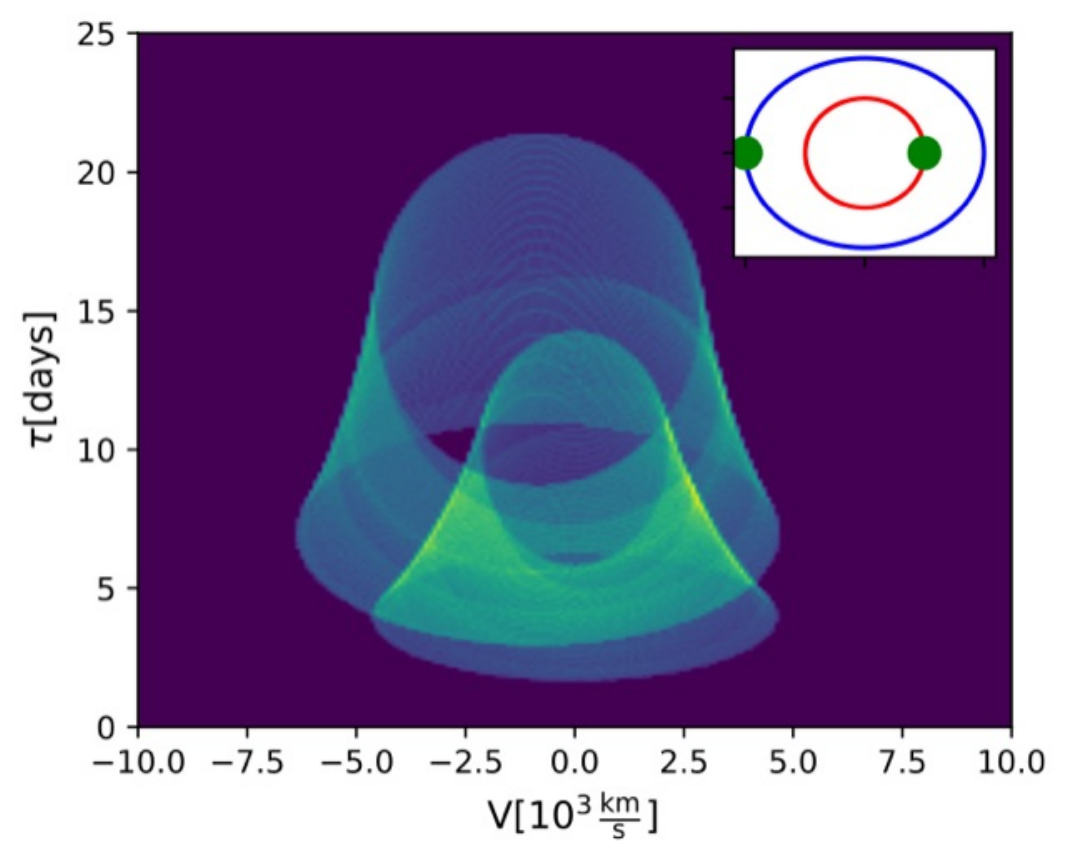}
    \caption{}
\end{subfigure}
\begin{subfigure}{6cm}
    \centering
    \includegraphics[width=5cm]{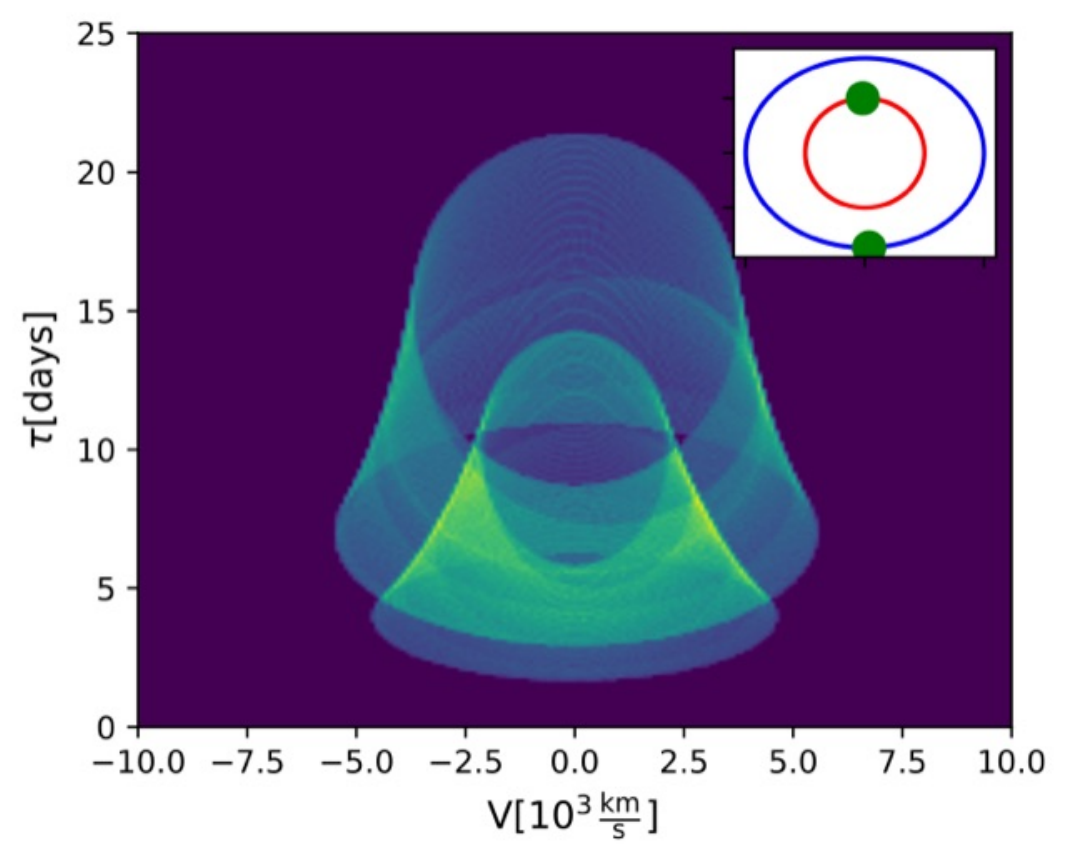}
    \caption{}
\end{subfigure}
\begin{subfigure}{6cm}
    \centering
    \includegraphics[width=5cm]{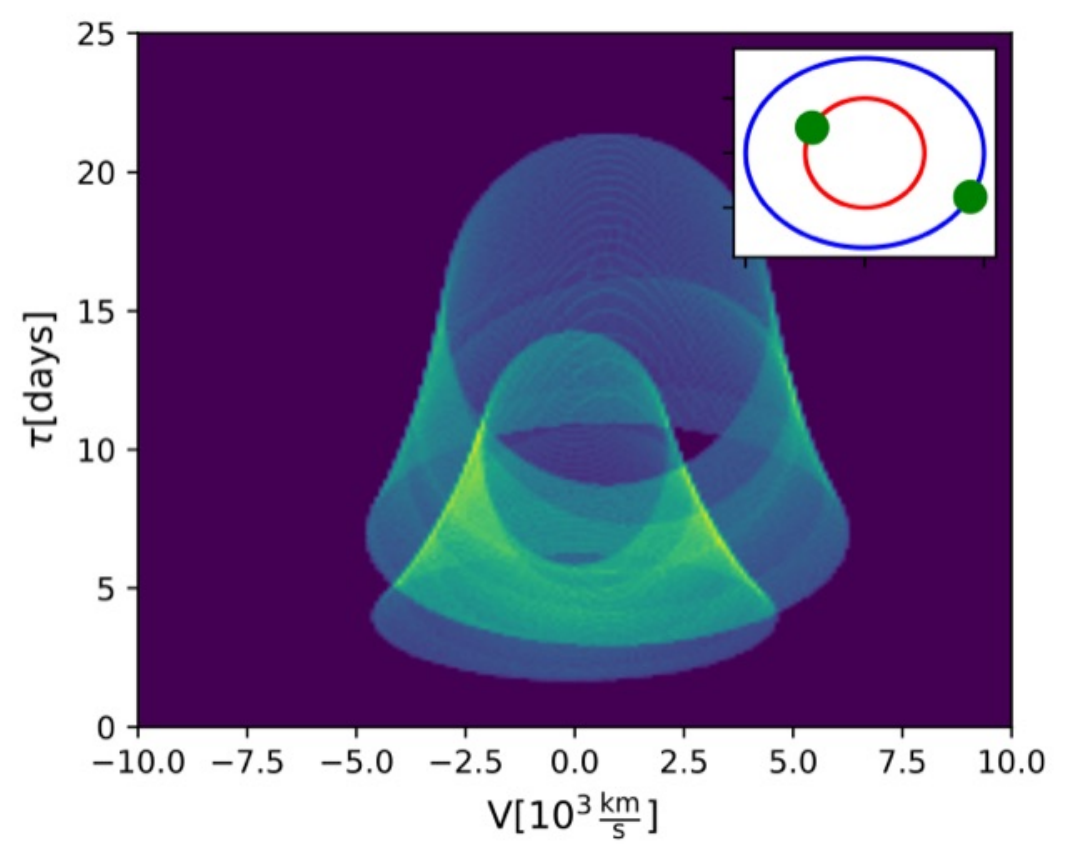}
    \caption{}
\end{subfigure}
\begin{subfigure}{6cm}
    \centering
    \includegraphics[width=5cm]{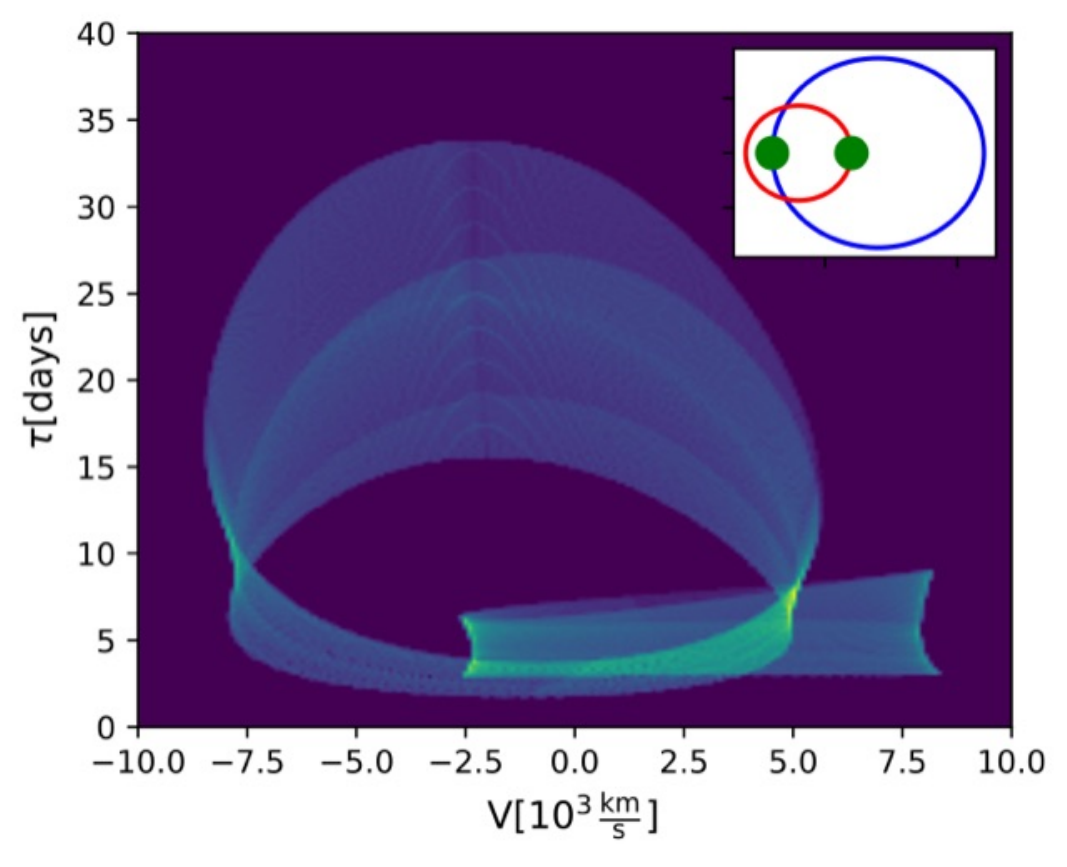}
    \caption{}
\end{subfigure}
\begin{subfigure}{6cm}
\centering
    \includegraphics[width=5cm]{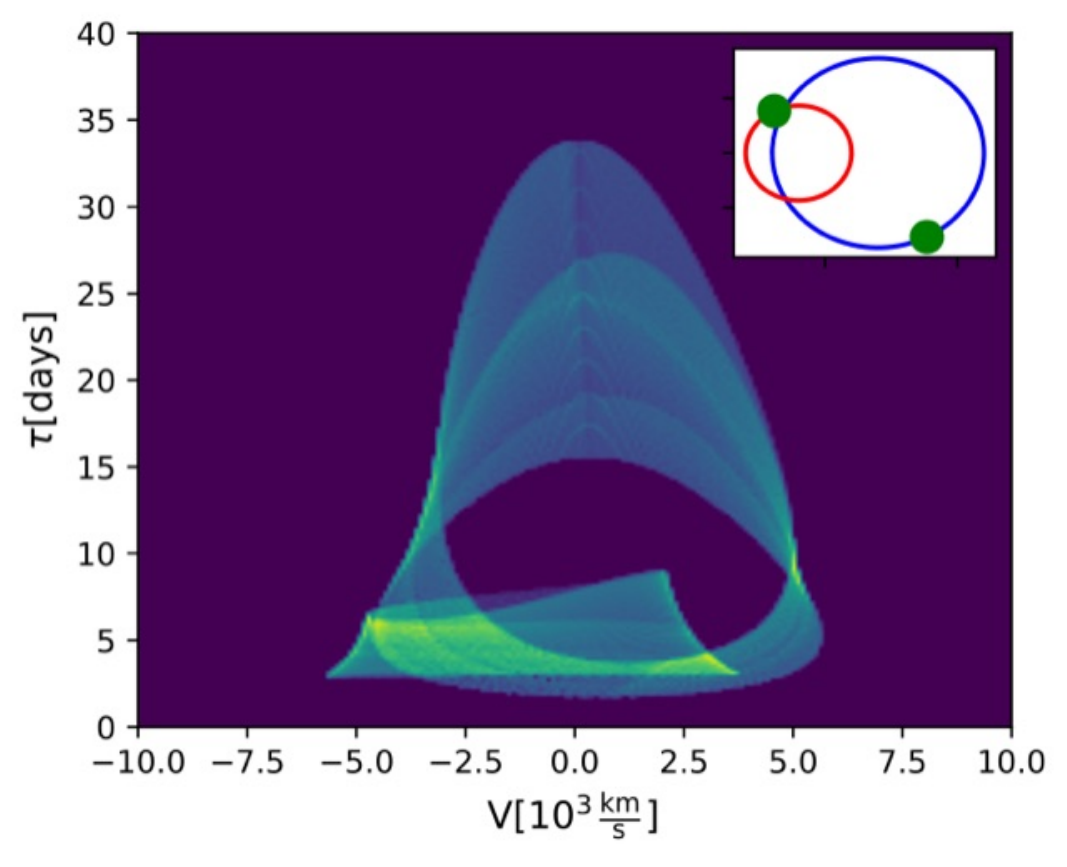}
    \caption{}
\end{subfigure}
\begin{subfigure}{6cm}
\centering
    \includegraphics[width=5cm]{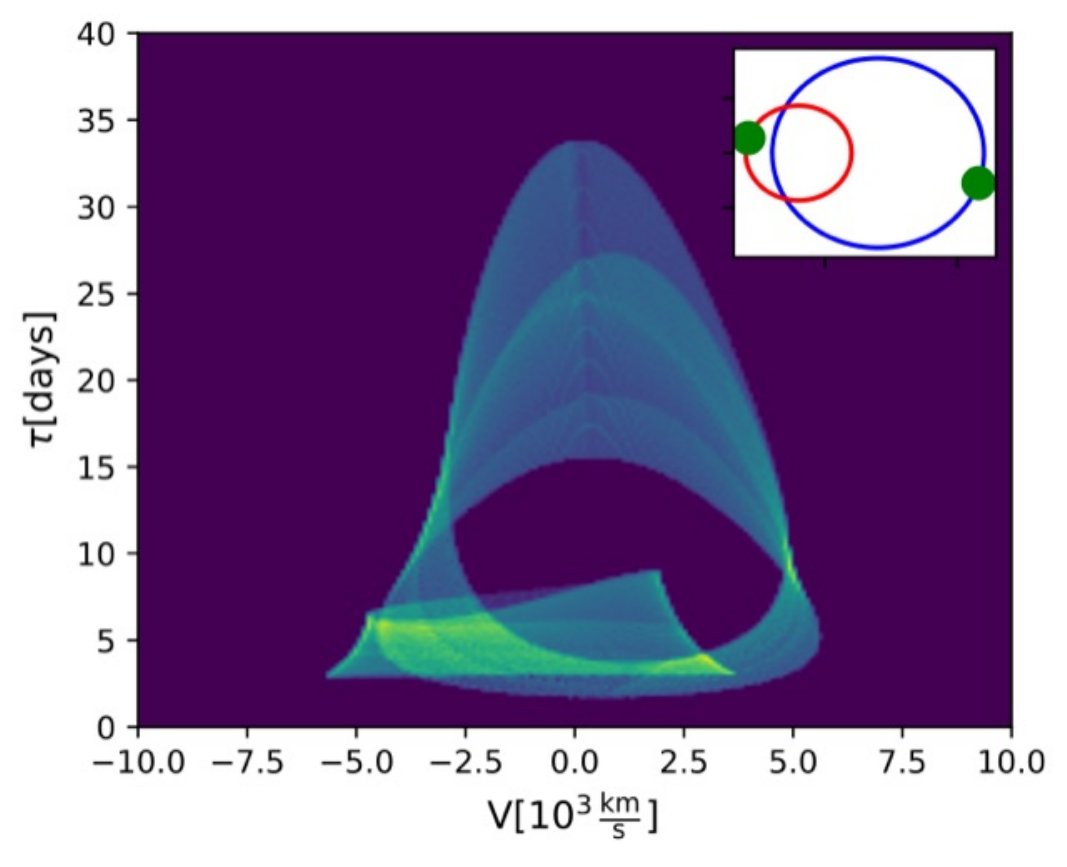}
    \caption{}
\end{subfigure}
\begin{subfigure}{6cm}
    \centering
    \includegraphics[width=5cm]{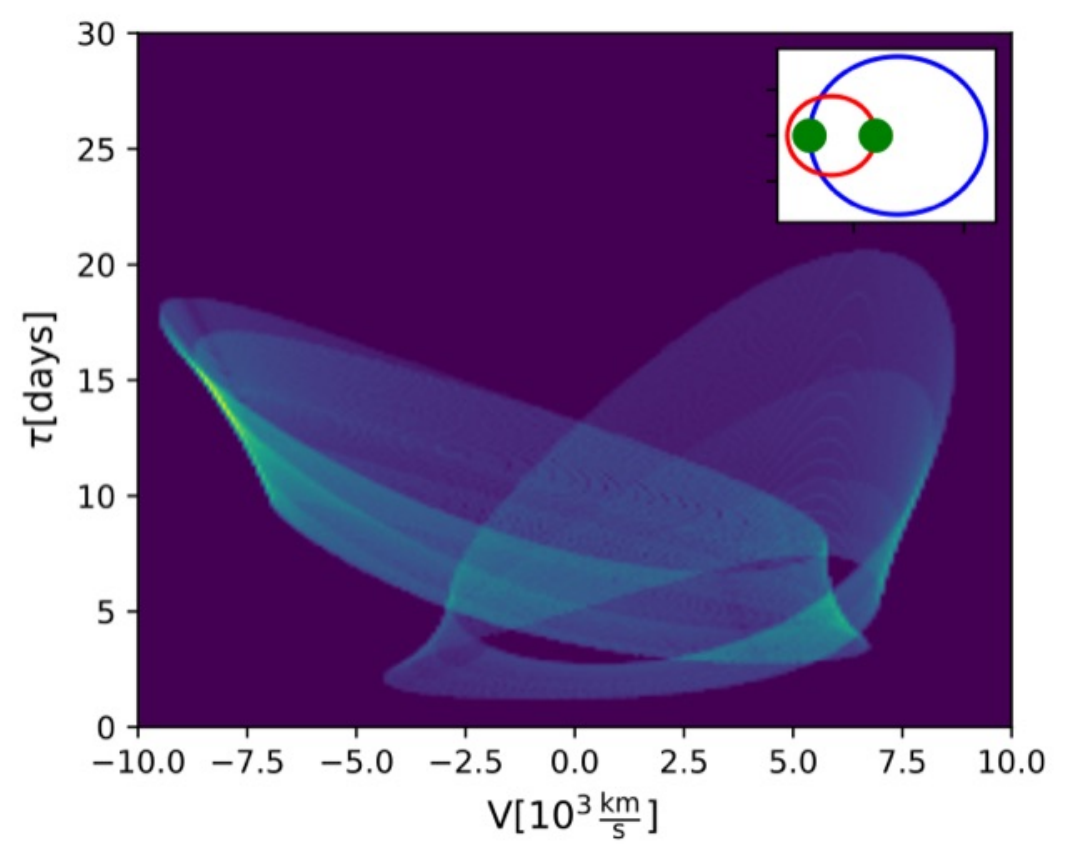}
    \caption{}
\end{subfigure}
\begin{subfigure}{6cm}
\centering
    \includegraphics[width=5cm]{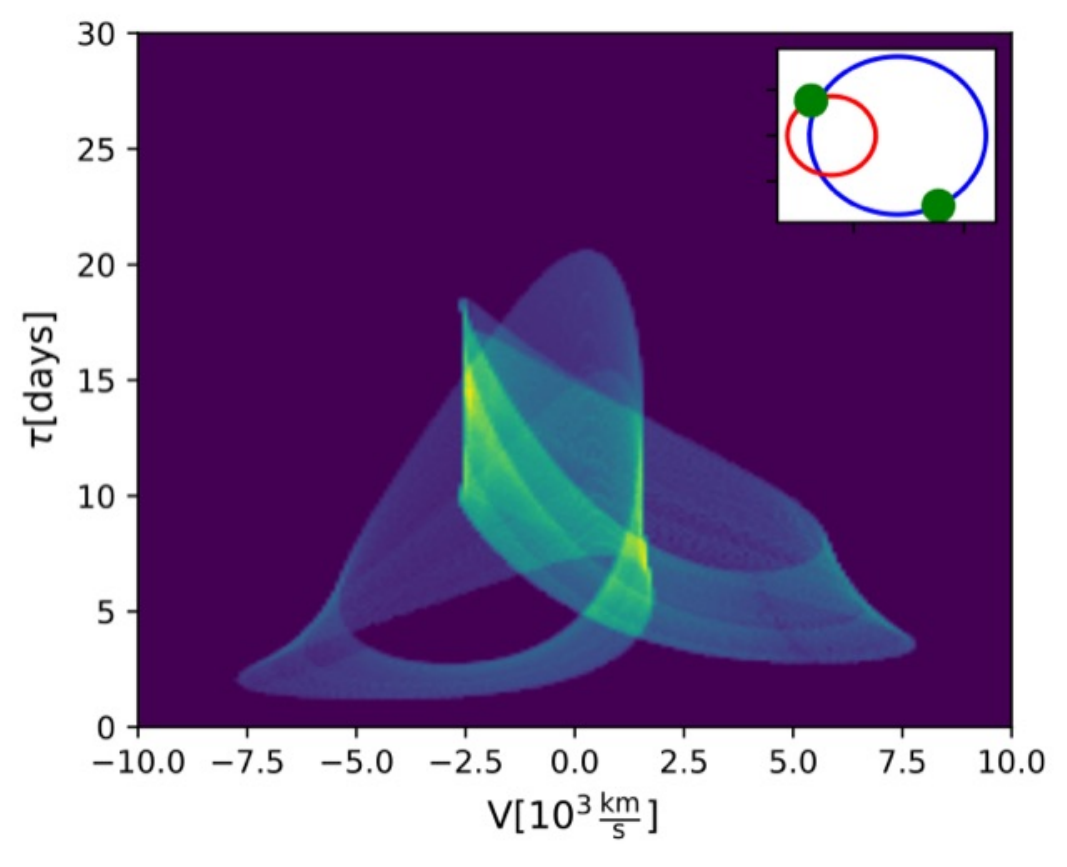}
    \caption{}
\end{subfigure}
\begin{subfigure}{6cm}
\centering
    \includegraphics[width=5cm]{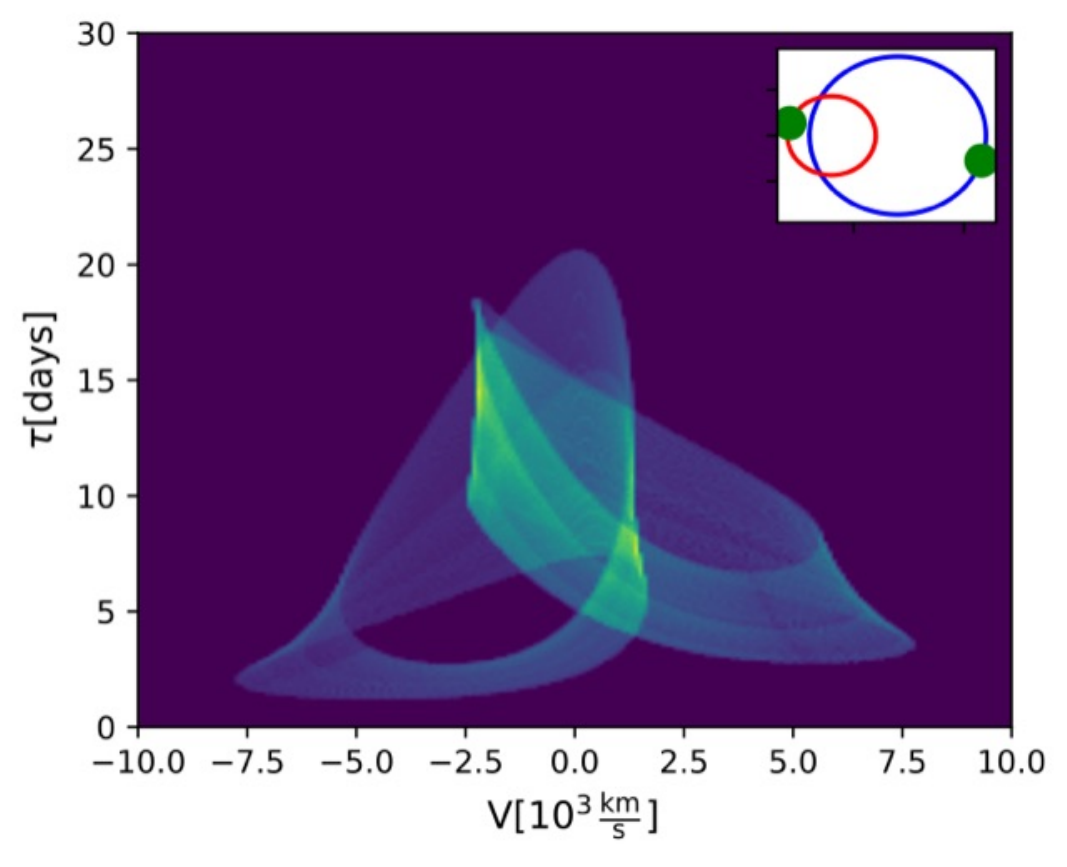}
    \caption{}
\end{subfigure}
\caption{2DTF  maps obtained for different geometries of a binary system with disc-like BLRs. The insets present the orbital phase of the binary system corresponding to the map. The direction of motion of the binary SMBH is anticlockwise. Panels (a)--(c) show a coplanar circular case obtained using the parameters from Paper I  (see their Fig. 2(a)); panels (d)-(f) show the same as panels (a)-(c), but    the inclination  of the smaller SMBH orbit is $30^{\circ}$; panels (g)-(i) show a coplanar elliptical binary system with  e=0.5, $\Omega_{ 1}=\Omega_{2}=0^{\circ}$, $\omega_{ 1}=0,\text{ and } \omega_{2}=180^{\circ}$, and orbital parameters of the { clouds} of $e_{\mathrm c1}=e_{\mathrm c2}=0.5, \Omega_{\mathrm c1}=\Omega_{\mathrm c2}=100^{\circ}, 
\omega_{\mathrm c1}=0^{\circ},\text{ and } \omega_{\mathrm c2}=180^{\circ}$; and panels (j)-(l) show the same orbital parameters for the binary system as in panels (g)-(i) , but  the orbital  parameters of the { clouds} are   $e_{\mathrm c1}=e_{\mathrm c2}=0.5, \Omega_{\mathrm c1}=\Omega_{\mathrm c2}=100^{\circ}, \text{and }\omega_{\mathrm c1}=110^{\circ}, \omega_{\mathrm c2}=290^{\circ}$.}
 \label{fig:tf2}
\end{figure*}

\begin{figure*}
\centering
\begin{subfigure}{6cm}
    \centering
    \includegraphics[width=5cm]{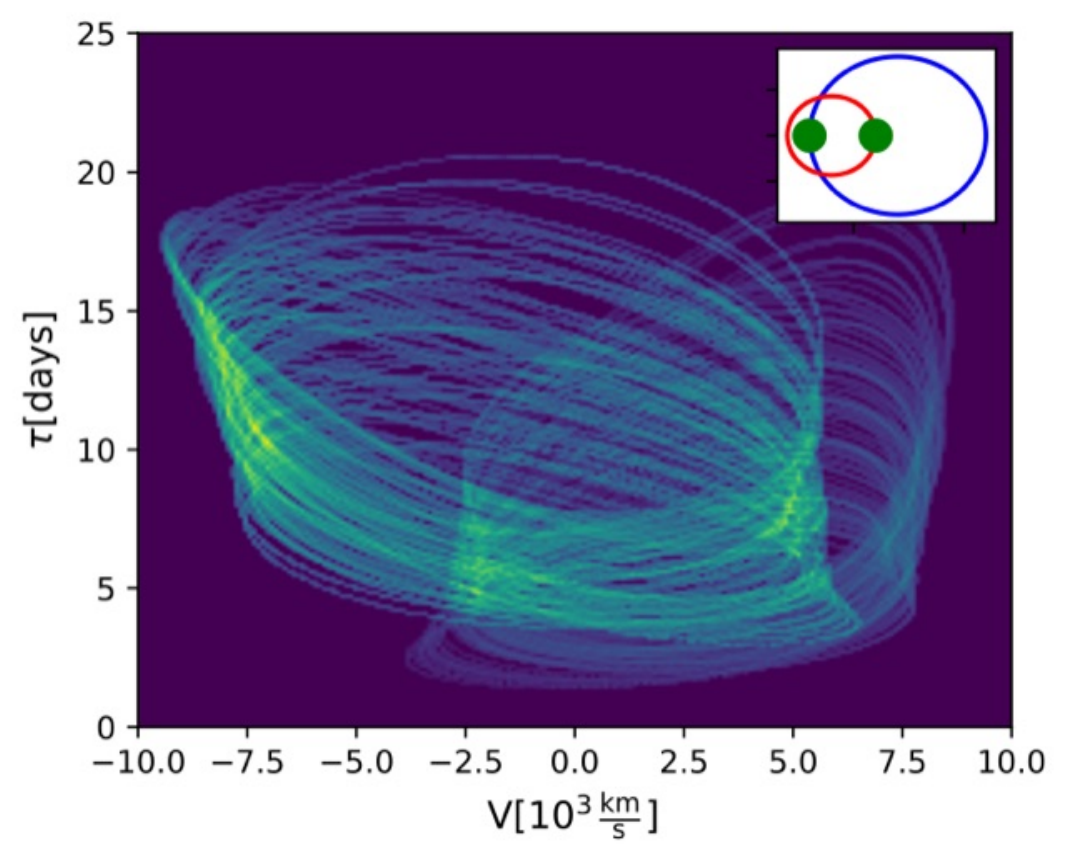}
    \caption{}
\end{subfigure}%
\begin{subfigure}{6cm}
    \centering
    \includegraphics[width=5cm]{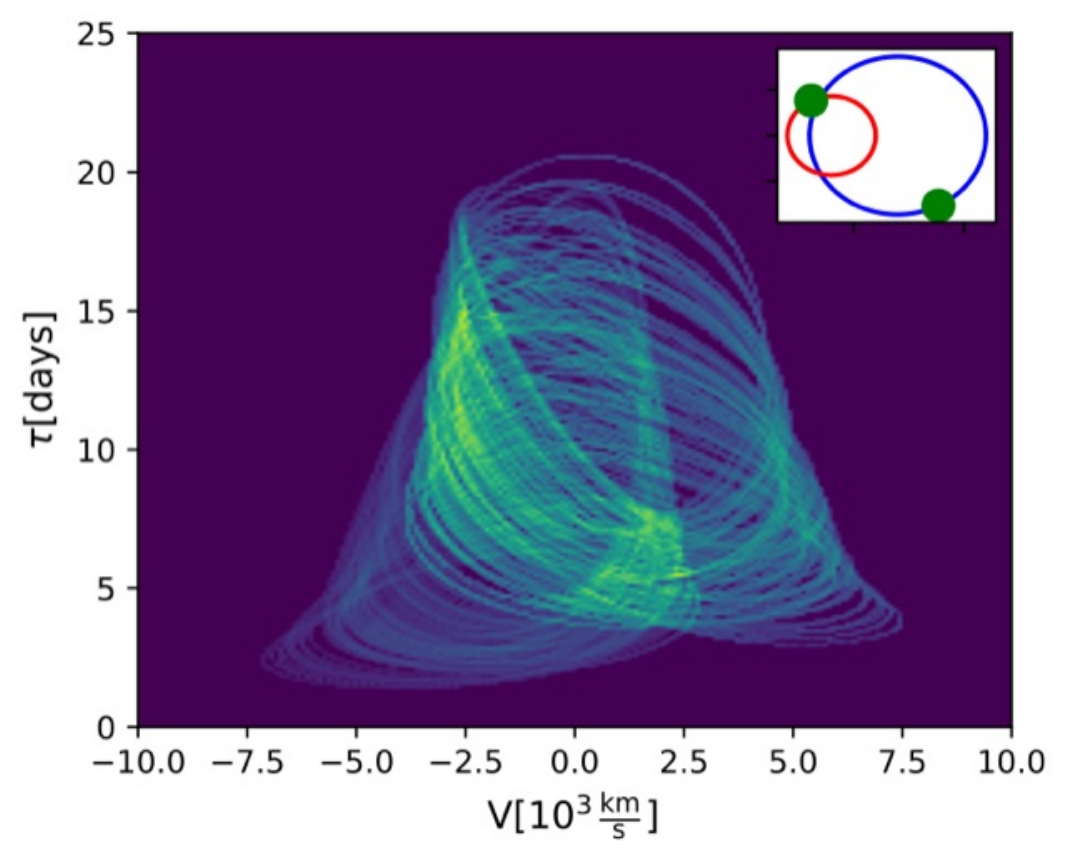}
    \caption{}
\end{subfigure}
\begin{subfigure}{6cm}
    \centering
    \includegraphics[width=5cm]{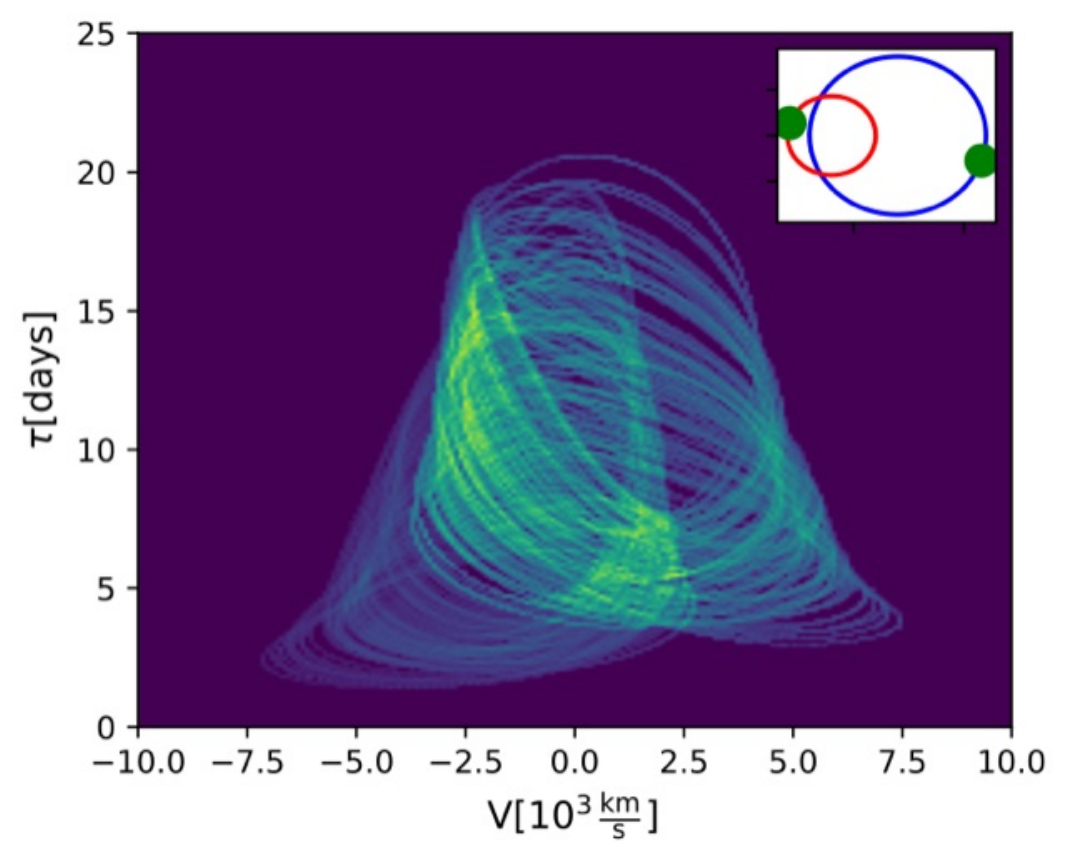}
    \caption{}
\end{subfigure}
\begin{subfigure}{6cm}
    \centering
    \includegraphics[width=5cm]{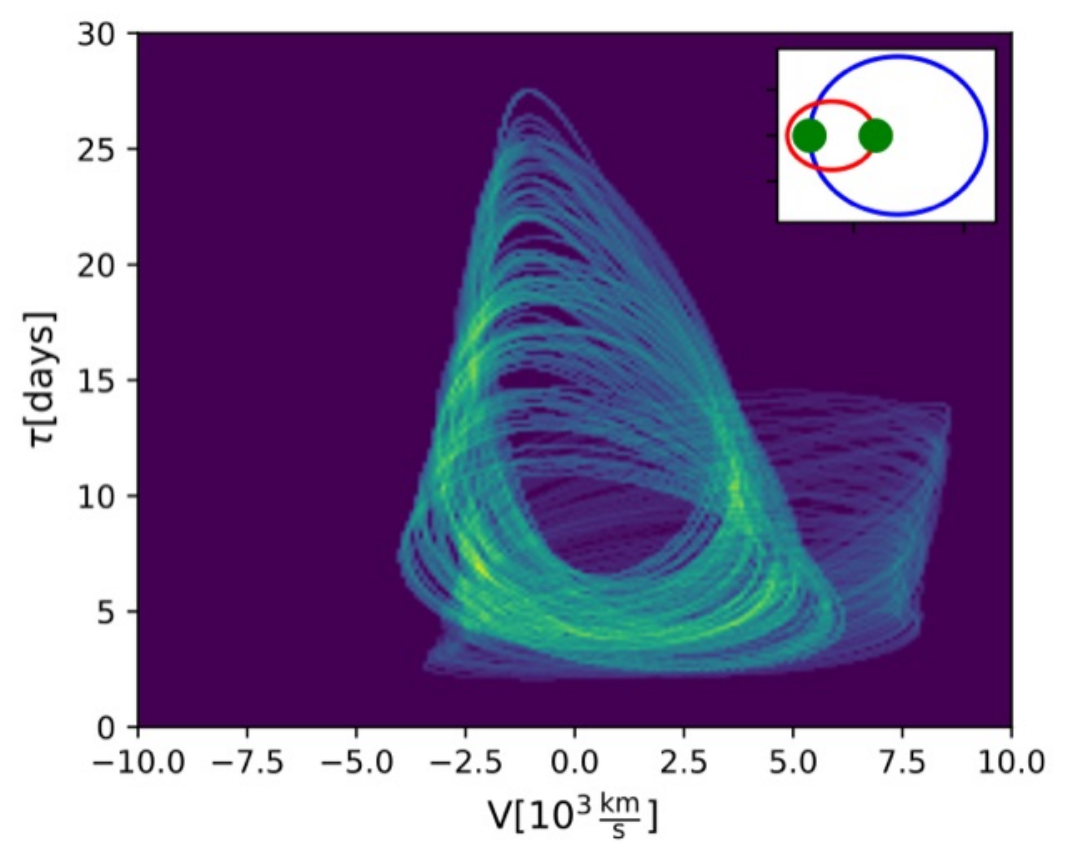}
    \caption{}
\end{subfigure}
\begin{subfigure}{6cm}
    \centering
    \includegraphics[width=5cm]{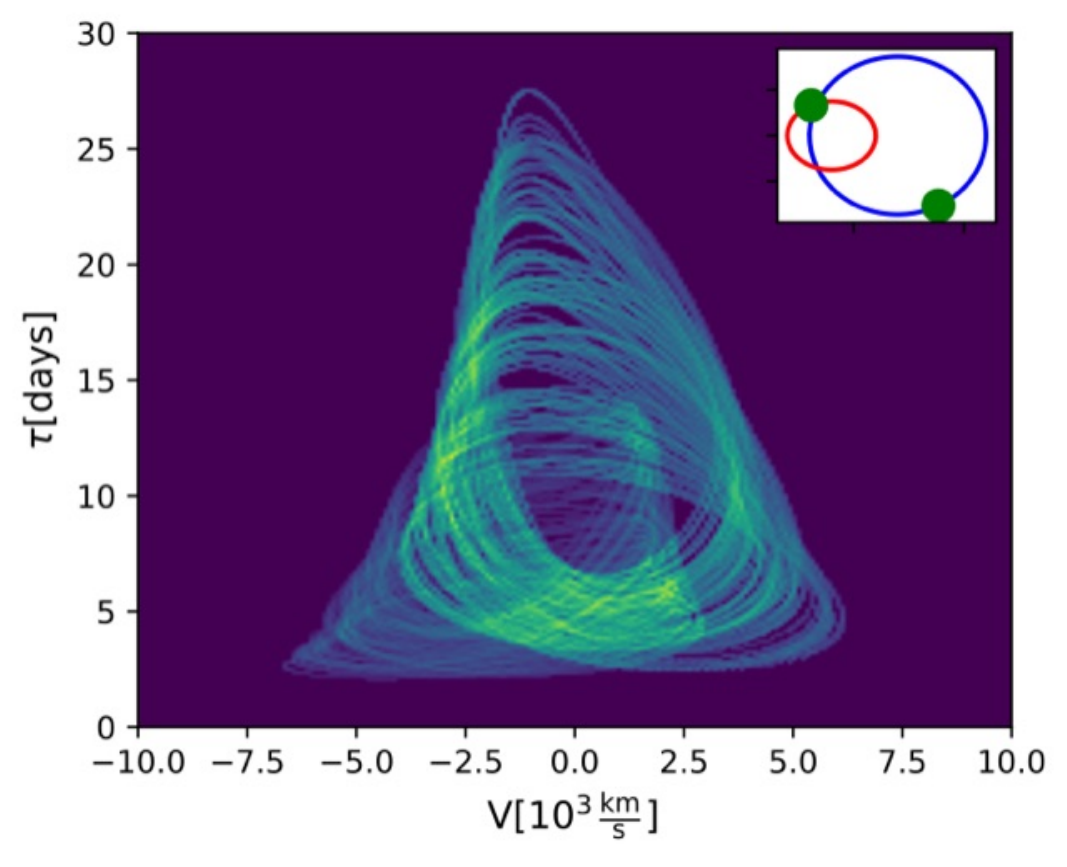}
    \caption{}
\end{subfigure}
\begin{subfigure}{6cm}
    \centering
    \includegraphics[width=5cm]{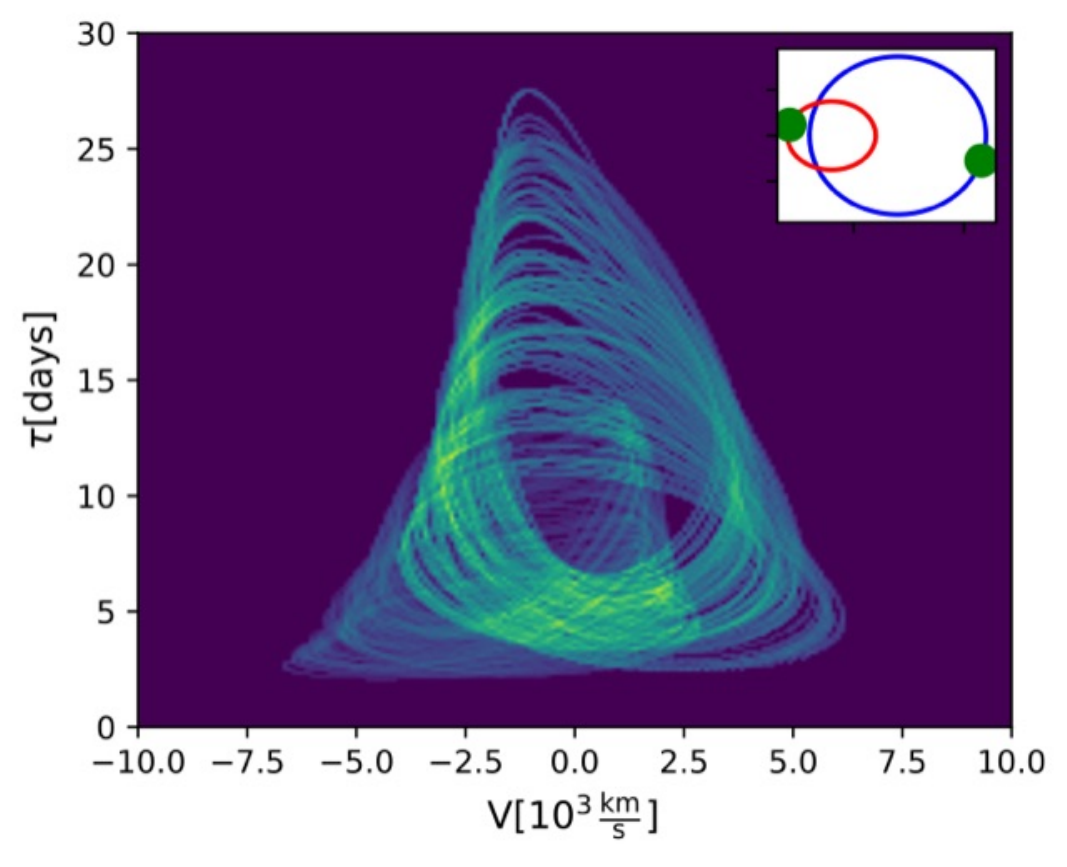}
    \caption{}
\end{subfigure}
\begin{subfigure}{6cm}
    \centering
    \includegraphics[width=5cm]{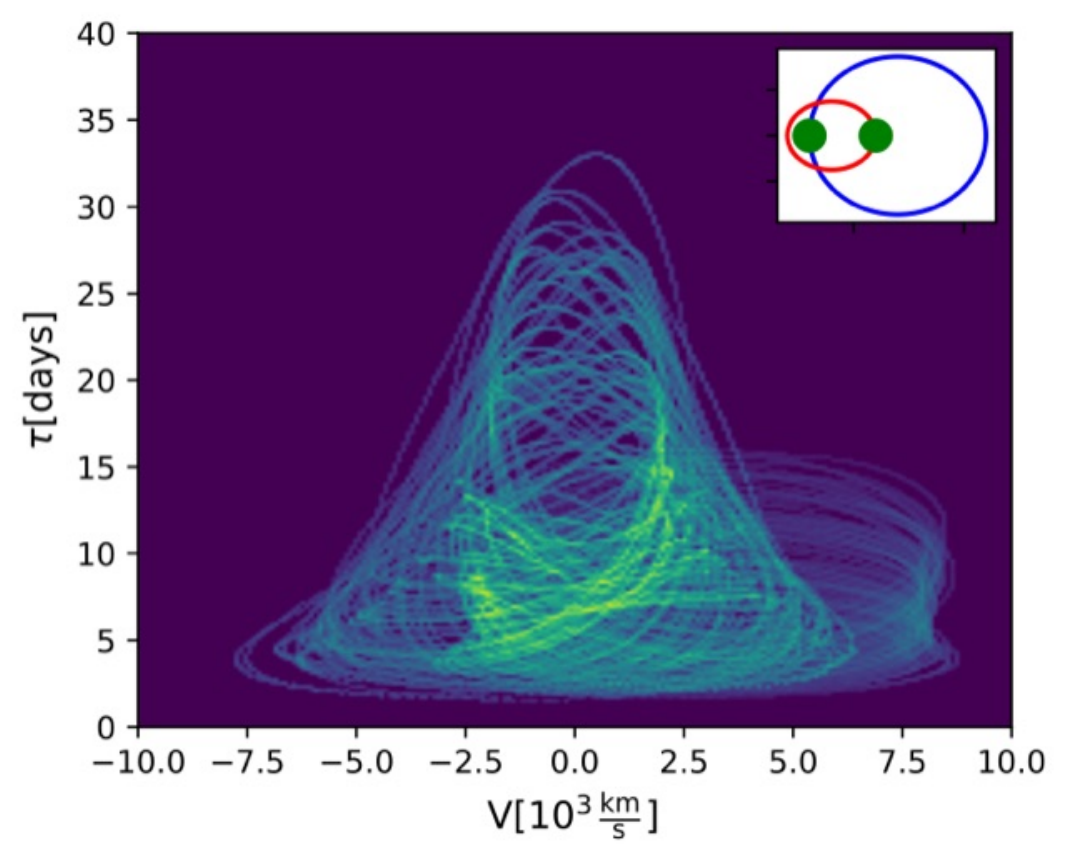}
    \caption{}
\end{subfigure}
\begin{subfigure}{6cm}
\centering
    \includegraphics[width=5cm]{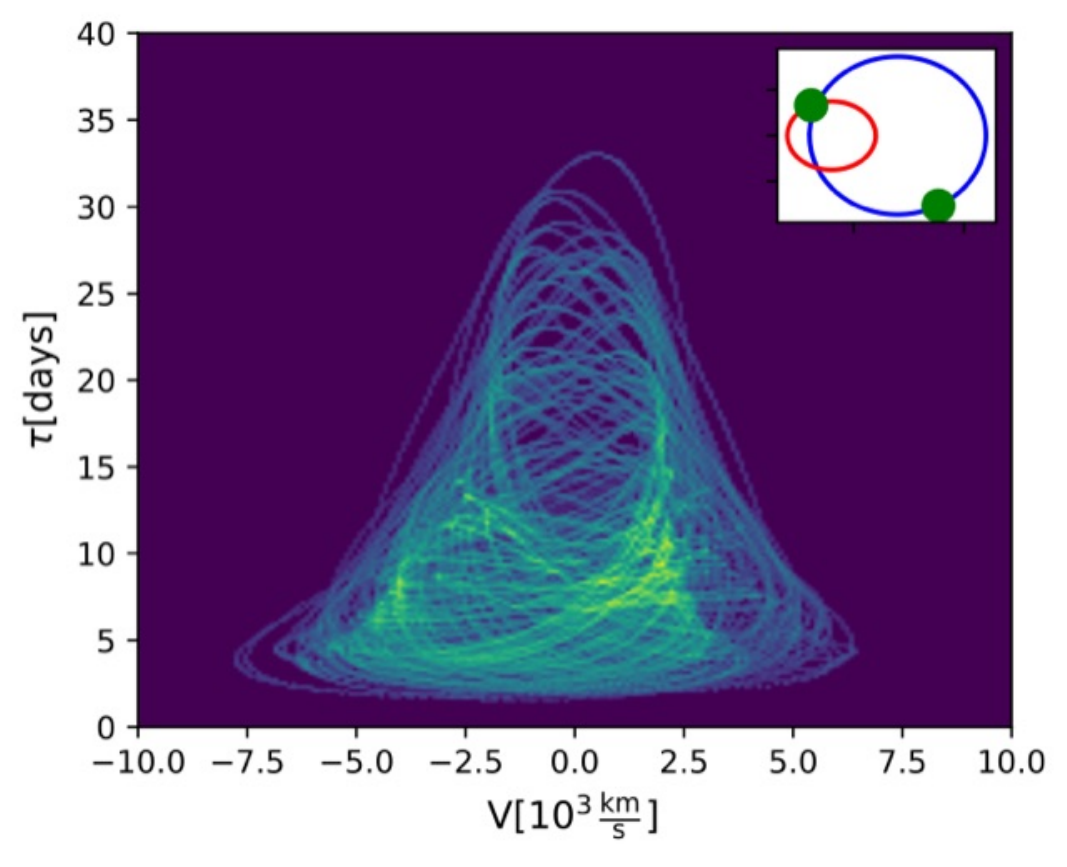}
    \caption{}
\end{subfigure}
\begin{subfigure}{6cm}
\centering
    \includegraphics[width=5cm]{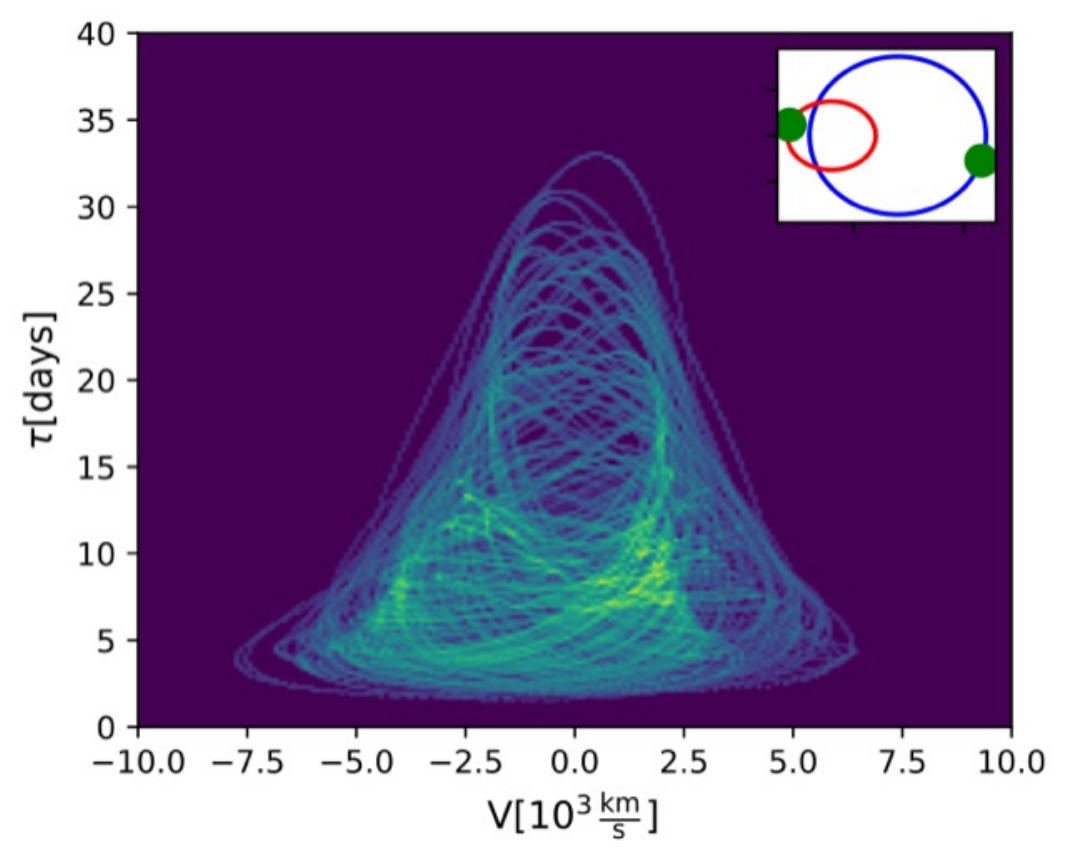}
    \caption{}
\end{subfigure}
\begin{subfigure}{6cm}
    \centering
    \includegraphics[width=5cm]{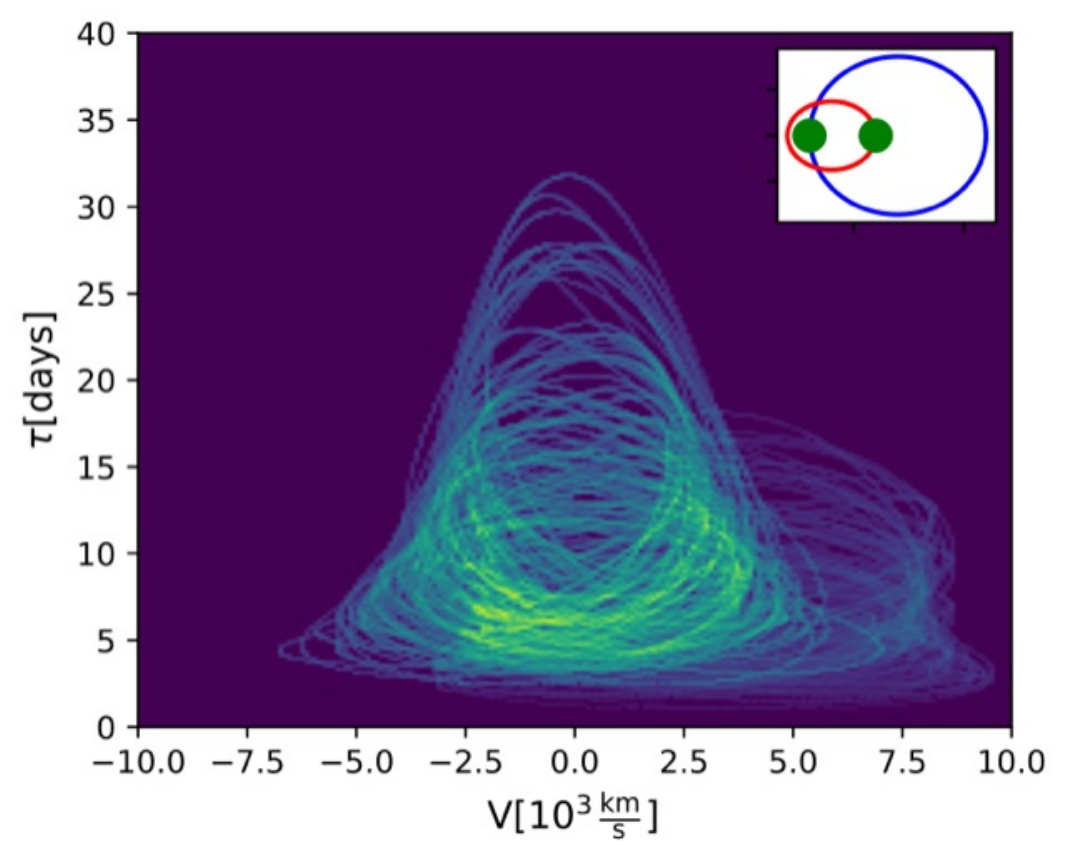}
    \caption{}
\end{subfigure}
\begin{subfigure}{6cm}
\centering
    \includegraphics[width=5cm]{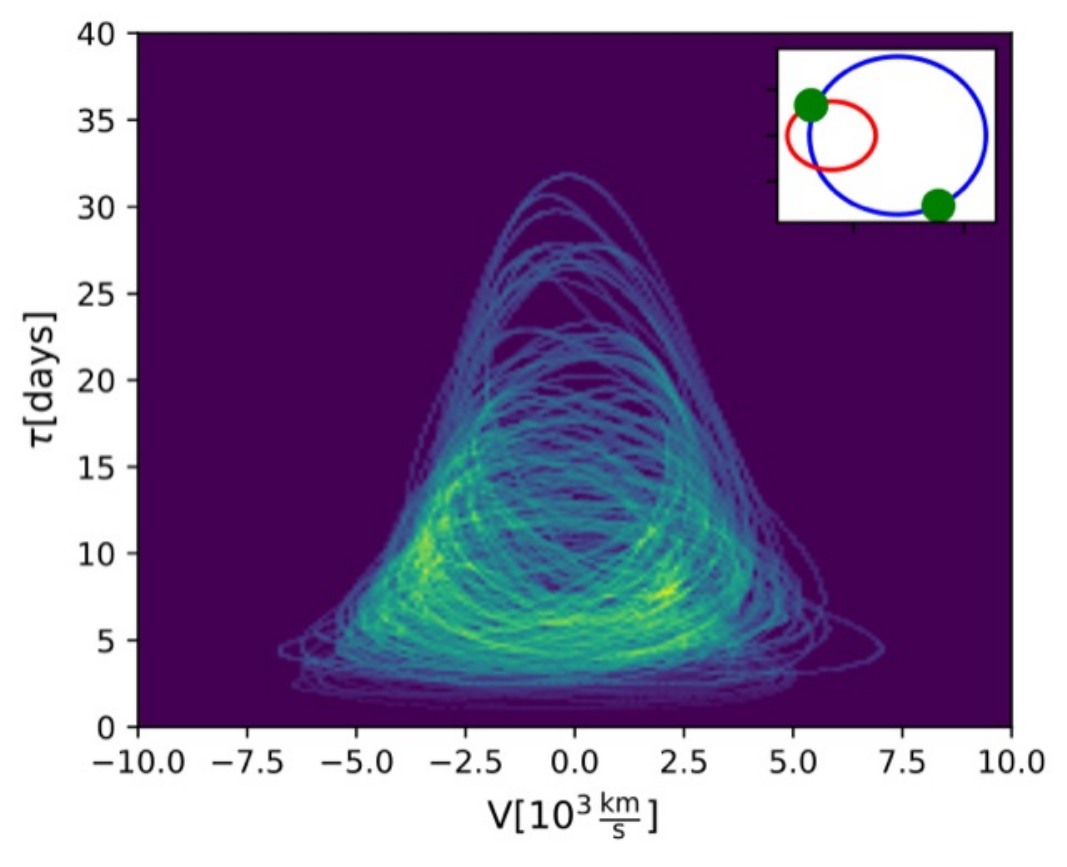}
    \caption{}
\end{subfigure}
\begin{subfigure}{6cm}
\centering
    \includegraphics[width=5cm]{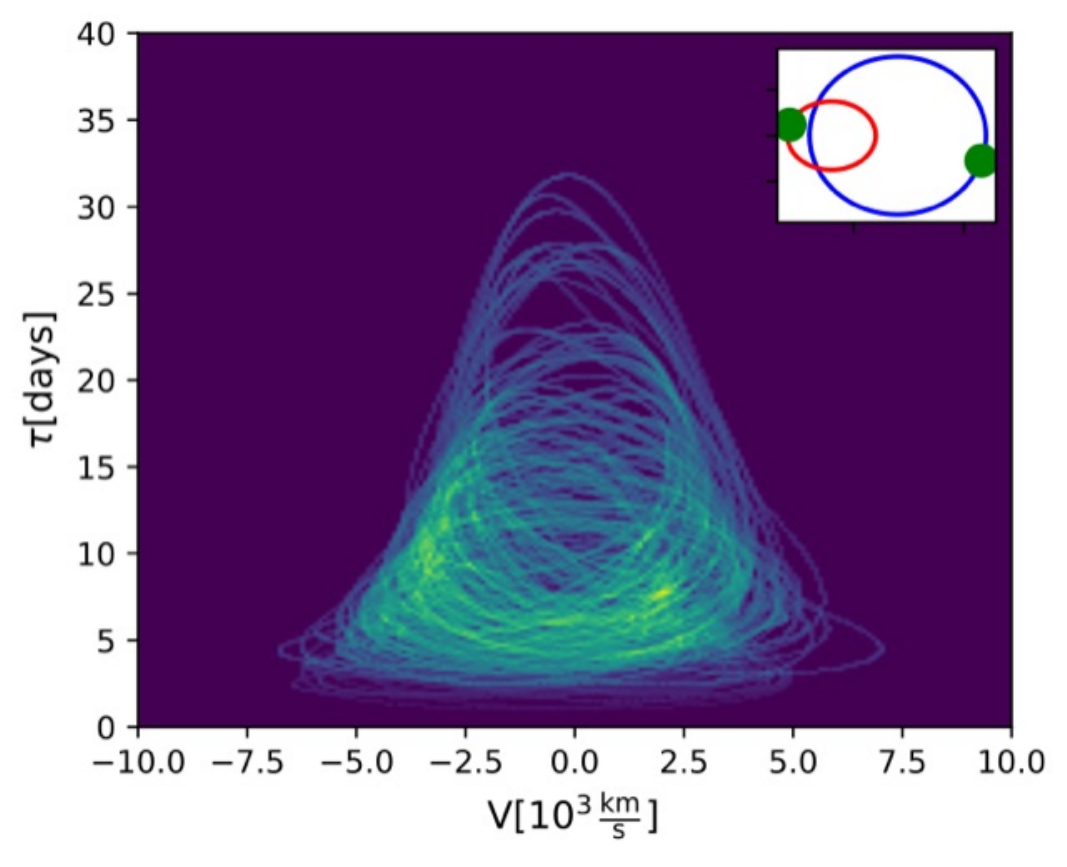}
    \caption{}
\end{subfigure}
\caption{Same as Fig. \ref{fig:tf2}. Panels (a)-(c) show a  coplanar elliptical binary system with $e=0.5$,  $\Omega_{ 1}=\Omega_{ 2}=0^{\circ}$, $\omega_{ 1}=0, \omega_{2}=180^{\circ}$,  { clouds orbits} in both BLRs  have random  eccentricities  and   $ \Omega_{\mathrm c1}=\Omega_{\mathrm c2}=100^{\circ}, \omega_{\mathrm c1}=110^{\circ}, \omega_{\mathrm c2}=290^{\circ}$;  panels (d)-(f) show that the plane of the more massive SMBH orbit is inclined by $30^{\circ}$, { clouds} in both  disc-like BLR  have an orientation   $ \omega_{\mathrm c1}=50^{\circ}, \omega_{\mathrm c2}=230^{\circ}$, and the remaining  orbital parameters are the same as in panels (a)--(c) and (g)--(i), except that the orbital eccentricities  of { the clouds} are $e_{\mathrm c1}=0.5\text{ and }e_{\mathrm c2}=0.1$ and their orientations are randomised, all other orbital parameters are the same as in panels (a)--(c); in panels (j)--(l), the binary  orbital configuration  is the same as for panels (d)--(f), but the orbital parameters   of { the clouds} are random in both BLRs. (The stability of elliptical orbits is not considered). }
\label{fig:tf3}
\end{figure*}

\begin{figure*}
\centering
\begin{subfigure}{6cm}
    \centering
    \includegraphics[width=5cm]{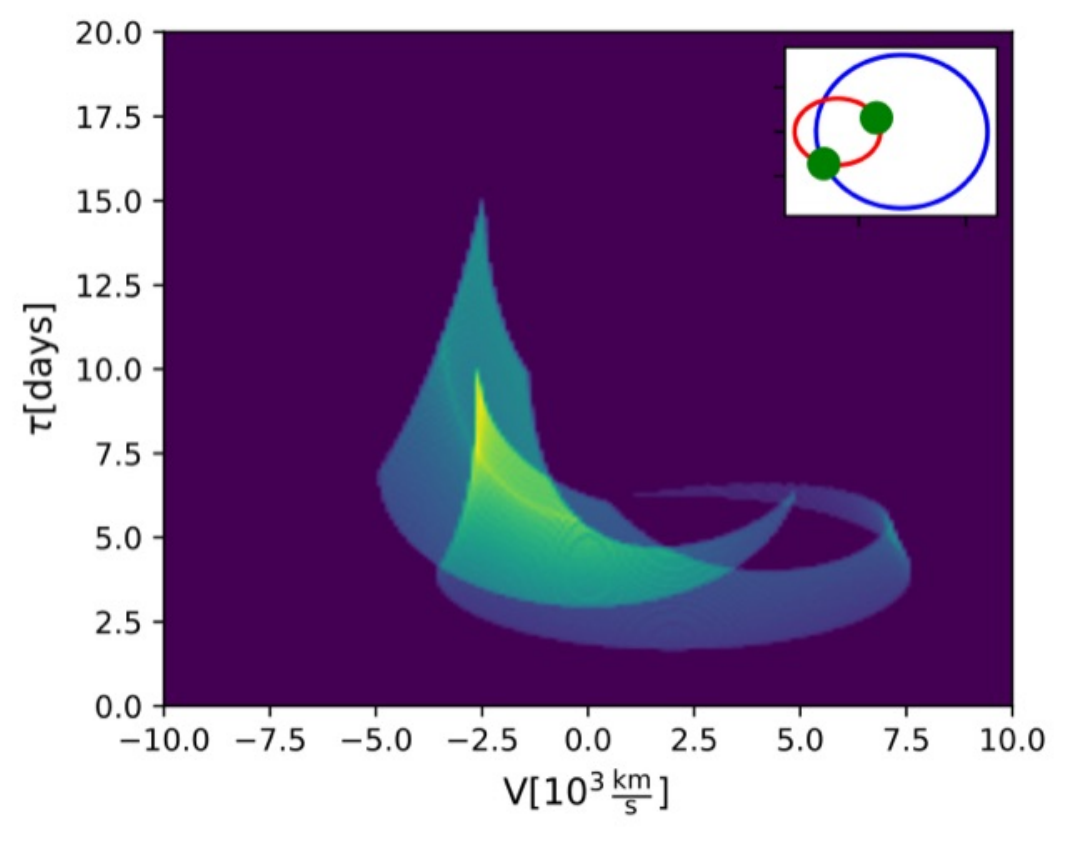}
    \caption{}
\end{subfigure}%
\begin{subfigure}{6cm}
    \centering
    \includegraphics[width=5cm]{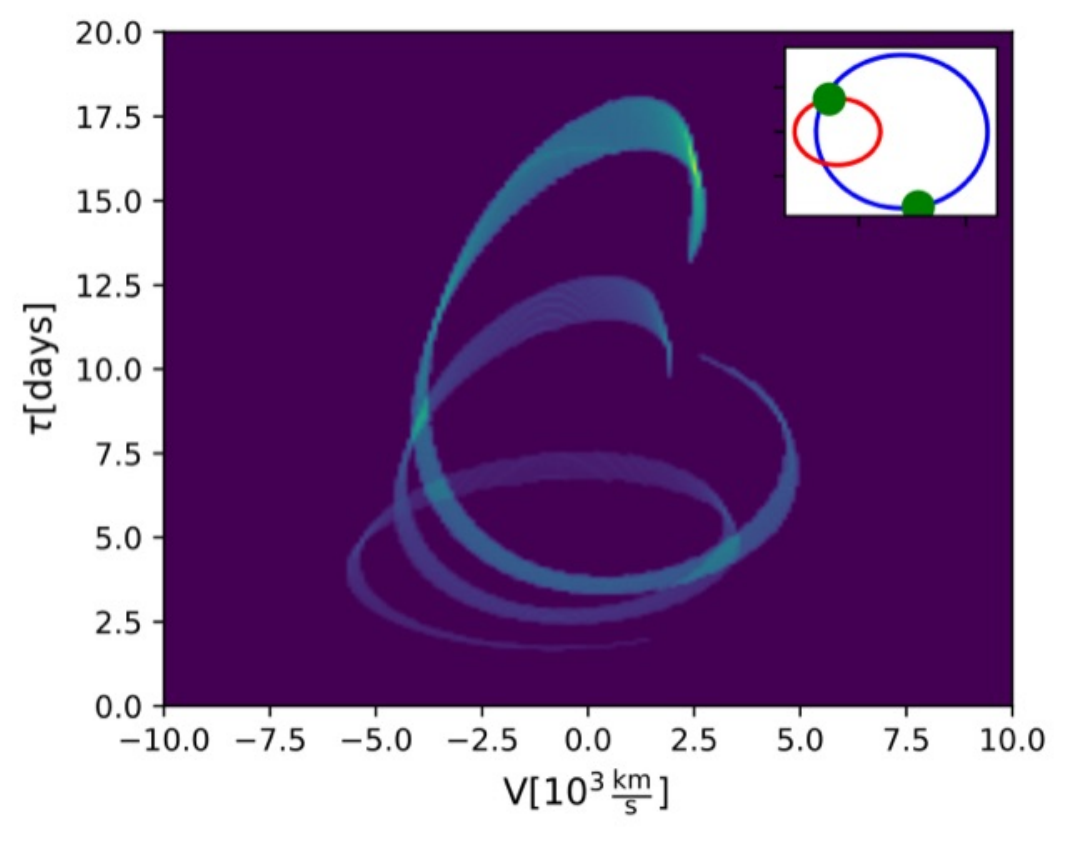}
    \caption{}
\end{subfigure}
\begin{subfigure}{6cm}
    \centering
    \includegraphics[width=5cm]{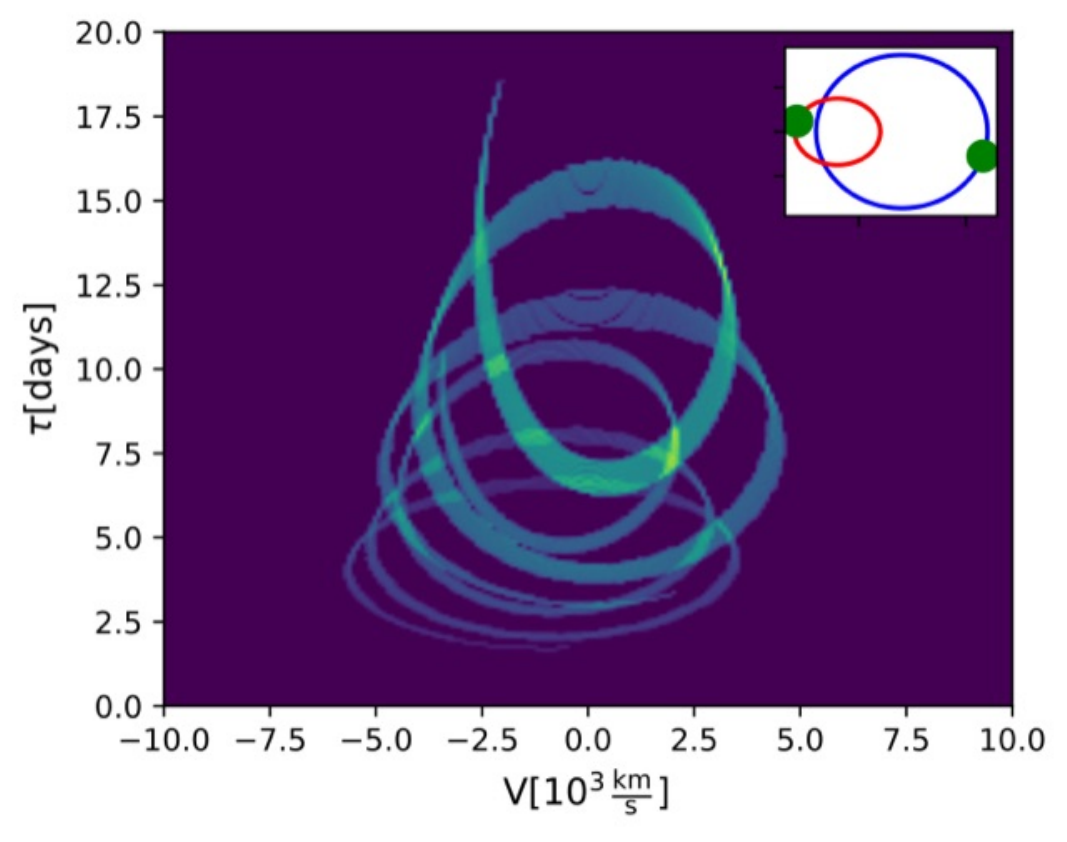}
    \caption{}
\end{subfigure}
\begin{subfigure}{6cm}
    \centering
    \includegraphics[width=5cm]{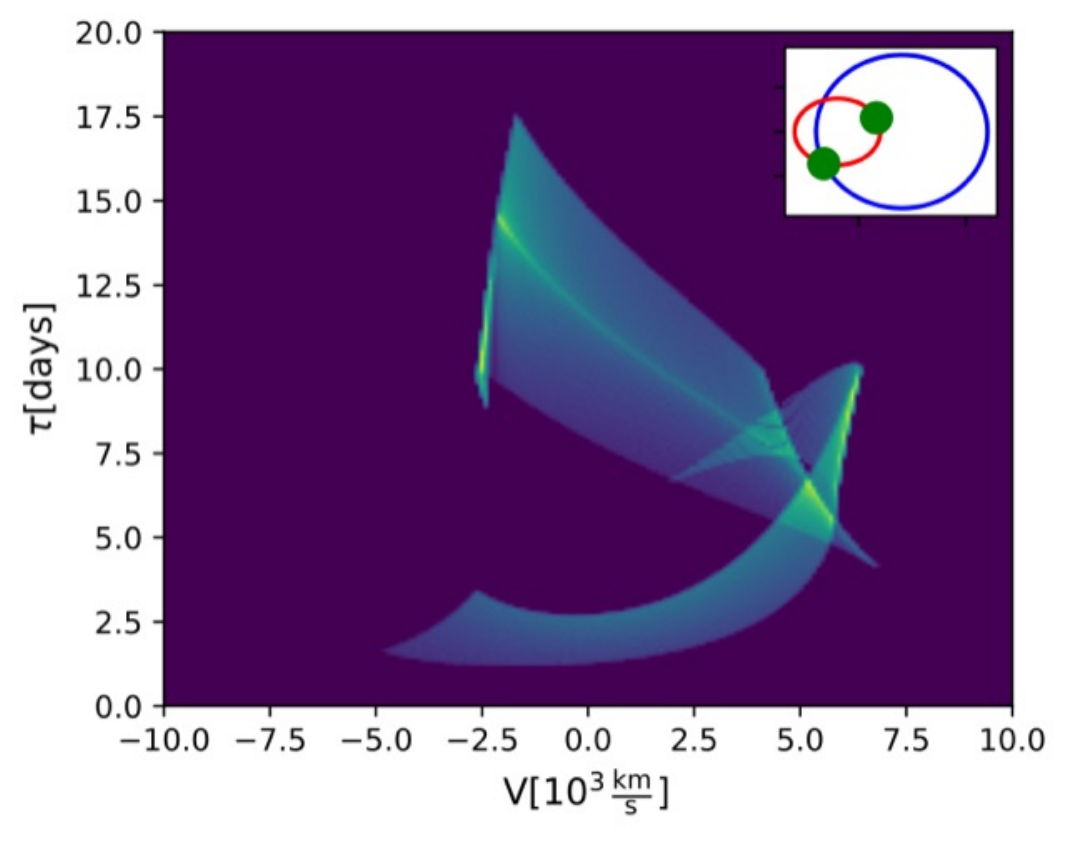}
    \caption{}
\end{subfigure}
\begin{subfigure}{6cm}
    \centering
    \includegraphics[width=5cm]{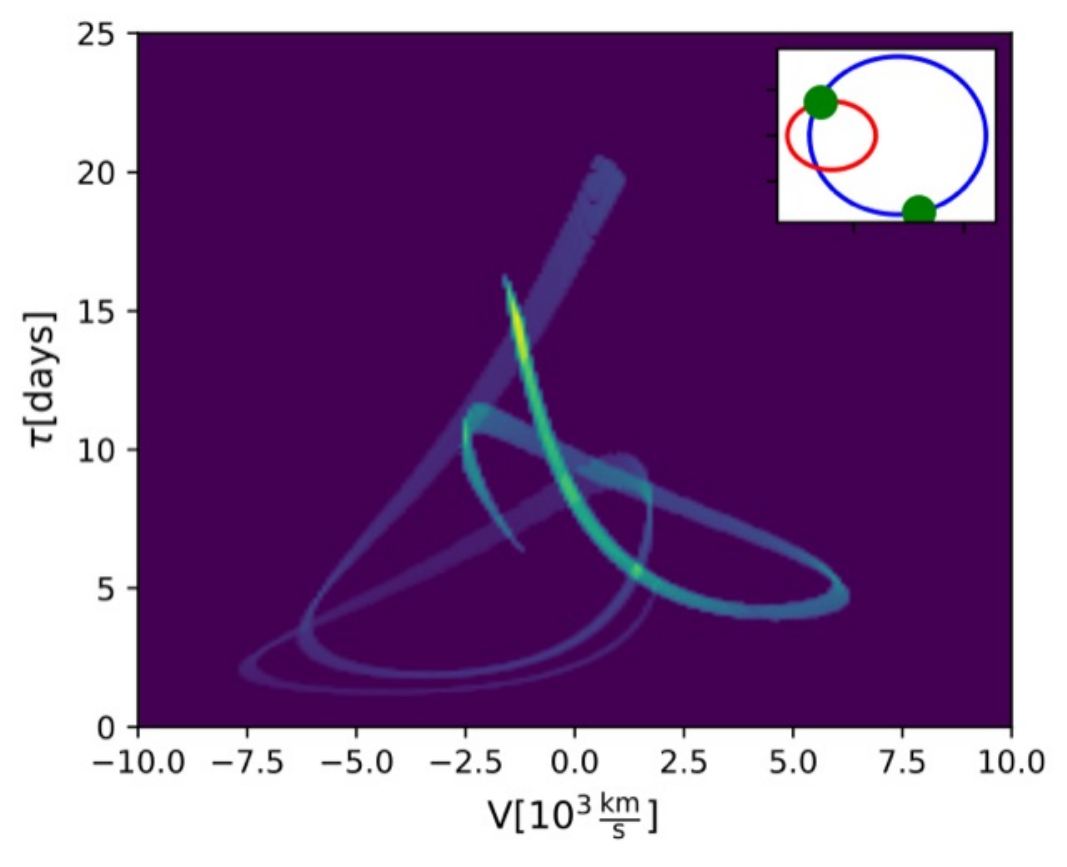}
    \caption{}
\end{subfigure}
\begin{subfigure}{6cm}
    \centering
    \includegraphics[width=5cm]{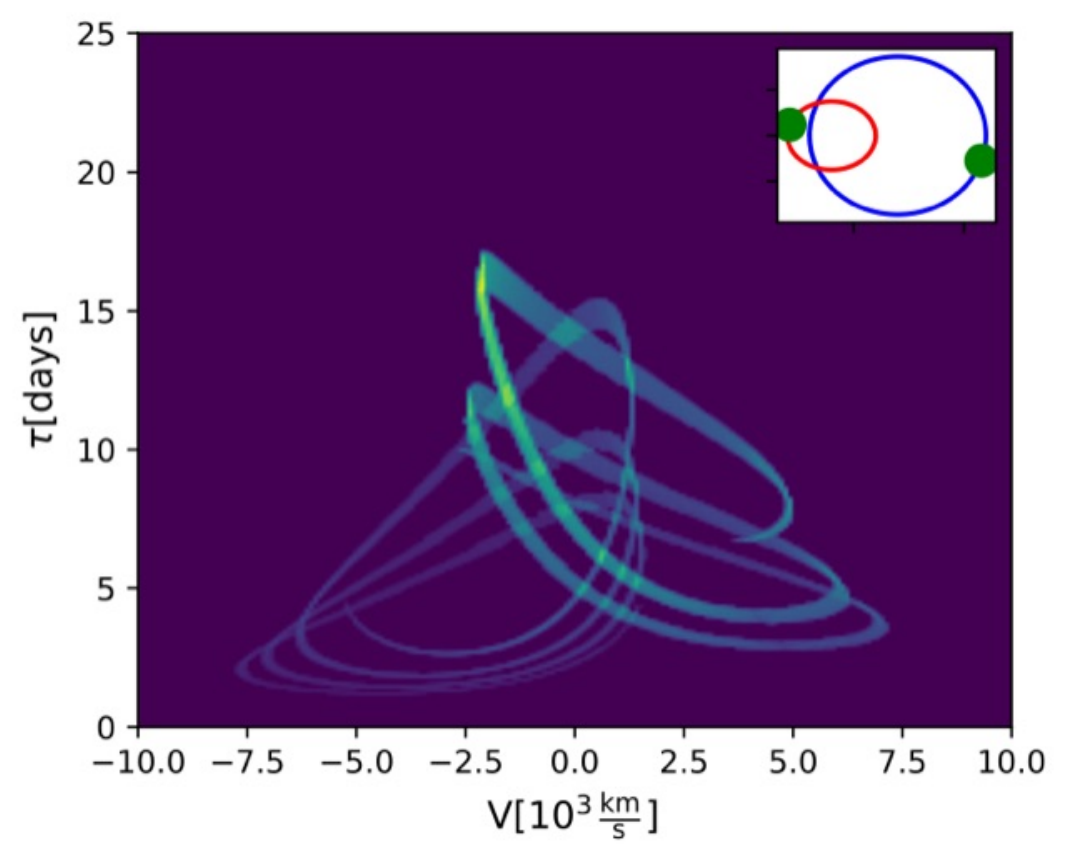}
    \caption{}
\end{subfigure}
\begin{subfigure}{6cm}
    \centering
    \includegraphics[width=5cm]{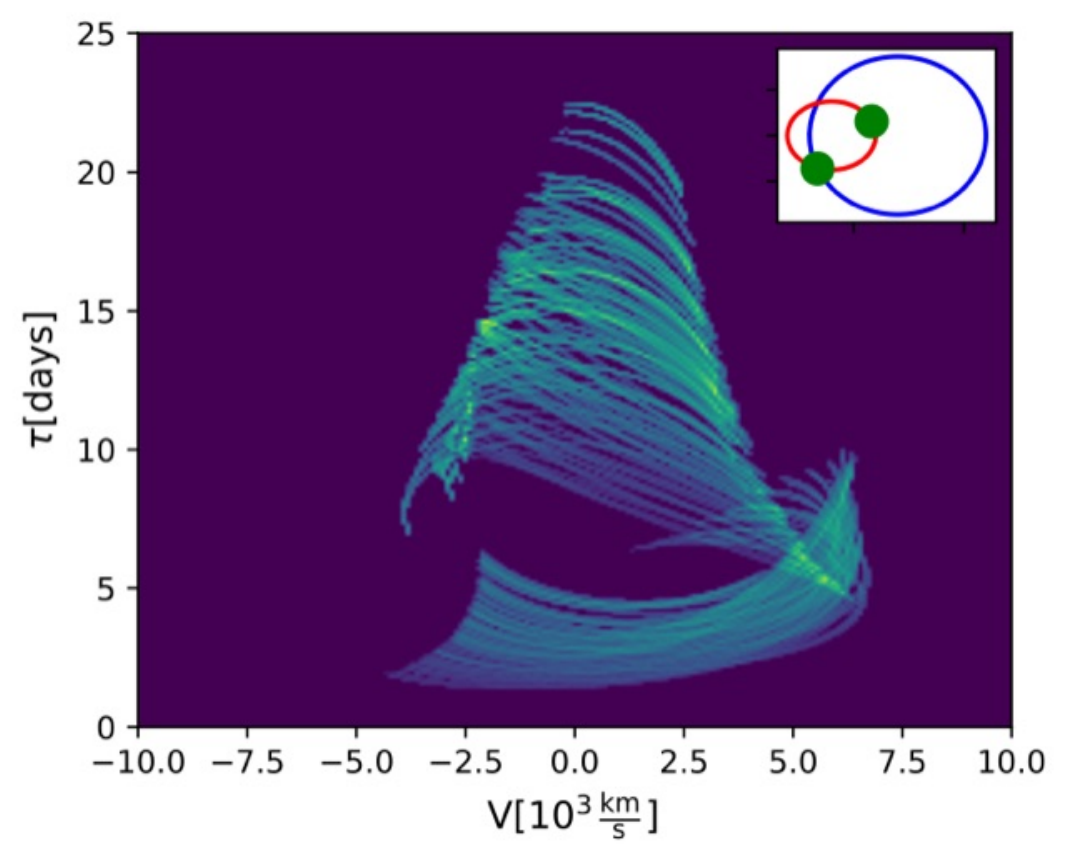}
    \caption{}
\end{subfigure}
\begin{subfigure}{6cm}
\centering
    \includegraphics[width=5cm]{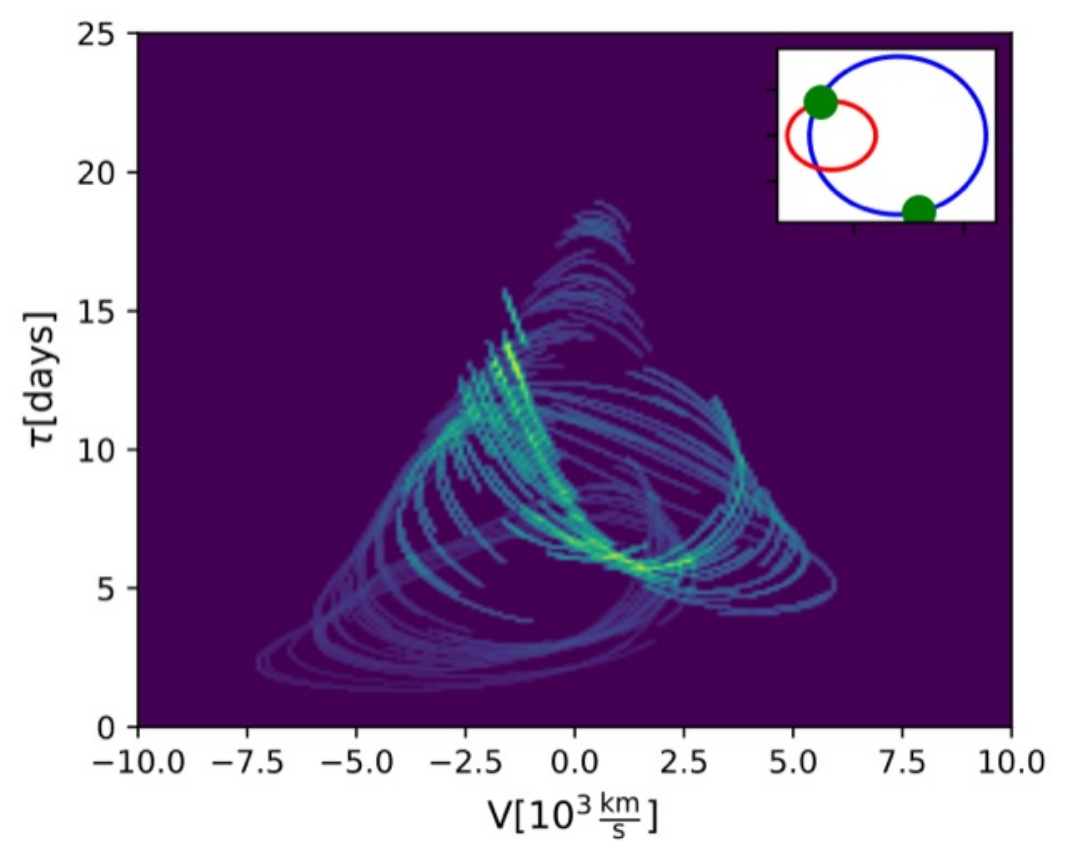}
    \caption{}
\end{subfigure}
\begin{subfigure}{6cm}
\centering
    \includegraphics[width=5cm]{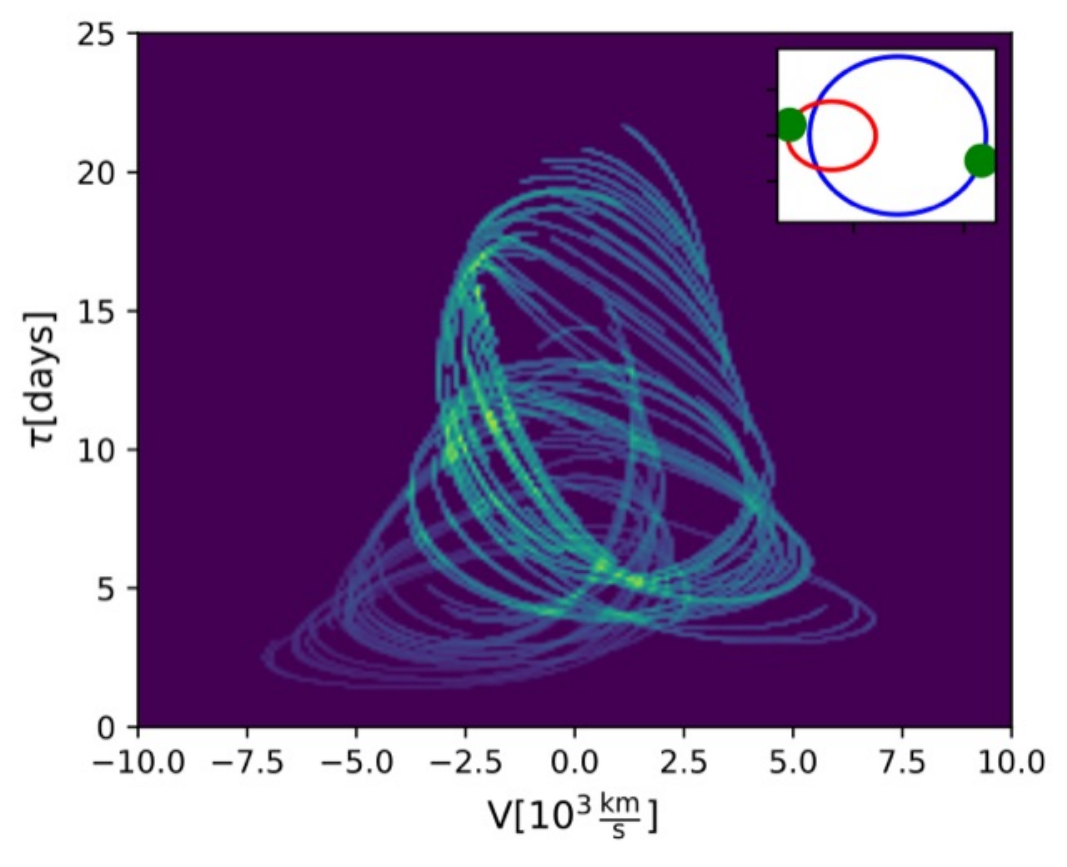}
    \caption{}
\end{subfigure}
\caption{Same as Fig. \ref{fig:tf2}, but   $10\%$ of { the clouds orbits} are visible.   From  left to  right,  the columns show orbital segments   $( k T_{\mathrm{c}ij}/10,   (k+1)T_{\mathrm{c}ij}/10)$ for $k\in {\{0,4,9\}}$  defined by the orbital period $T_{{\mathrm c}ij}$ of{  the clouds}. From the top down, panels (a)-(c) show an elliptical binary system in which the orbital plane of  the more massive SMBH  is inclined by $30^{\circ}$, $e=0.5, \Omega_ {1}=\Omega_{2}=0^{\circ}$, $\omega_{ 1}=0,\text{ and } \omega_{2}=180^{\circ}$, and both discs are circular; panels (d)-(f) show the same   orbital configuration of the binary system as panels (a)-(c), but their  disc-like BLRs  parameters are  $e_{\mathrm {c}1}=e_{\mathrm {c}2}=0.5, \Omega_{\mathrm {c}1}=\Omega_{\mathrm {c}2}=100^{\circ}, \omega_{\mathrm {c}1}=110^{\circ},\text{ and } \omega_{\mathrm {c}2}=290^{\circ}$; and panels (g)--(i) show the same  configuration of the  binary system as panels (a)--(c), and  the orbital { configurations} of the BLR { clouds} are the same  as in panels (d)--(f), except  that their eccentricities are random. The stability of the  elliptical orbits in randomised motion is not considered.}
\label{fig:tf4}
\end{figure*}

Novel 2DTF maps of an elliptical binary system with elliptical disc-like BLRs are shown in Figs. \ref{fig:tf2}- \ref{fig:tf4}.  Each row  of panels displays the 2DTF maps that correspond to the barycentric  orbital configuration of the SMBH binary system given in the insets.

 The typical   2DTF  maps of clouds with specific orbital parameters are  shown in Figure \ref{fig:tf2}. The first row presents the  maps for  a circular SMBH system with circular disc-like BLRs as given in Paper I, in order  to check the consistency of our binary SMBH model. 
 
The second row depicts a non-coplanar SMBH system with circular BLRs  with the same orbital parameters as above, but the orbital plane of the smaller SMBH is inclined by $30^{\circ}$. The orbital inclination of the SMBH shrinks the size and   the shape of the 2DTF core of the smaller SMBH, which is evident in comparison with the previous case. 
Although they appear distinctive, the nuances between a coplanar and a non-coplanar circular binary  would be difficult to discern without highly reliable reverberation mapping observations.  

The third and forth row emphasize the effects of eccentricity and orientation angles  of the orbits of the clouds and   the coplanar SMBH.  The maps in the third row suggest a significant degeneration of the bell shape when the argument of pericenter of the cloud is $180^{\circ}$. This means that the argument of pericenter of the cloud lies on the line of nodes and thus within the vicinity of the binary reference plane.

The effects  of the inclined orbit of the larger SMBH in the binary system  are shown in  Figure \ref{fig:tf3}. For comparison,  Fig.  \ref{fig:tf3} (a)-(c) shows the orbits of the coplanar   elliptical binary  SMBH. 
The    hollow   bell-shape  of the more massive component caused by its inclined orbit  is shown in the second row.
The third and forth row show the effects of different combinations of randomised orbital cloud parameters (we did not consider the stability of elliptical orbits in the randomised motion). These  maps illustrate a topology   that is  complex, as expected from similar  maps in Fig. \ref{fig:tf1}.

However, the orbital period of the binary might be too long to be easily observed \citep{1980Natur.287..307B}. Thus observations might cover only some portion of the full range of the period of the clouds and the binary motion. For these reasons, we analysed the case when clouds  are observed for just 10$\%$ of their orbits, see  Figure \ref{fig:tf4}.  For example, the cloud takes about ten years to complete one orbit around the SMBH, but the monitoring campaign  lasts just one year.
 Thus, orbital arcs of  about one year   in both BLRs would be covered by observations. The orbital arc  starts  at $ k T_{\mathrm {c} ij}/10$  and ends at $ (k+1)T_{\mathrm {c}ij}/10$ for $k\in {\{0,4,9\}}$, defined by
 the orbital period $T_{\mathrm {c}ij}$ of the cloud.

The insets present the orbital configuration of an SMBH system at  the instances of the right edge of the orbital  arcs of the cloud  ($(k+1)T_{{\mathrm c}ij}/10$).
The first row of panels  show  maps for   elliptical  SMBH orbits $ e=0.5$, while the orbital plane of  the more massive SMBH is inclined by $30^{\circ}$,  but  both disc-like BLRs are circular. The remaining panels  show maps for  an elliptical SMBH with elliptical disc-like BLRs, but the third row contains the effects of random eccentricities of cloud trajectories in both discs  (stability of elliptical orbits is not considered).   When the orbital portion of each BLR  cloud is observed for only one year, the orbit  has a different structure. More compact structures are  obtained in an orbital  range of the clouds of ($0, T_{{\mathrm c}ij}/10),$  and stripe-like structures  are observed for ($4T_{{\mathrm c}ij}/10, T_{{\mathrm c}ij}/2$) and ($9T_{{\mathrm c}ij}/10, T_{{\mathrm c}ij}$). The circular (Figure \ref{fig:tf4} (a)) and  elliptic (Figure \ref{fig:tf4} (d),(g))   compact structures differ among themselves. The circular case (Figure \ref{fig:tf4} (b), (c)) is similar to the simulated 49-day-long  monitoring  of  the single circular BLR case in  \citet[][see their Fig. 3 (b)]{2004PASP..116..465H}.
The  randomisation of the eccentricity of the cloud trajectories deforms the stripe-like structures strikingly. 

All  2DTFs  presented here  are strongly asymmetric with respect to the zero value of the radial velocity. In general, this means that  either the blue or the red wing responds most rapidly.

\subsection{Time-averaged spectral lines}

\begin{figure*}
\centering
\begin{subfigure}{6cm}
    \centering
    \includegraphics[width=5cm]{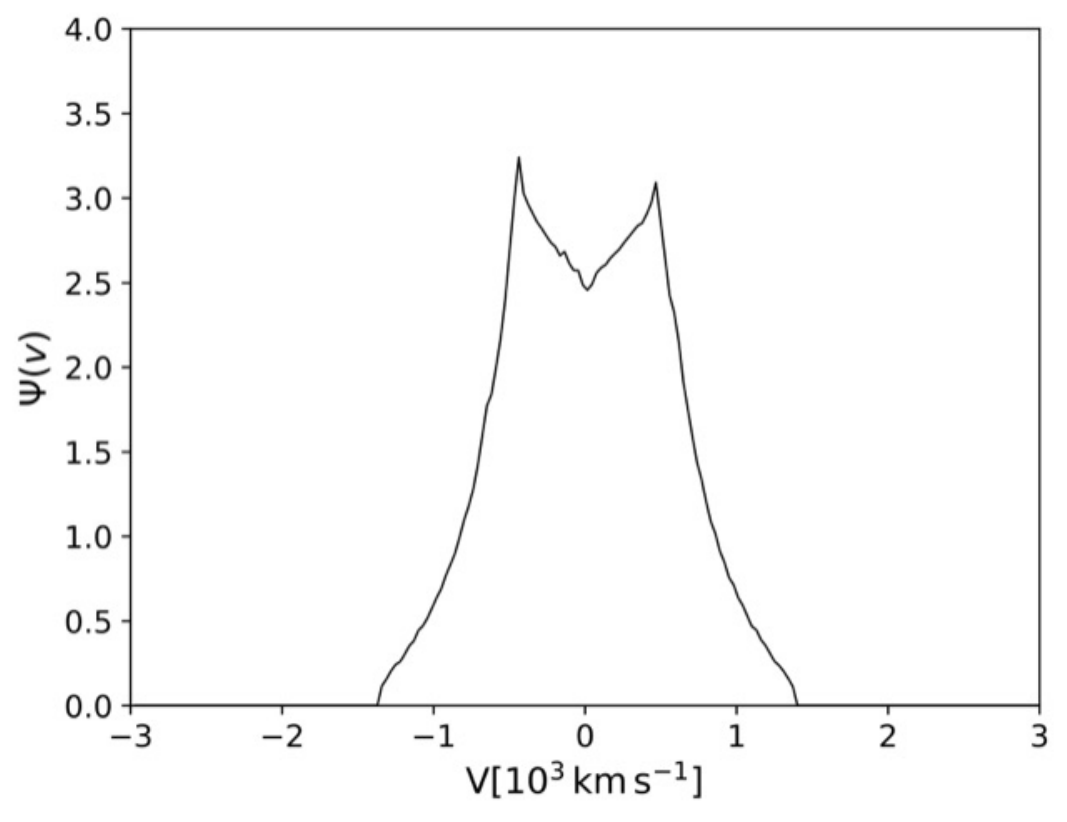}
    \caption{}
\end{subfigure}%
\begin{subfigure}{6cm}
    \centering
    \includegraphics[width=5cm]{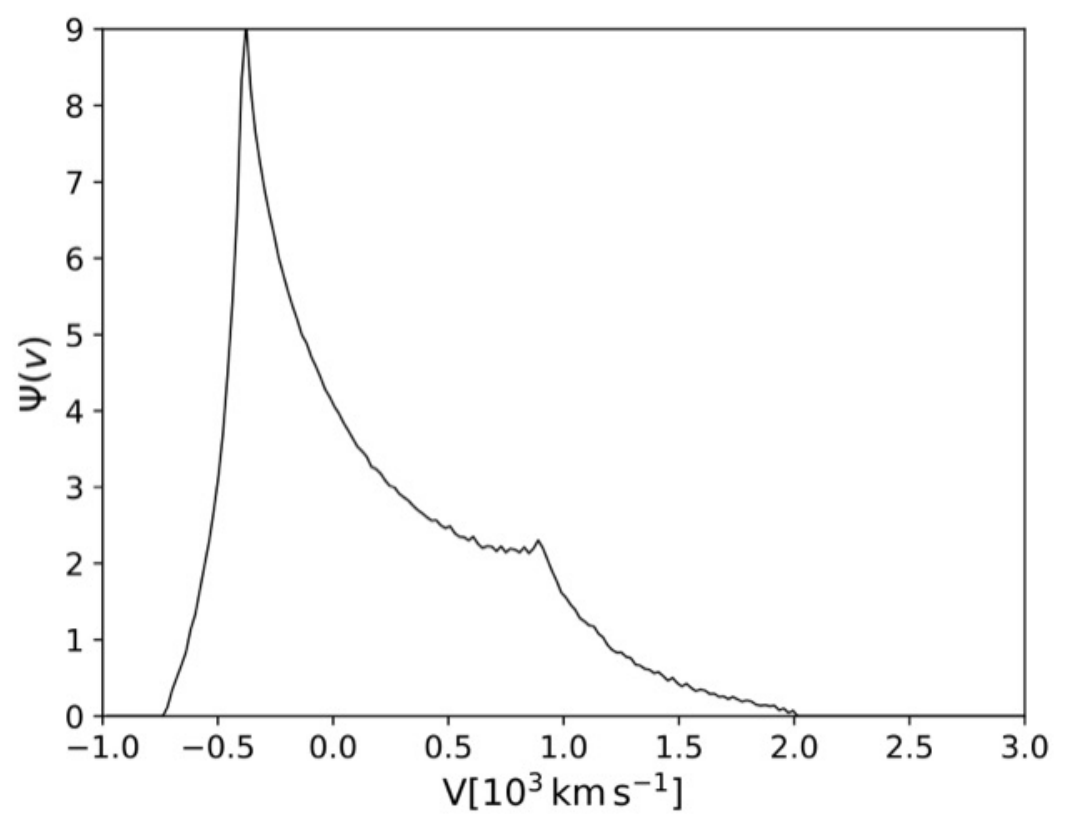}
    \caption{}
\end{subfigure}
\begin{subfigure}{6cm}
    \centering
    \includegraphics[width=5cm]{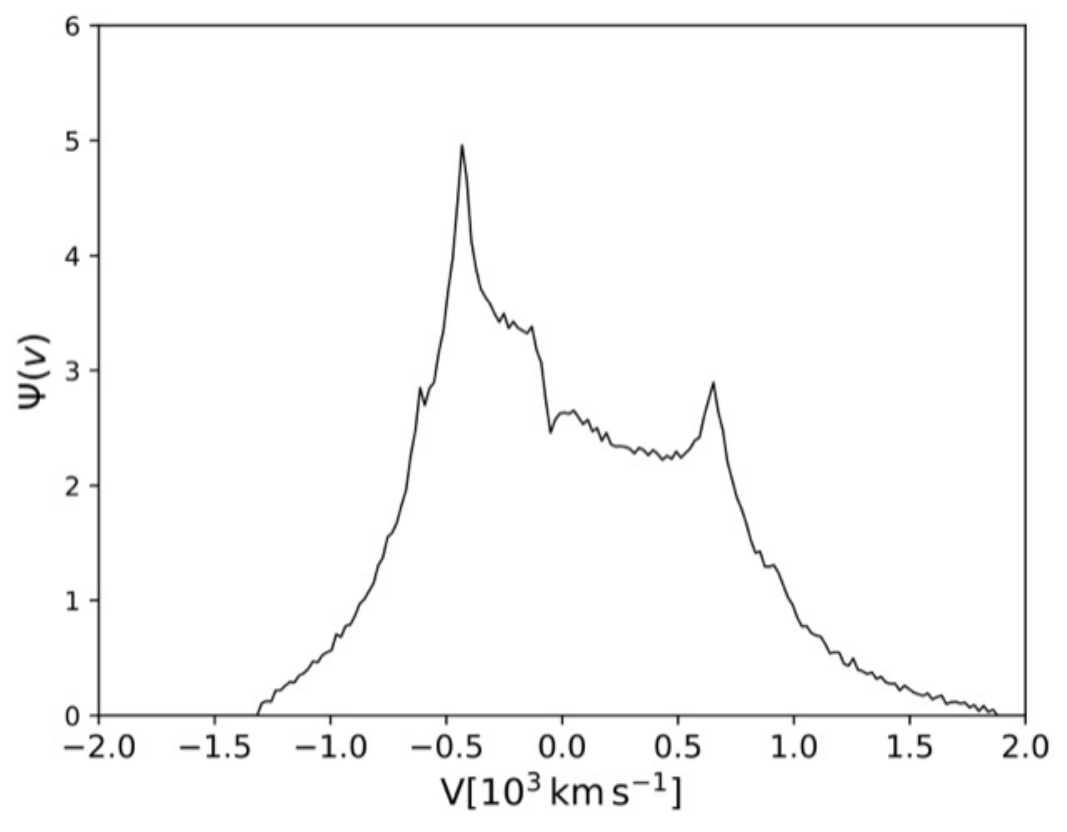}
    \caption{}
\end{subfigure}
\begin{subfigure}{6cm}
    \centering
    \includegraphics[width=5cm]{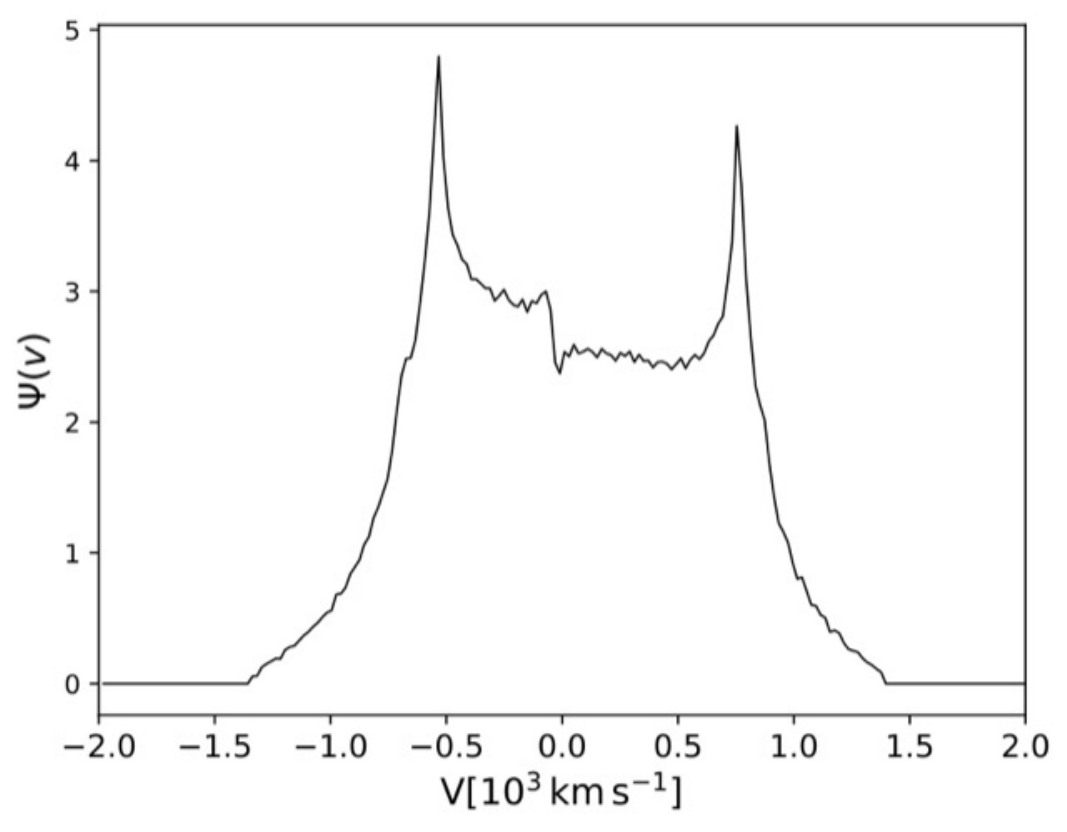}
    \caption{}
\end{subfigure}
\begin{subfigure}{6cm}
    \centering
    \includegraphics[width=5cm]{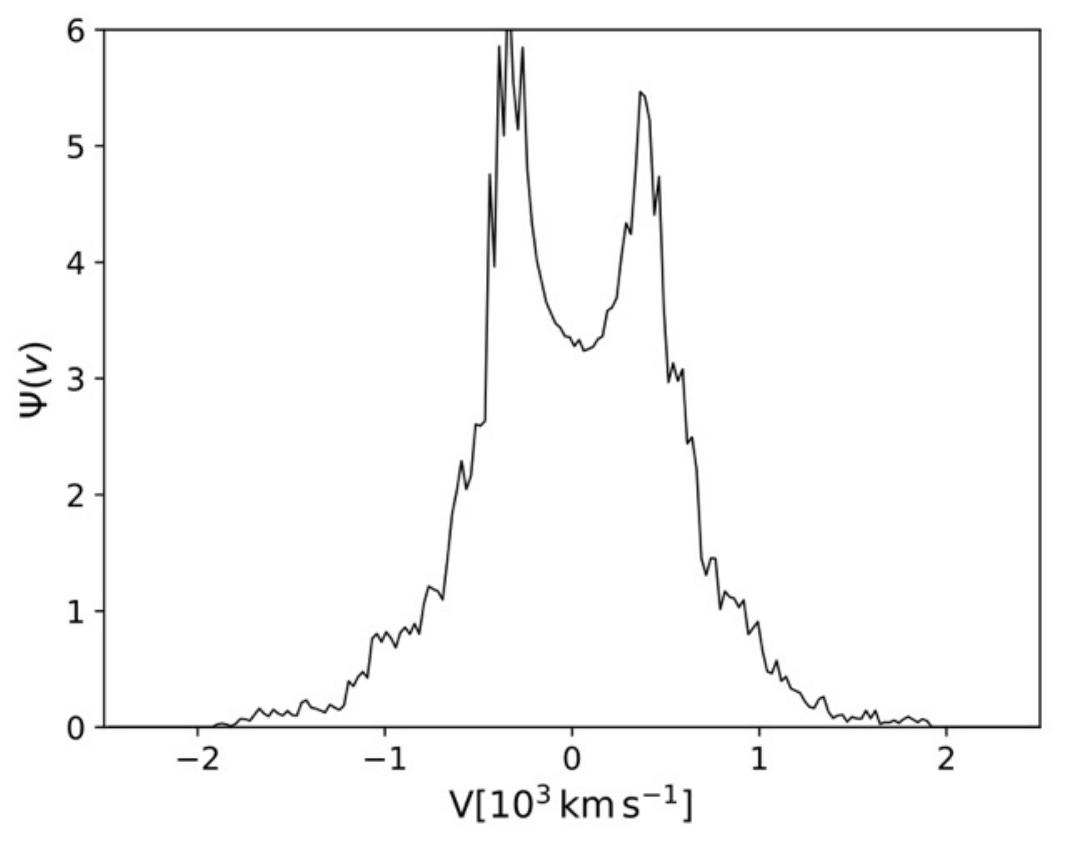}
    \caption{}
\end{subfigure}
\begin{subfigure}{6cm}
    \centering
    \includegraphics[width=5cm]{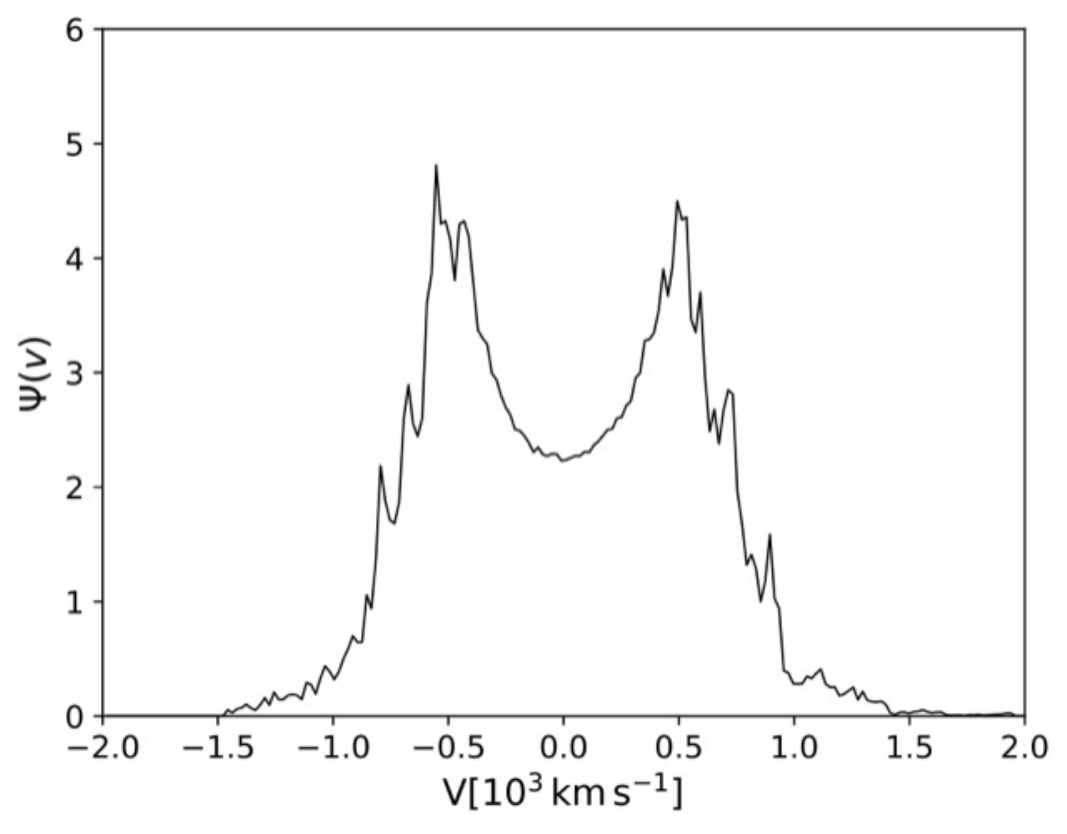}
    \caption{}
\end{subfigure}
\begin{subfigure}{6cm}
    \centering
    \includegraphics[width=5cm]{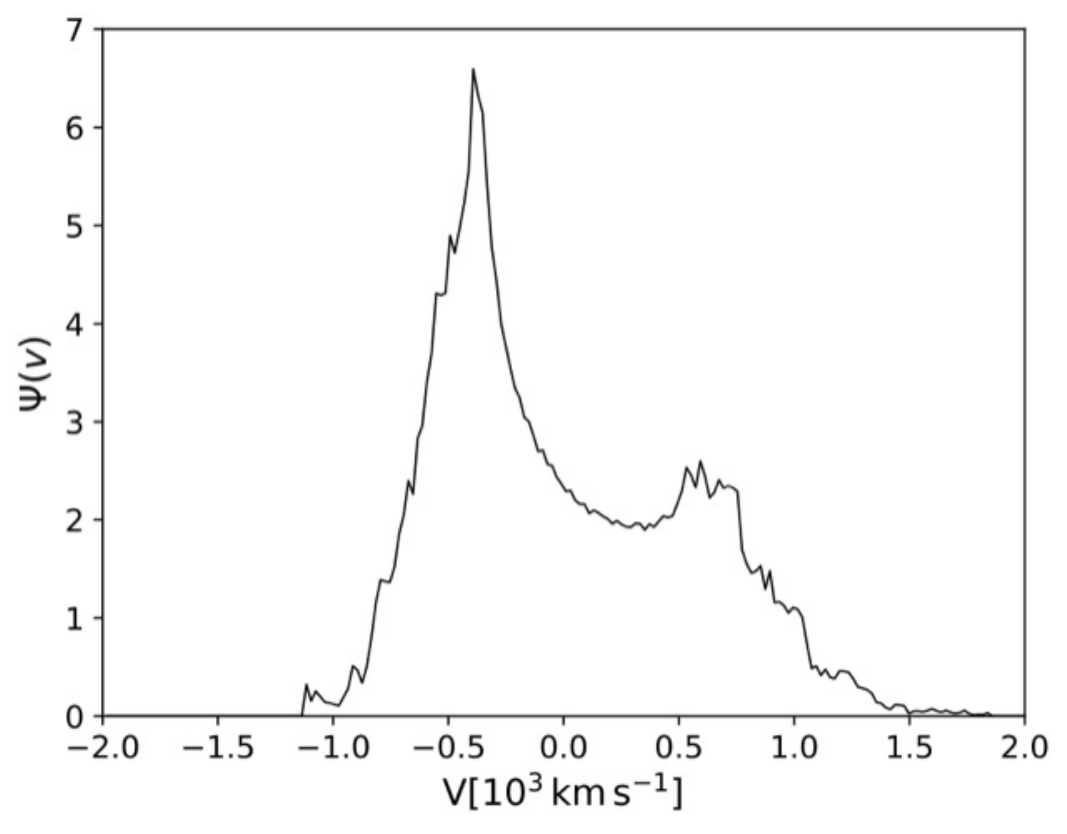}
    \caption{}
\end{subfigure}
\begin{subfigure}{6cm}
\centering
    \includegraphics[width=5cm]{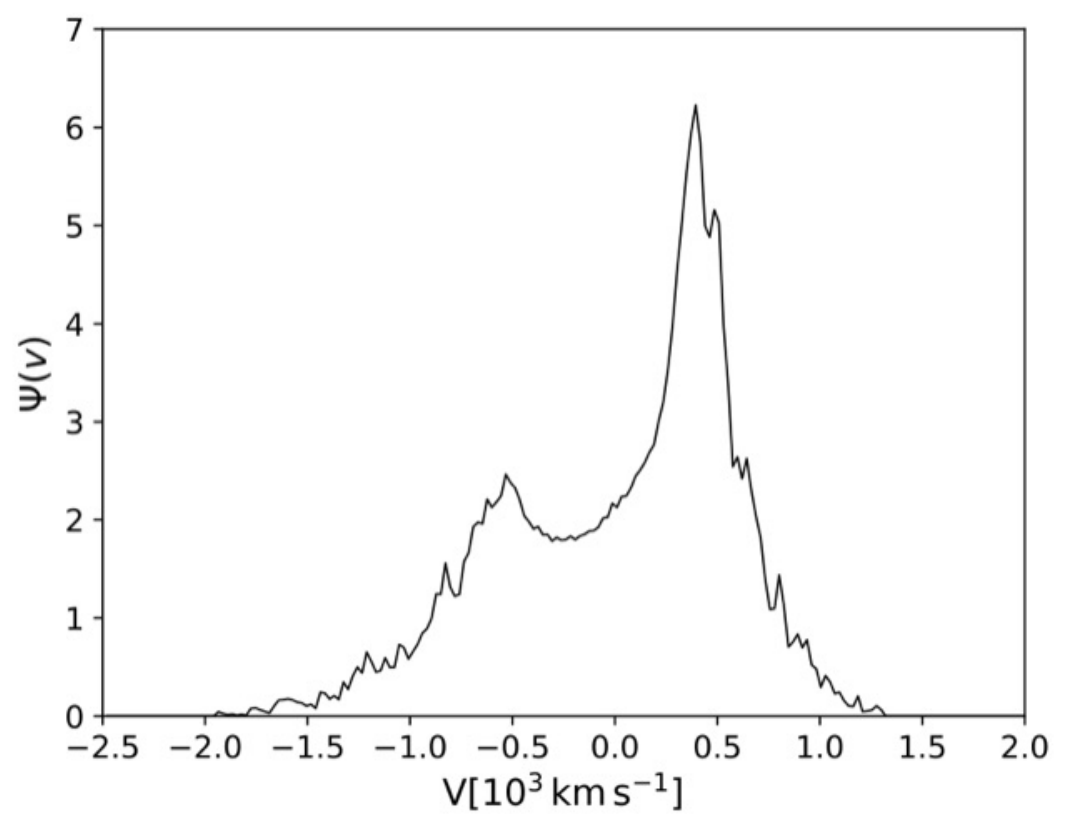}
    \caption{}
\end{subfigure}
\begin{subfigure}{6cm}
  \hspace*{\fill}%
  \end{subfigure}
\caption{Spectrum corresponding to theoretical 2DTF  maps obtained for different geometries of a single BLR in Fig. \ref{fig:tf1}.}
 \label{fig:ttf1}
\end{figure*}

\begin{figure*}
\centering
\begin{subfigure}{6cm}
    \centering
    \includegraphics[width=5cm]{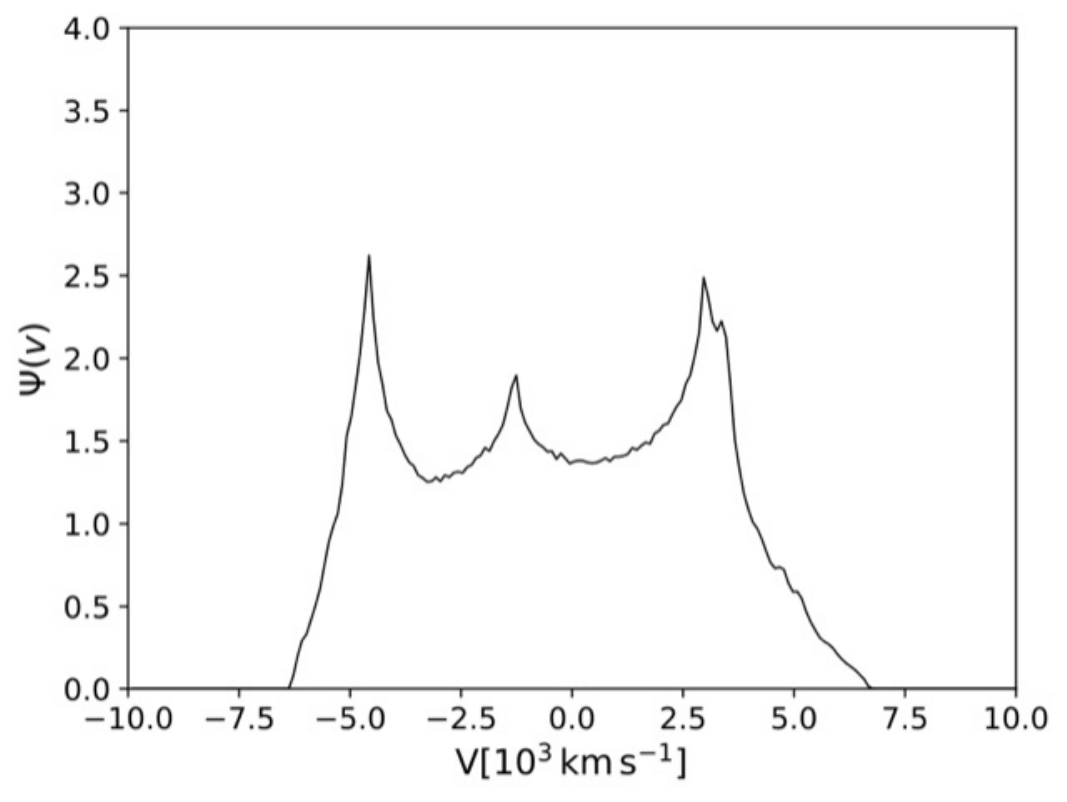}
    \caption{}
\end{subfigure}%
\begin{subfigure}{6cm}
    \centering
    \includegraphics[width=5cm]{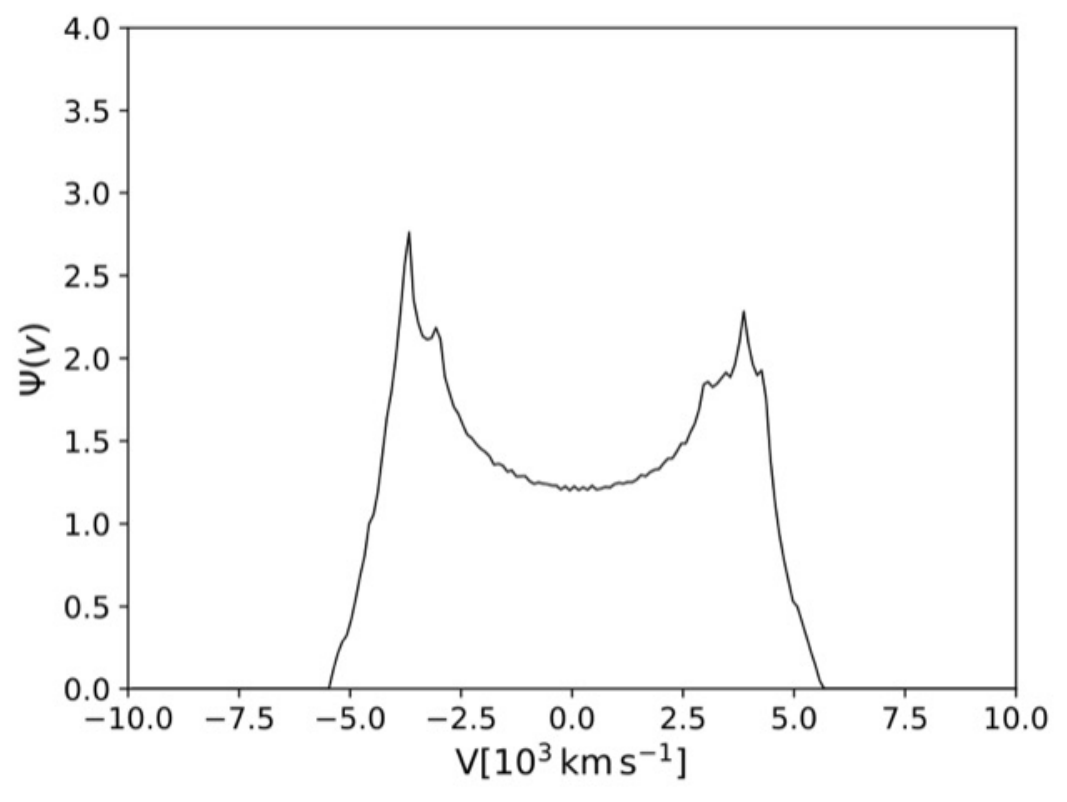}
    \caption{}
\end{subfigure}
\begin{subfigure}{6cm}
    \centering
    \includegraphics[width=5cm]{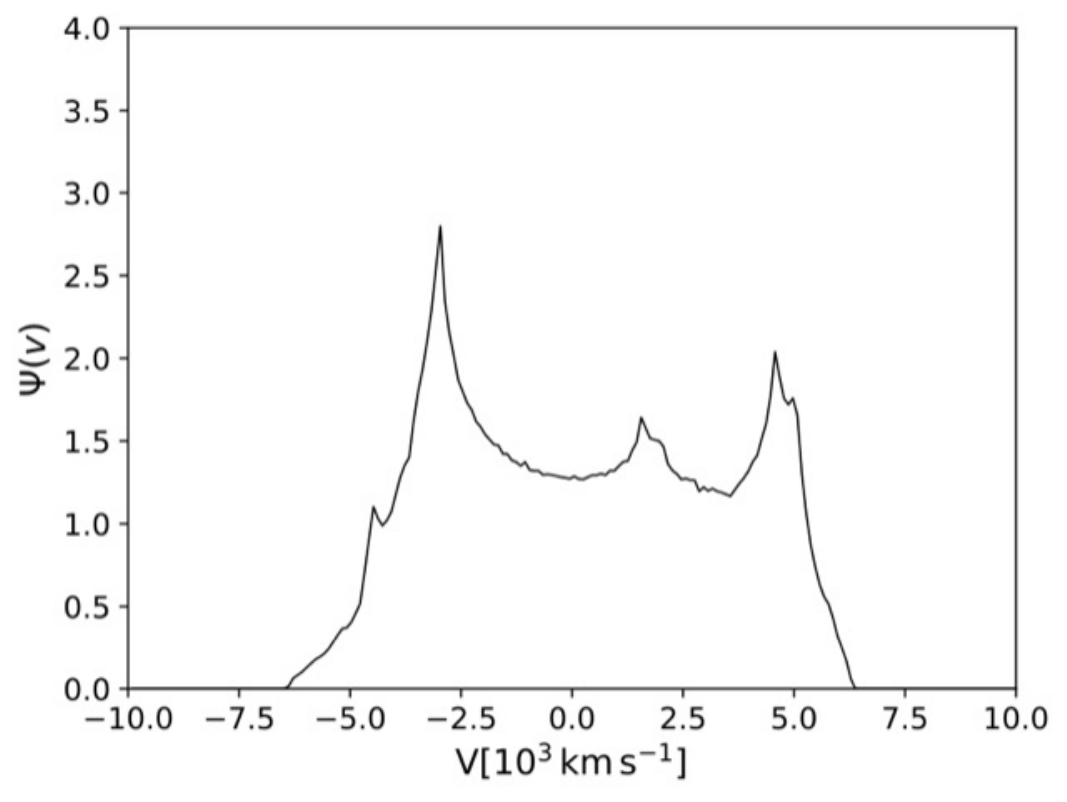}
    \caption{}
\end{subfigure}
\begin{subfigure}{6cm}
    \centering
    \includegraphics[width=5cm]{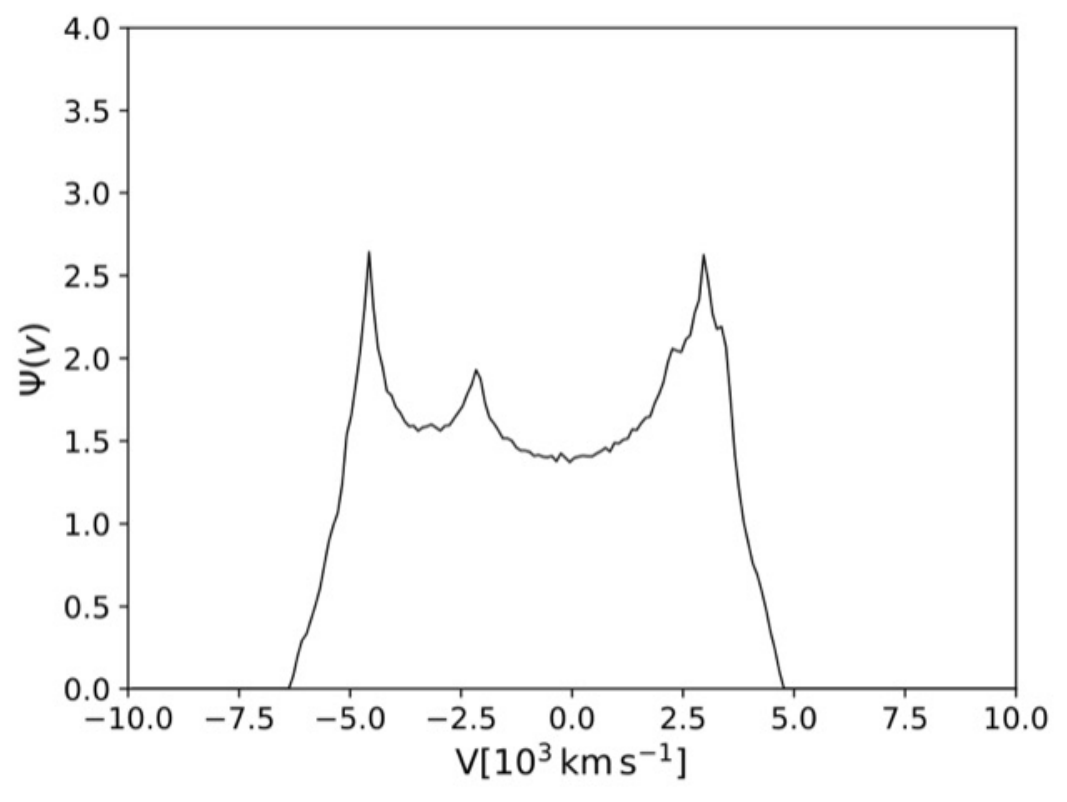}
    \caption{}
\end{subfigure}
\begin{subfigure}{6cm}
    \centering
    \includegraphics[width=5cm]{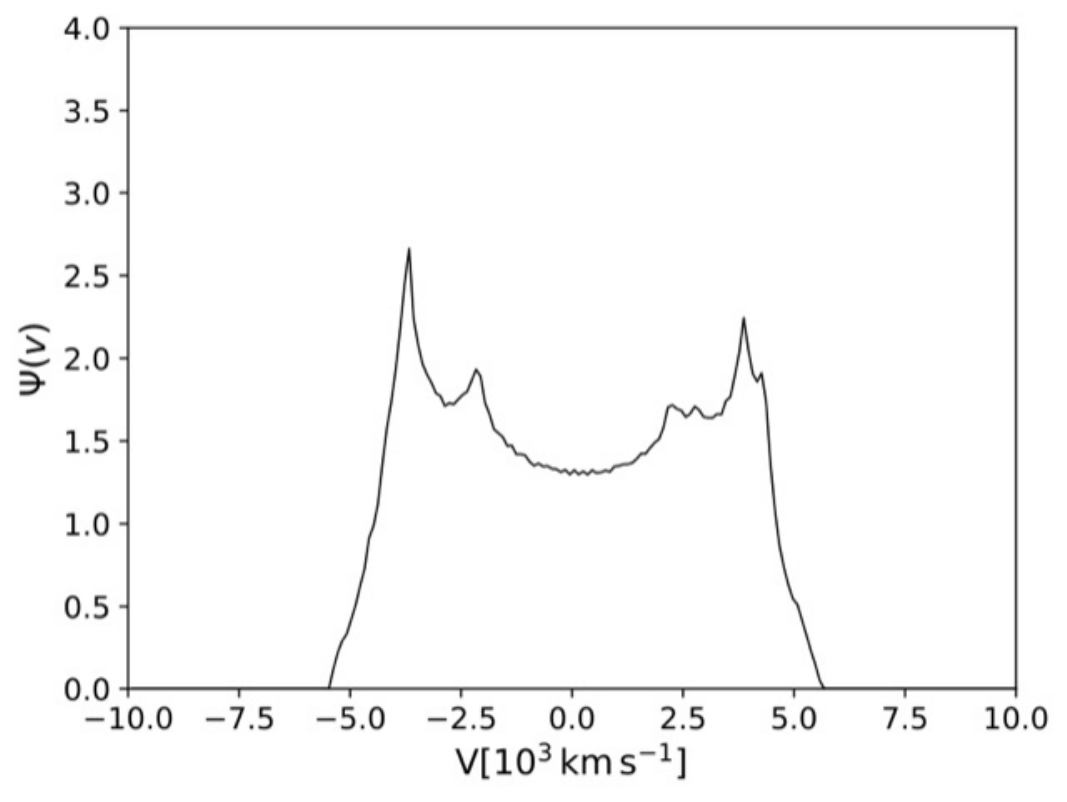}
    \caption{}
\end{subfigure}
\begin{subfigure}{6cm}
    \centering
    \includegraphics[width=5cm]{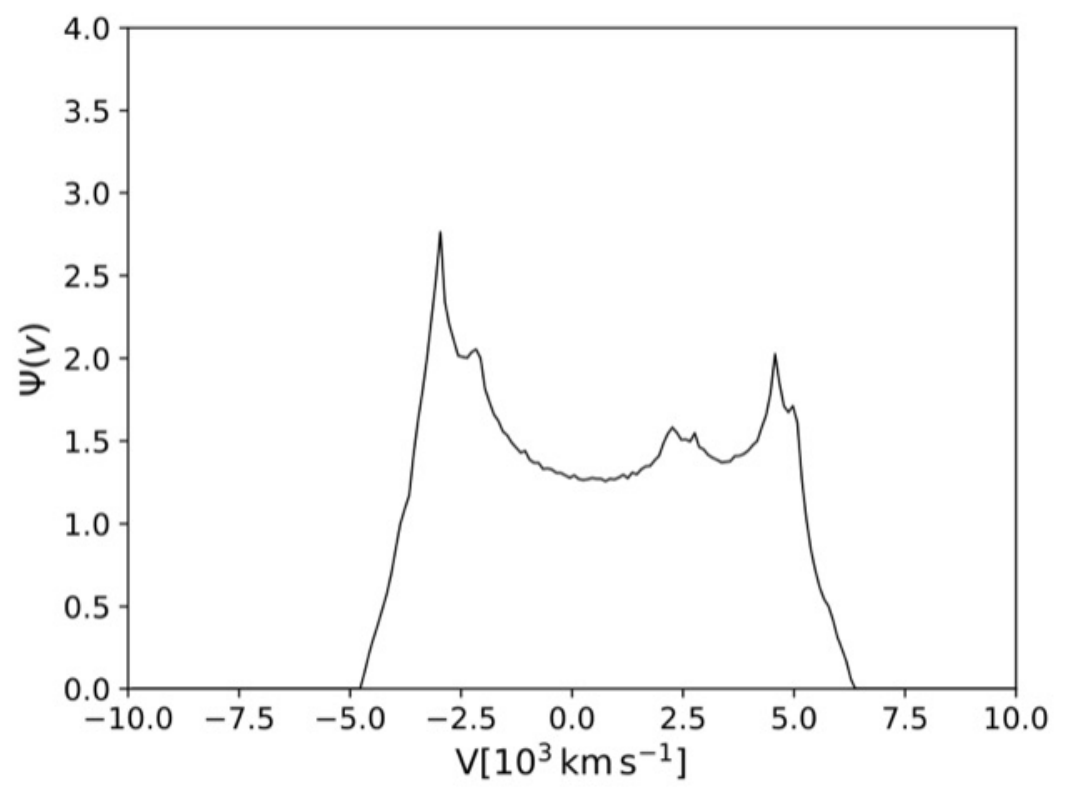}
    \caption{}
\end{subfigure}
\begin{subfigure}{6cm}
    \centering
    \includegraphics[width=5cm]{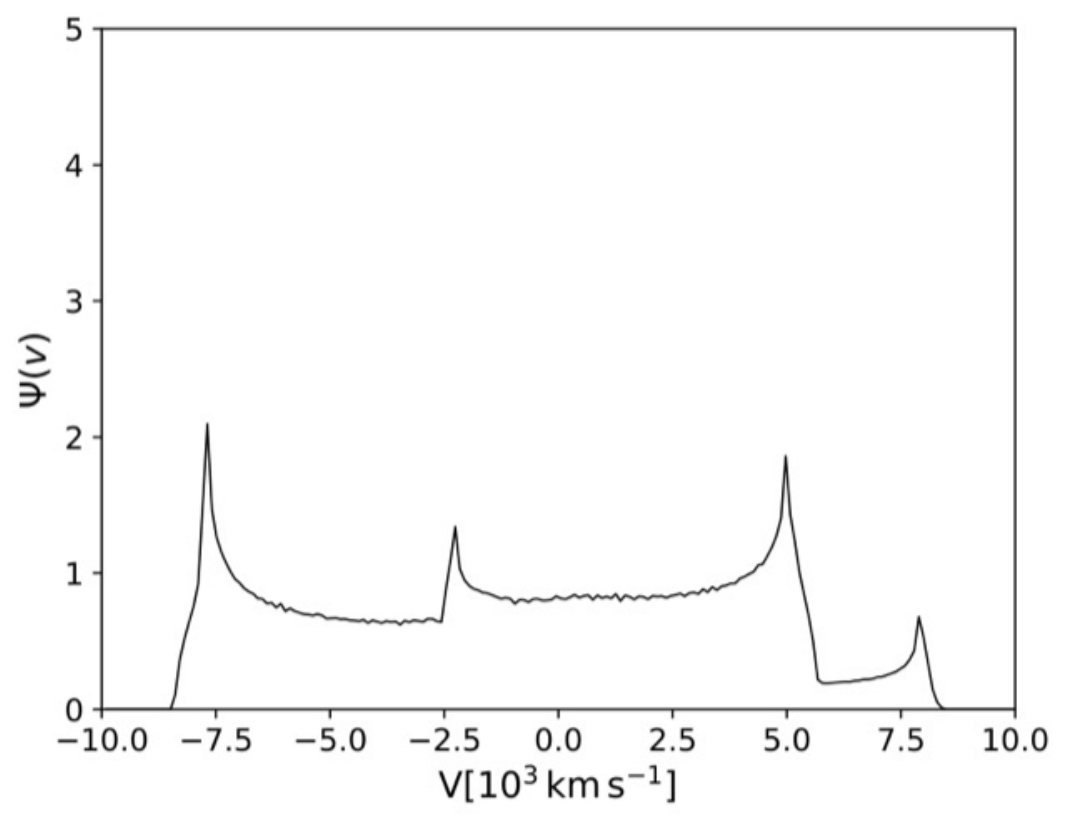}
    \caption{}
\end{subfigure}
\begin{subfigure}{6cm}
\centering
    \includegraphics[width=5cm]{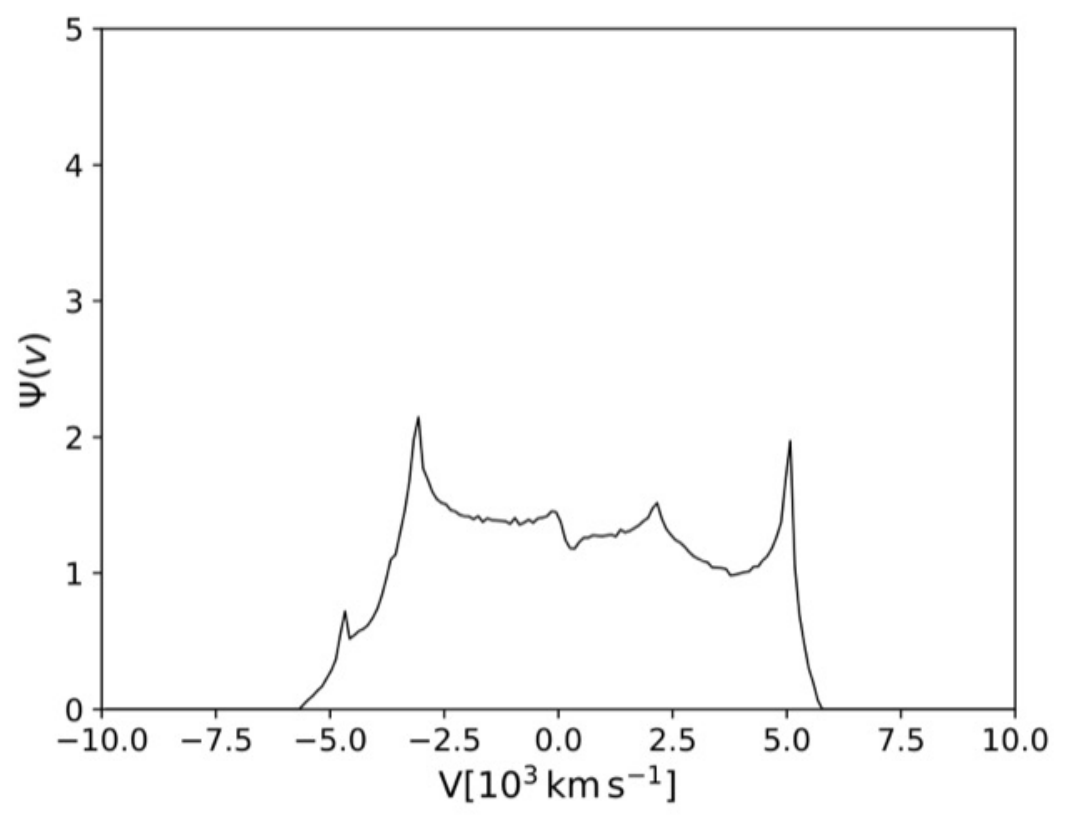}
    \caption{}
\end{subfigure}
\begin{subfigure}{6cm}
\centering
    \includegraphics[width=5cm]{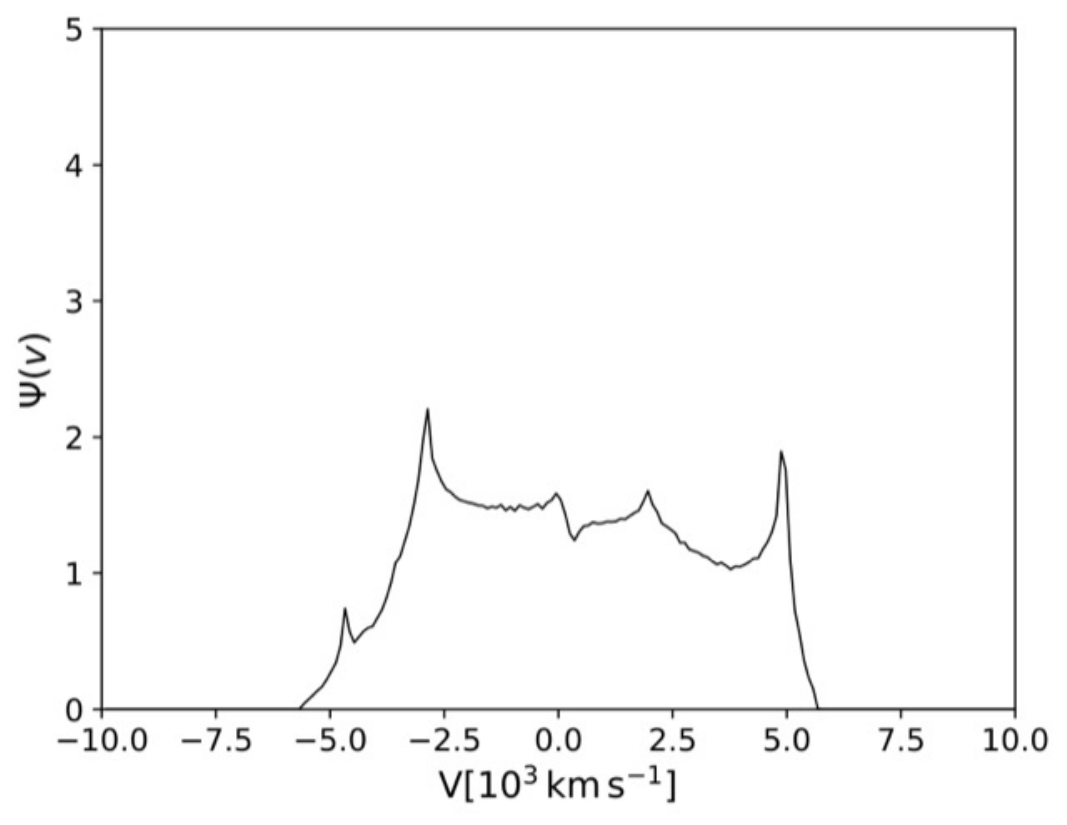}
    \caption{}
\end{subfigure}
\begin{subfigure}{6cm}
    \centering
    \includegraphics[width=5cm]{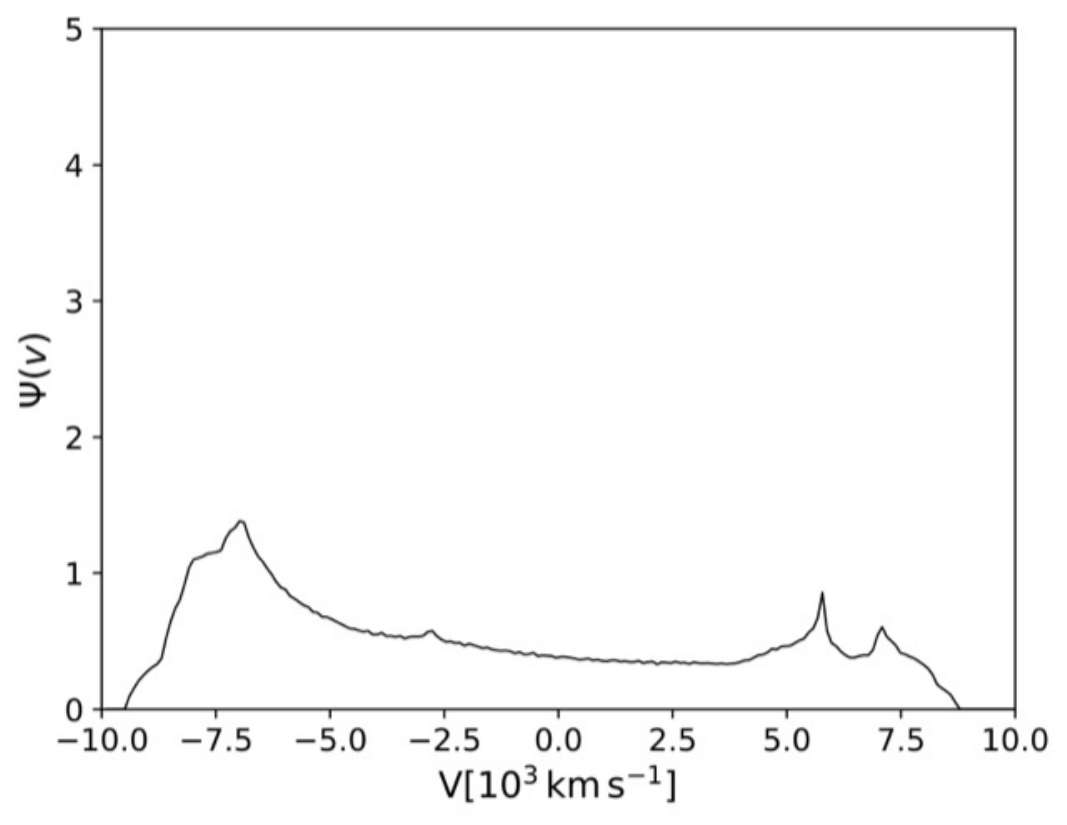}
    \caption{}
\end{subfigure}
\begin{subfigure}{6cm}
\centering
    \includegraphics[width=5cm]{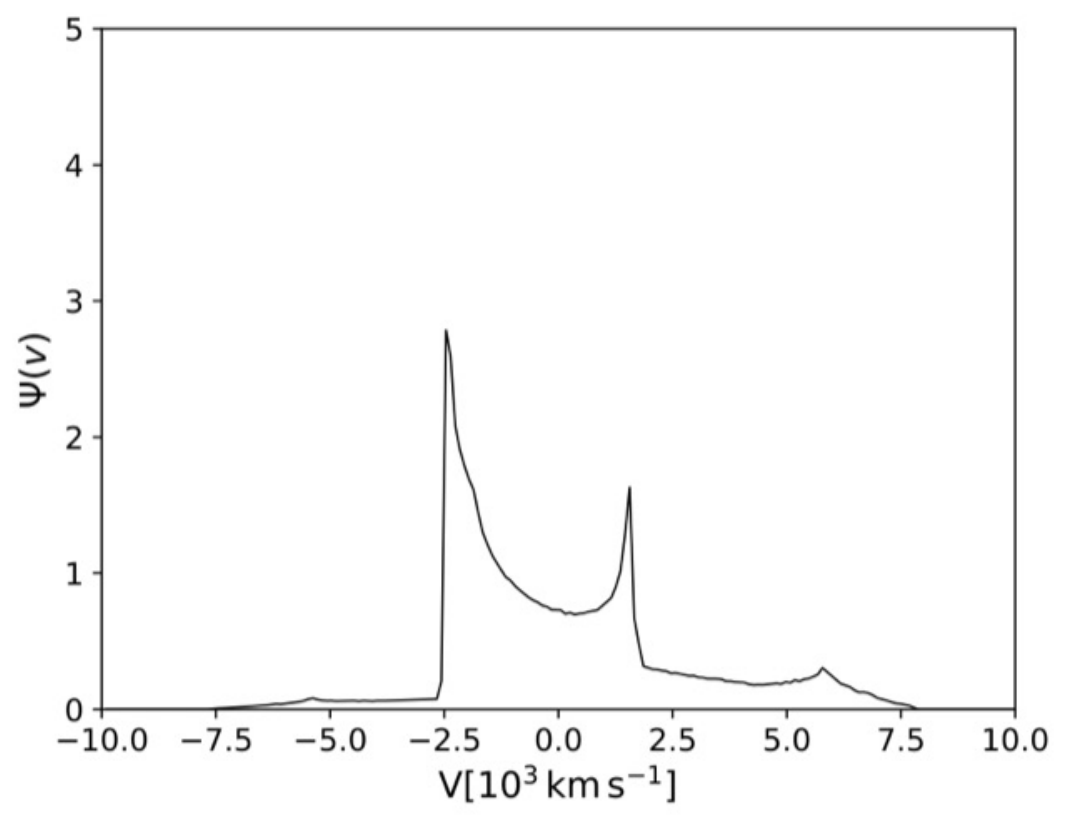}
    \caption{}
\end{subfigure}
\begin{subfigure}{6cm}
\centering
    \includegraphics[width=5cm]{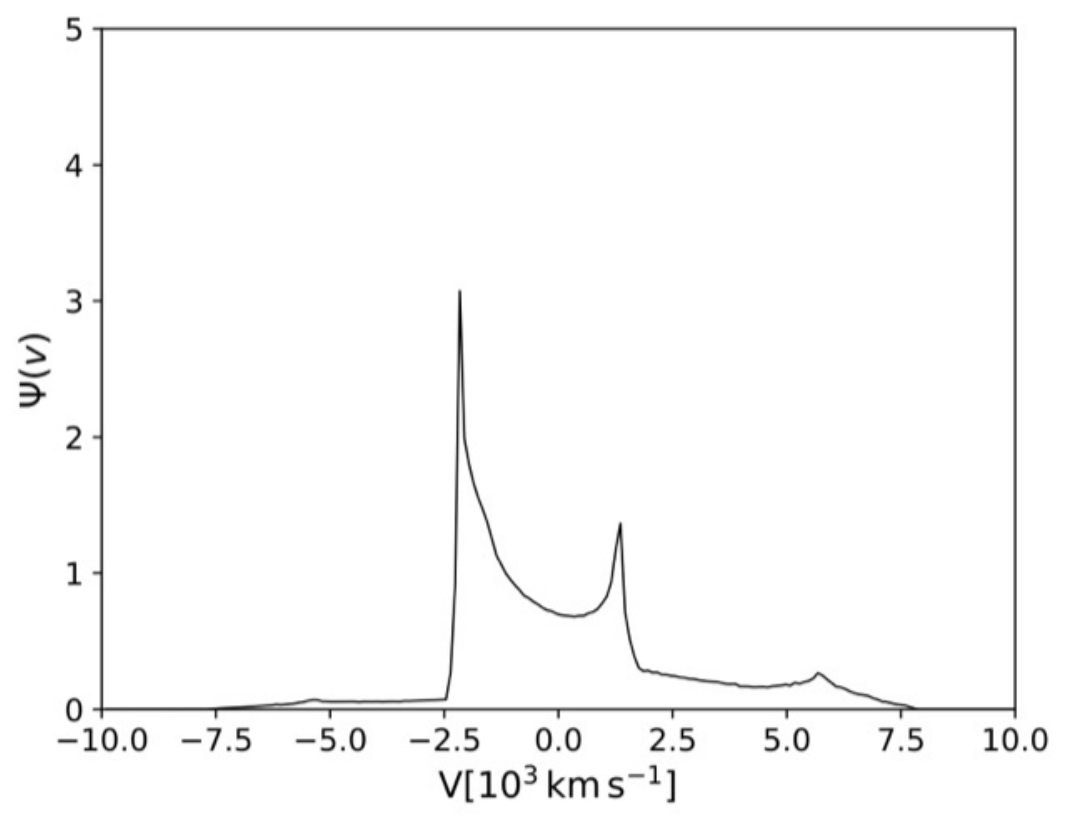}
    \caption{}
\end{subfigure}
 \caption{Spectrum corresponding to theoretical 2DTF  maps obtained for different geometries of the binary disc-like  BLRs given in  Fig. \ref{fig:tf2}.}
 \label{fig:ttf2}
\end{figure*}

\begin{figure*}
\centering
\begin{subfigure}{6cm}
    \centering
    \includegraphics[width=5cm]{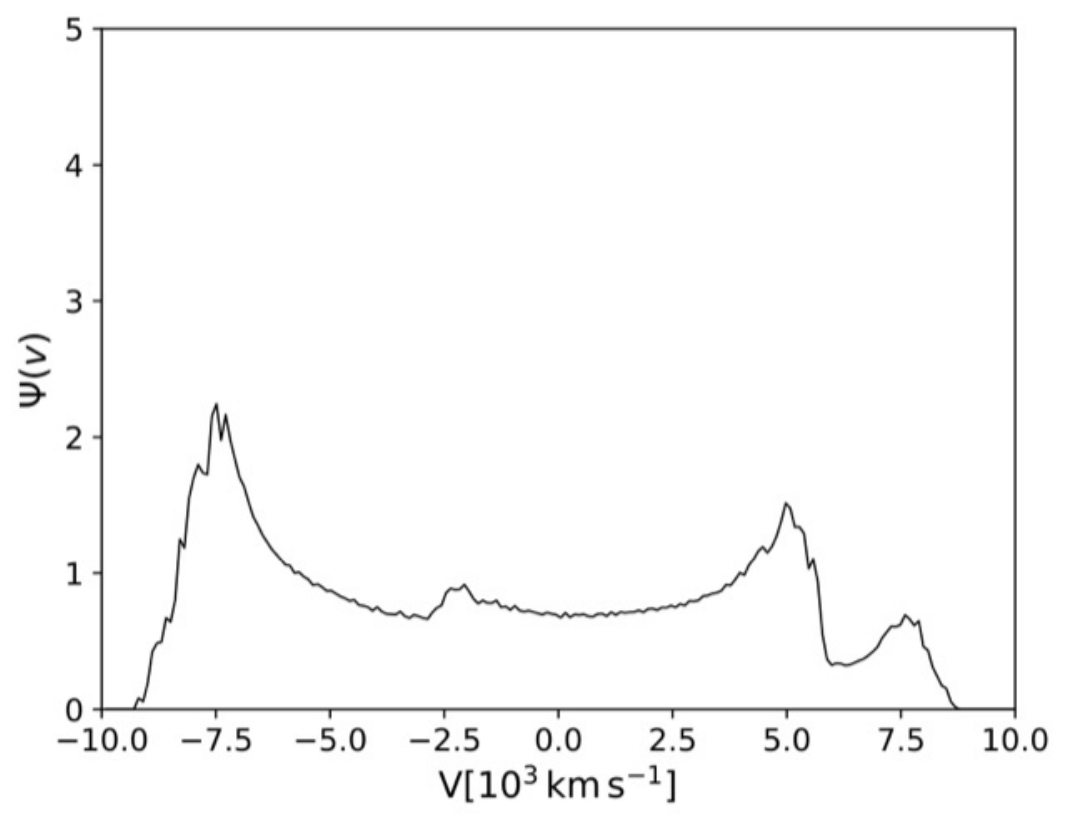}
    \caption{}
\end{subfigure}%
\begin{subfigure}{6cm}
    \centering
    \includegraphics[width=5cm]{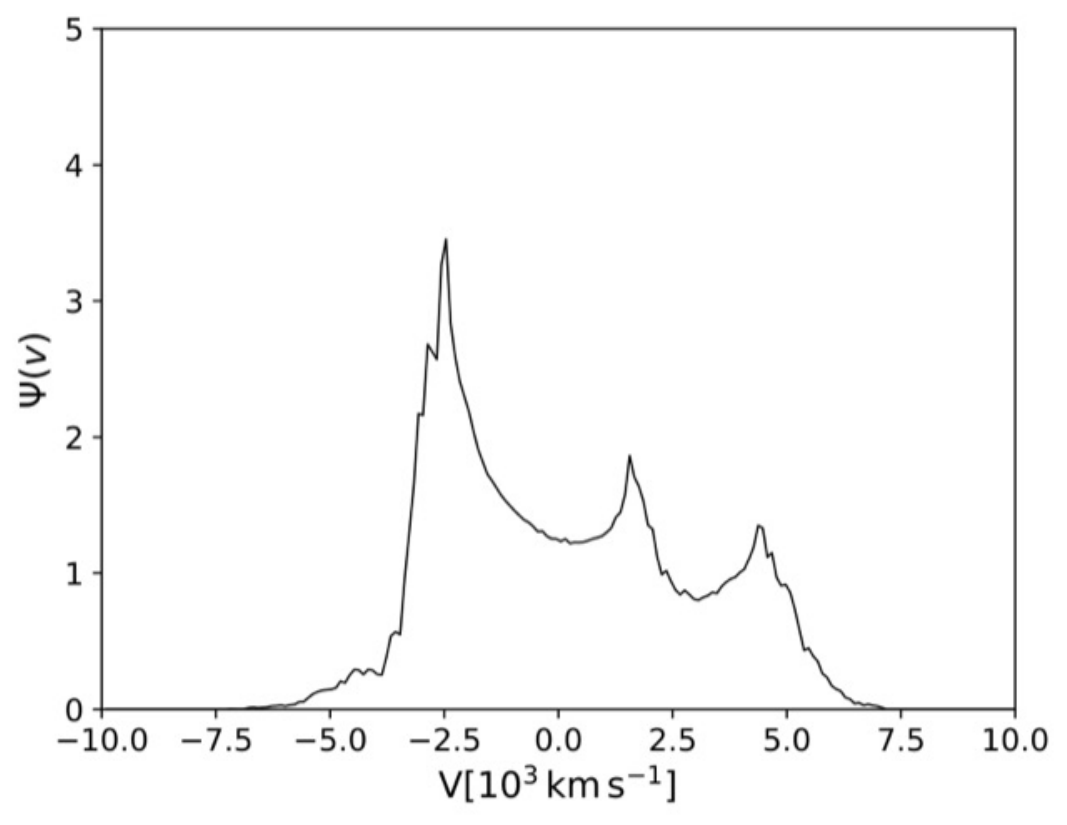}
    \caption{}
\end{subfigure}
\begin{subfigure}{6cm}
    \centering
    \includegraphics[width=5cm]{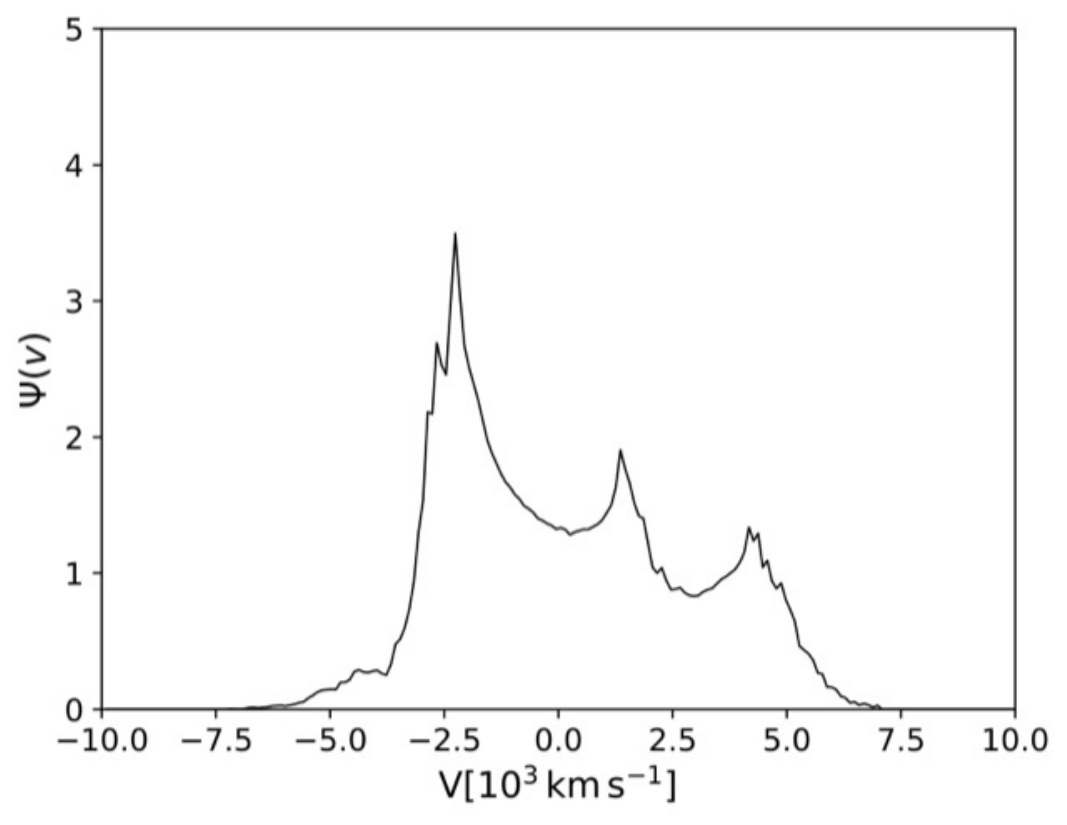}
    \caption{}
\end{subfigure}
\begin{subfigure}{6cm}
    \centering
    \includegraphics[width=5cm]{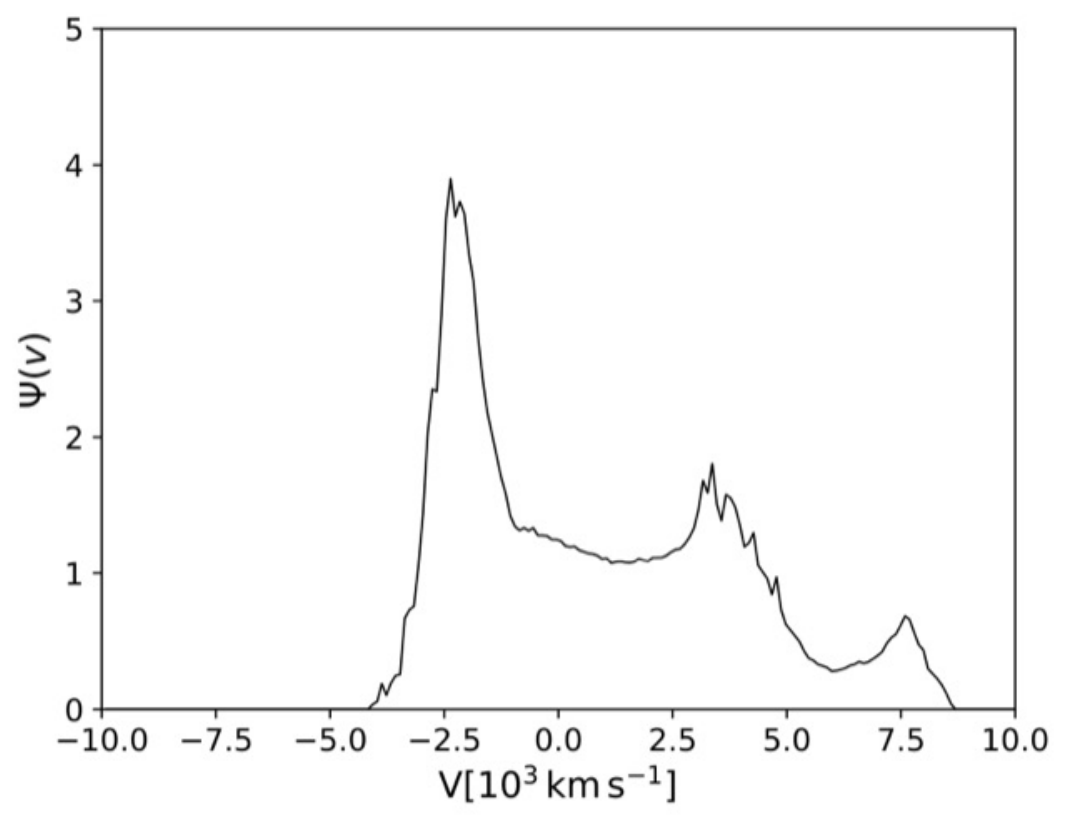}
    \caption{}
\end{subfigure}
\begin{subfigure}{6cm}
    \centering
    \includegraphics[width=5cm]{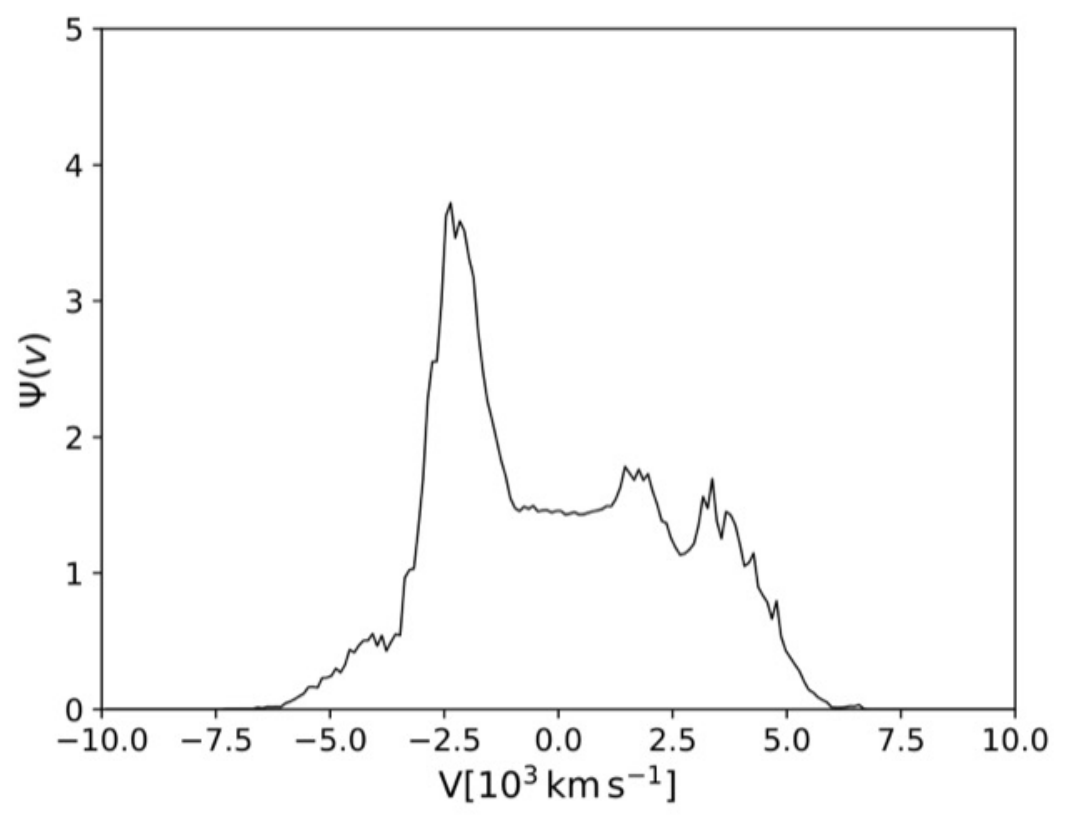}
    \caption{}
\end{subfigure}
\begin{subfigure}{6cm}
    \centering
    \includegraphics[width=5cm]{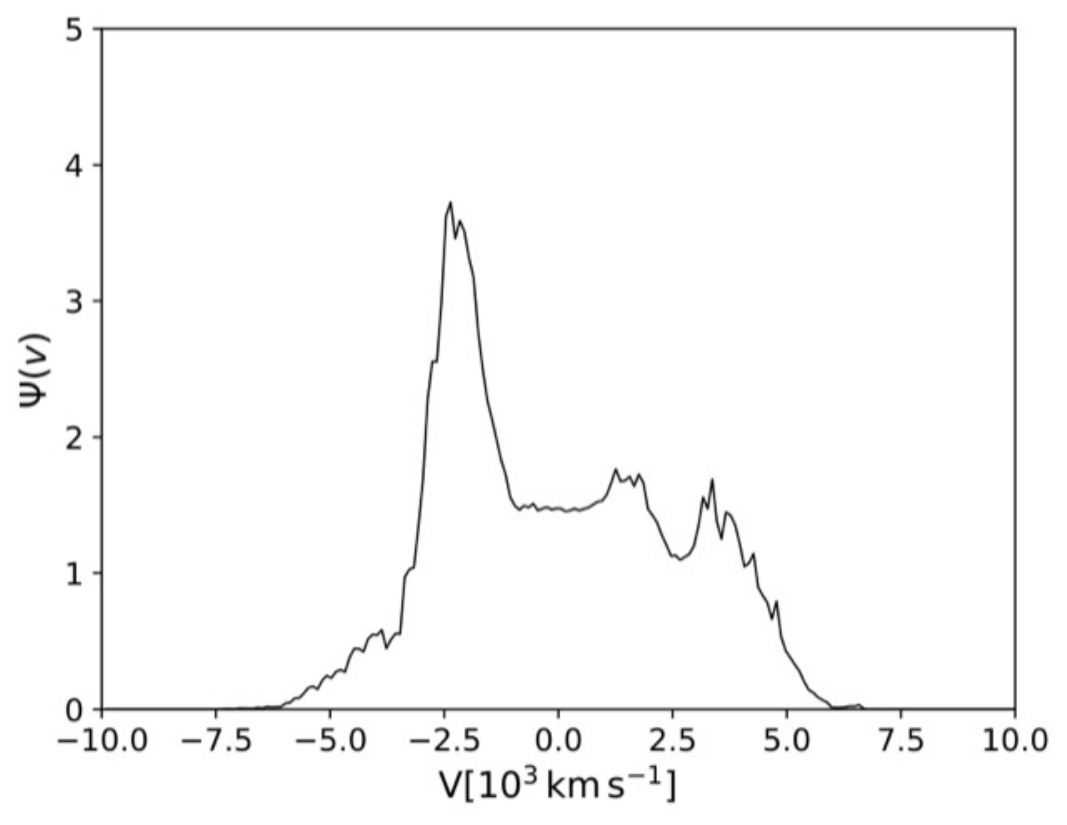}
    \caption{}
\end{subfigure}
\begin{subfigure}{6cm}
    \centering
    \includegraphics[width=5cm]{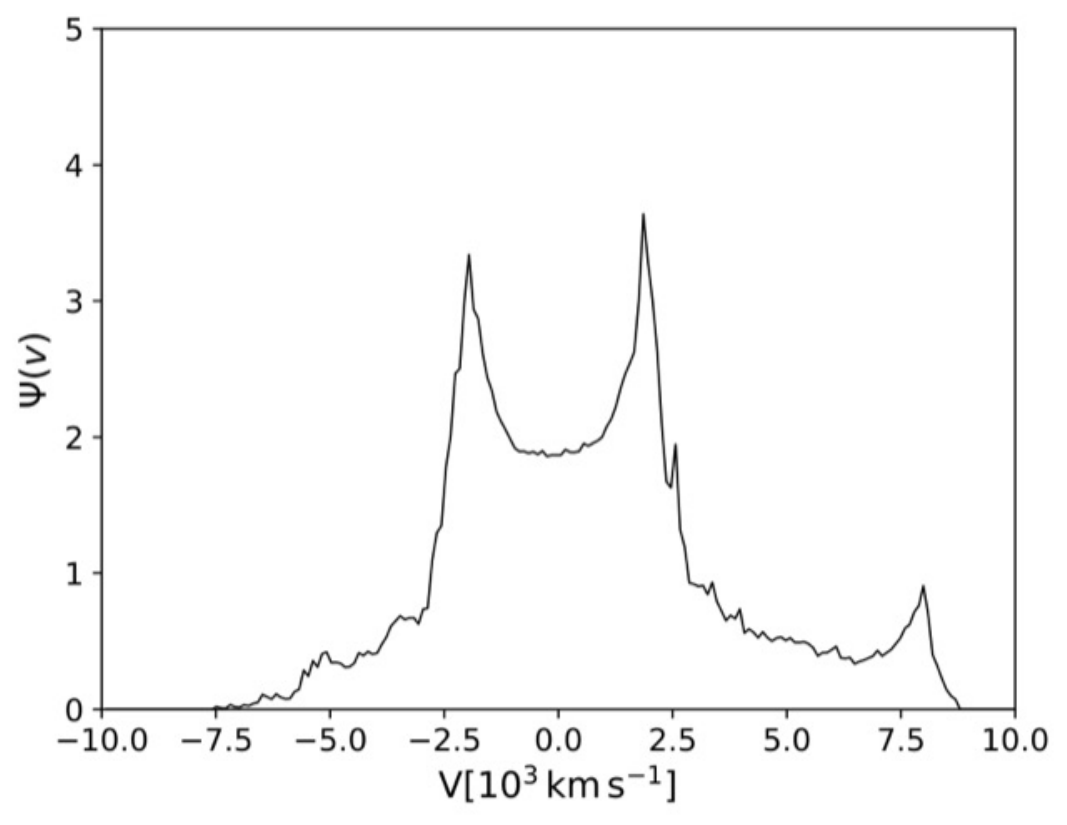}
    \caption{}
\end{subfigure}
\begin{subfigure}{6cm}
\centering
    \includegraphics[width=5cm]{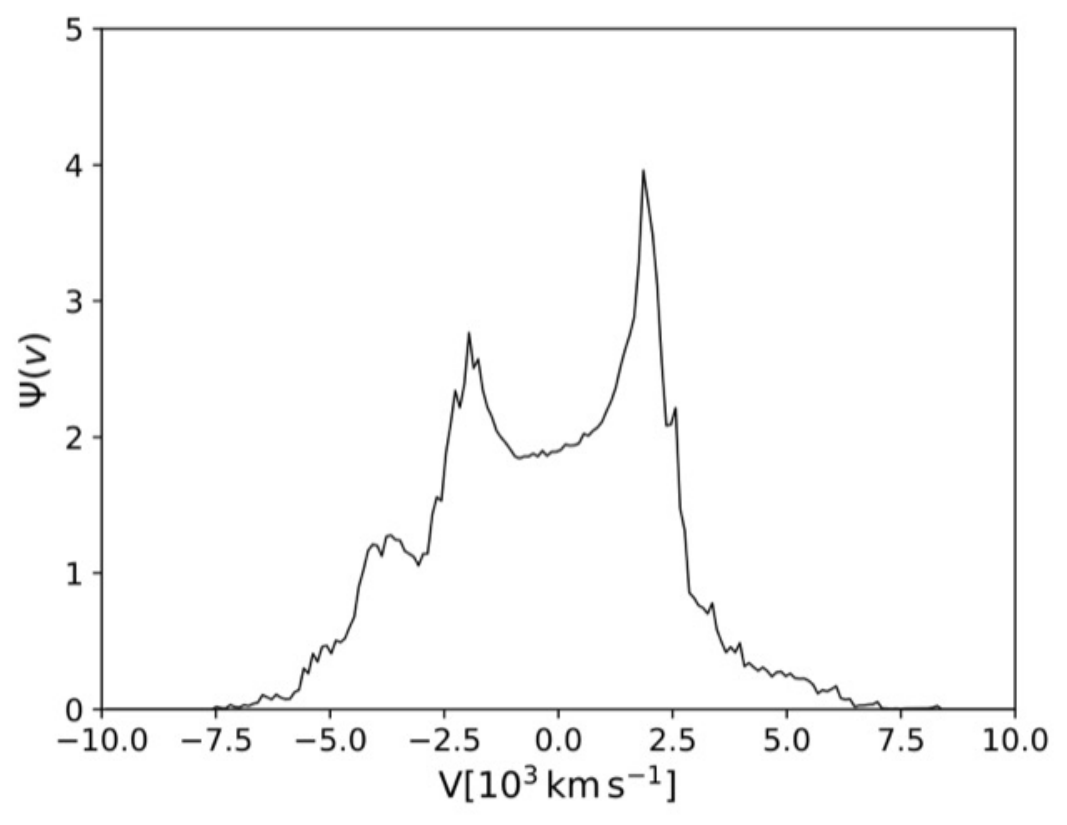}
    \caption{}
\end{subfigure}
\begin{subfigure}{6cm}
\centering
    \includegraphics[width=5cm]{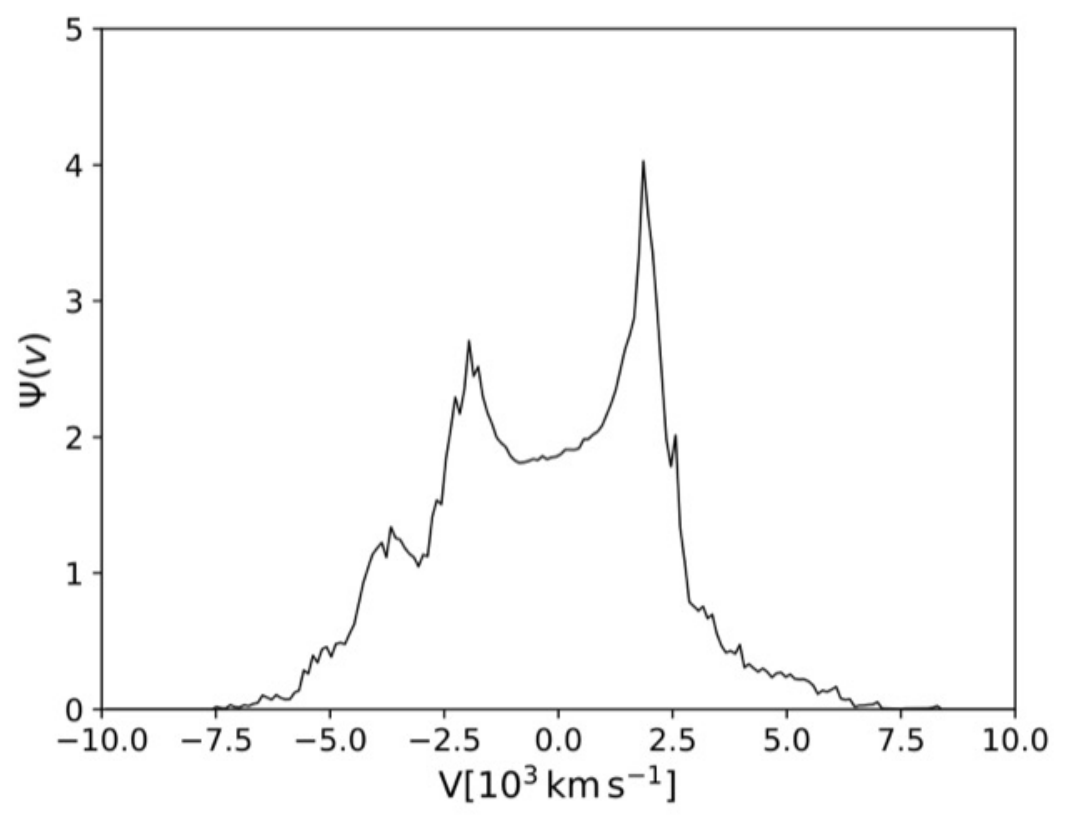}
    \caption{}
\end{subfigure}
\begin{subfigure}{6cm}
    \centering
    \includegraphics[width=5cm]{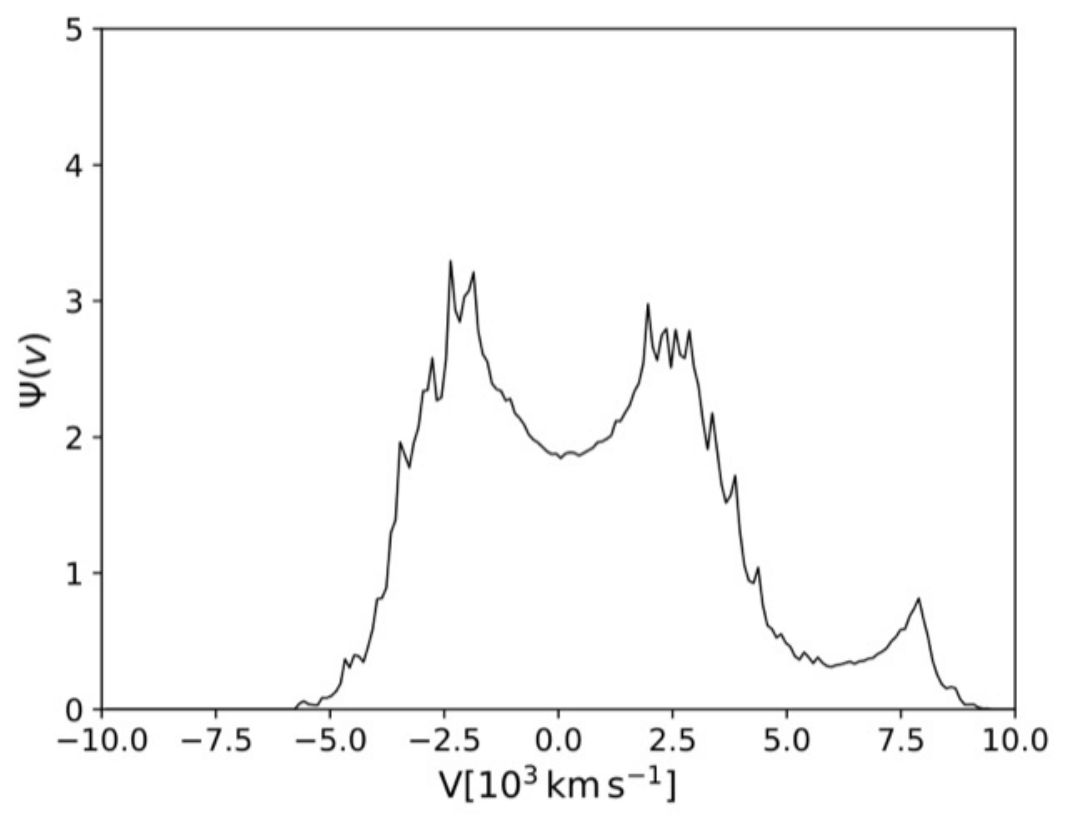}
    \caption{}
\end{subfigure}
\begin{subfigure}{6cm}
\centering
    \includegraphics[width=5cm]{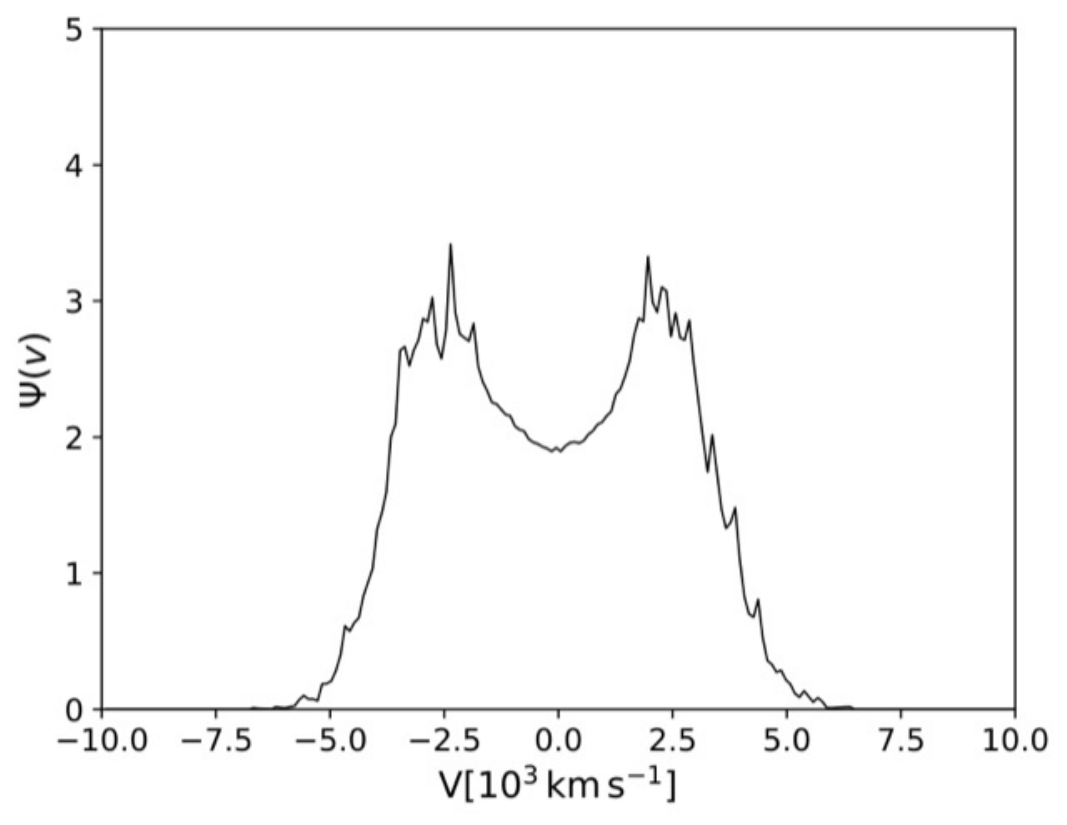}
    \caption{}
\end{subfigure}
\begin{subfigure}{6cm}
\centering
    \includegraphics[width=5cm]{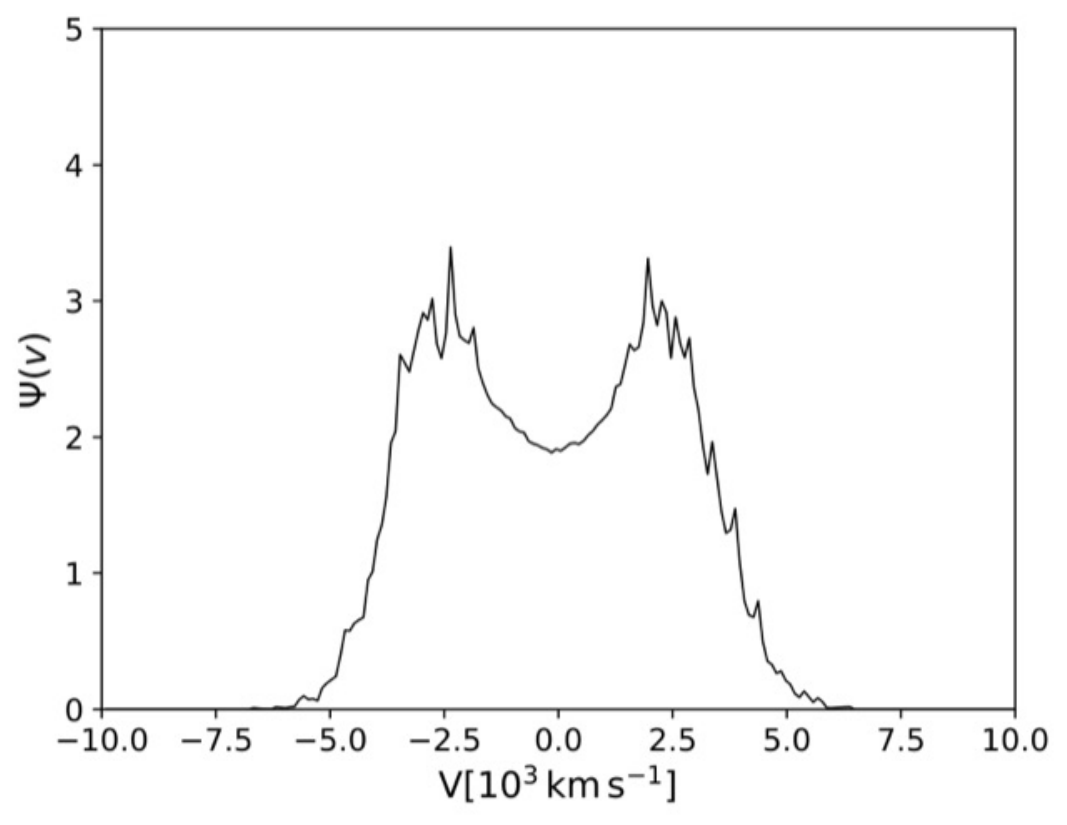}
    \caption{}
\end{subfigure}         
\caption{Spectrum corresponding to theoretical 2DTF maps obtained for different geometries of the binary disc-like  BLRs given in Fig. \ref{fig:tf3}.}
 \label{fig:ttf3}
\end{figure*}

\begin{figure*}
\centering
\begin{subfigure}{6cm}
    \centering
    \includegraphics[width=5cm]{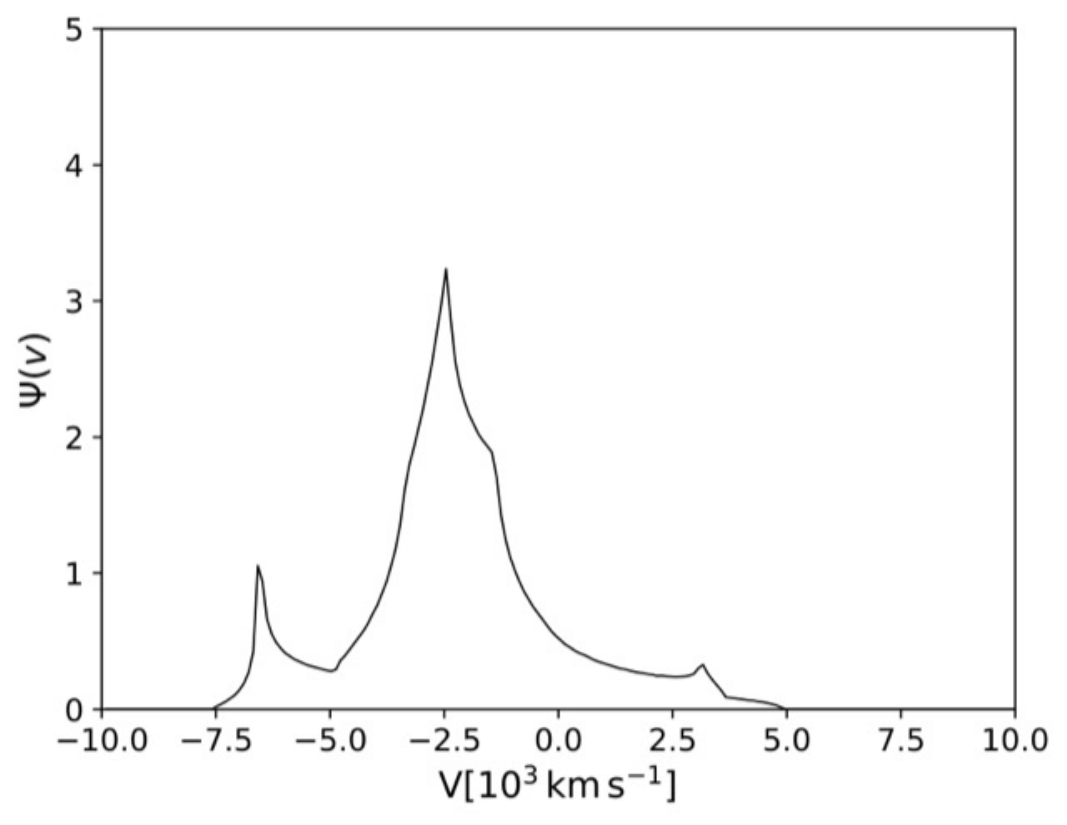}
    \caption{}
\end{subfigure}%
\begin{subfigure}{6cm}
    \centering
    \includegraphics[width=5cm]{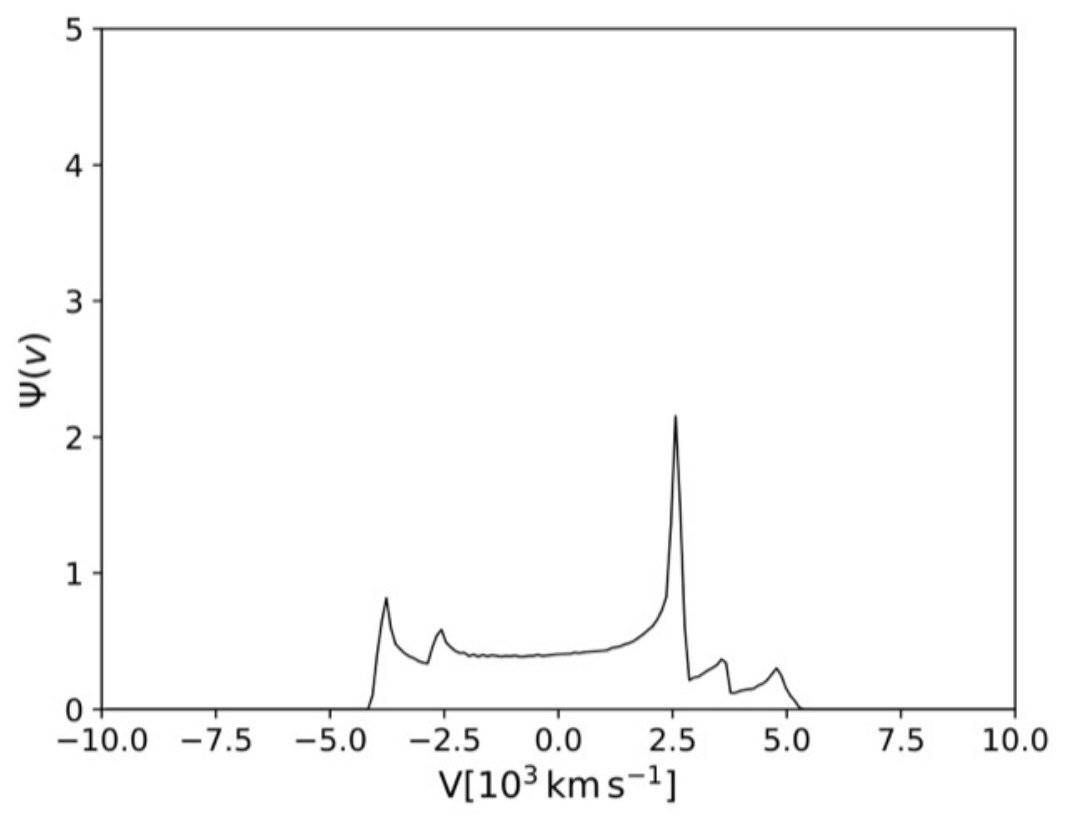}
    \caption{}
\end{subfigure}
\begin{subfigure}{6cm}
    \centering
    \includegraphics[width=5cm]{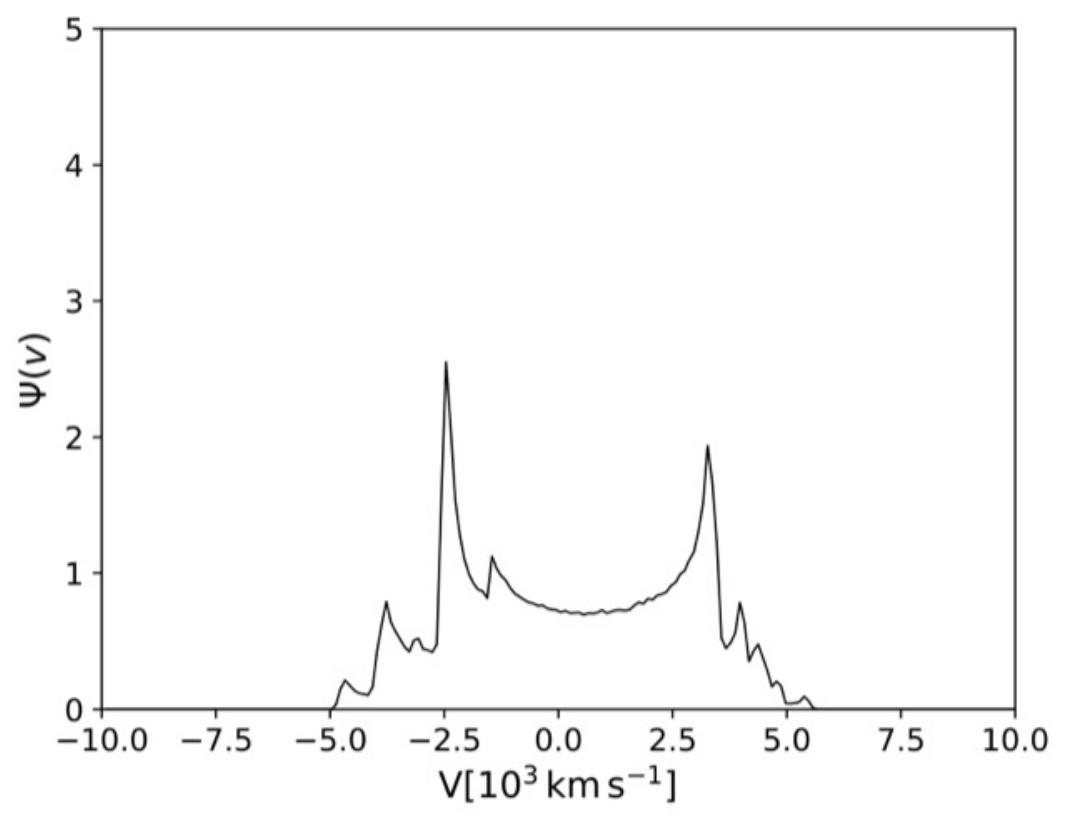}
    \caption{}
\end{subfigure}
\begin{subfigure}{6cm}
    \centering
    \includegraphics[width=5cm]{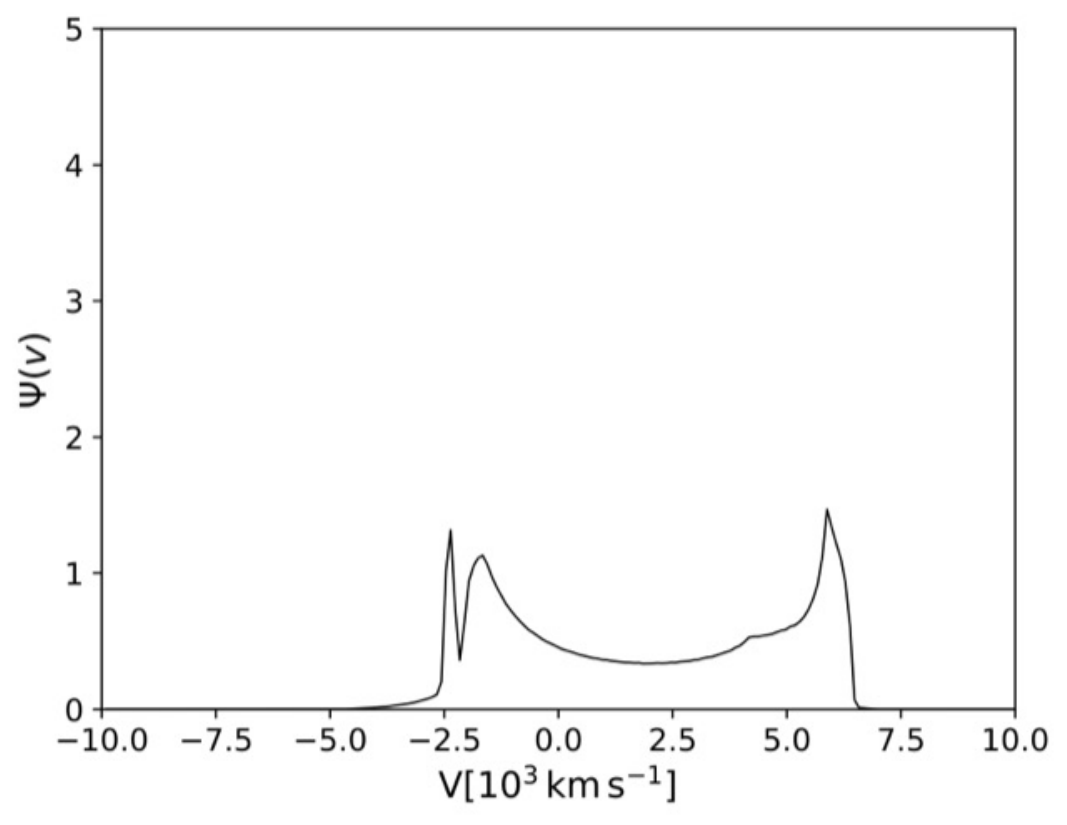}
    \caption{}
\end{subfigure}
\begin{subfigure}{6cm}
    \centering
    \includegraphics[width=5cm]{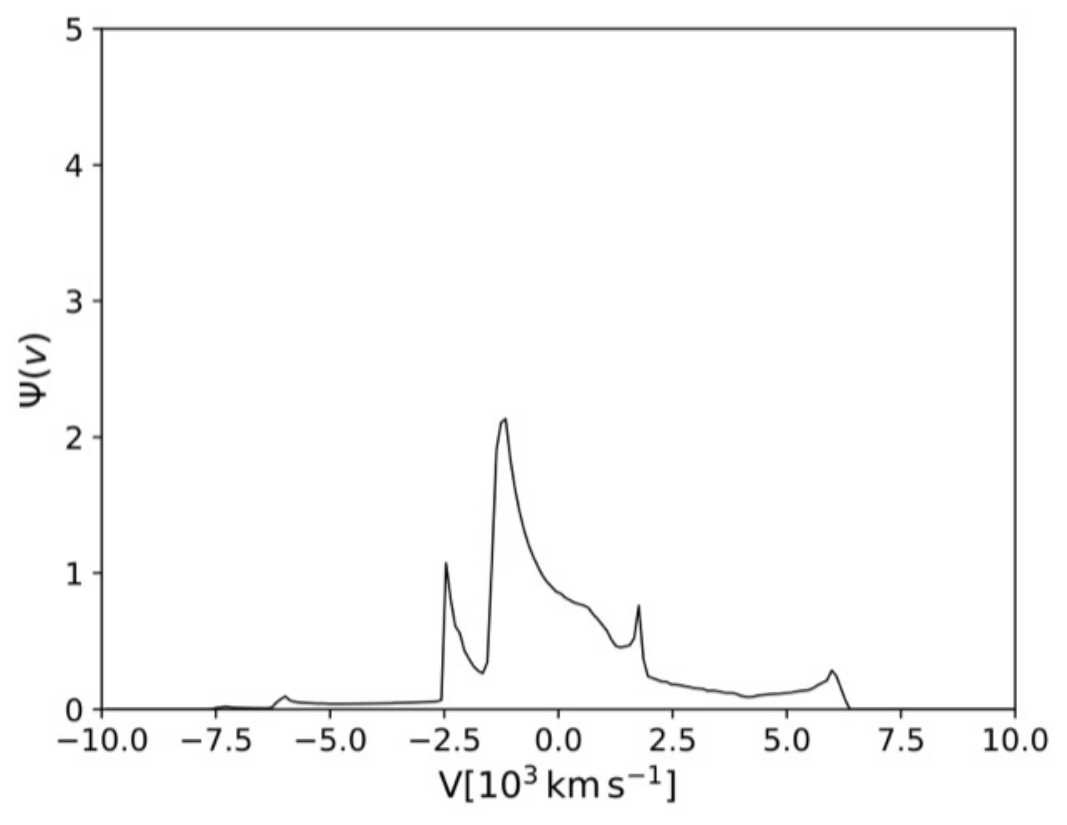}
    \caption{}
\end{subfigure}
\begin{subfigure}{6cm}
    \centering
    \includegraphics[width=5cm]{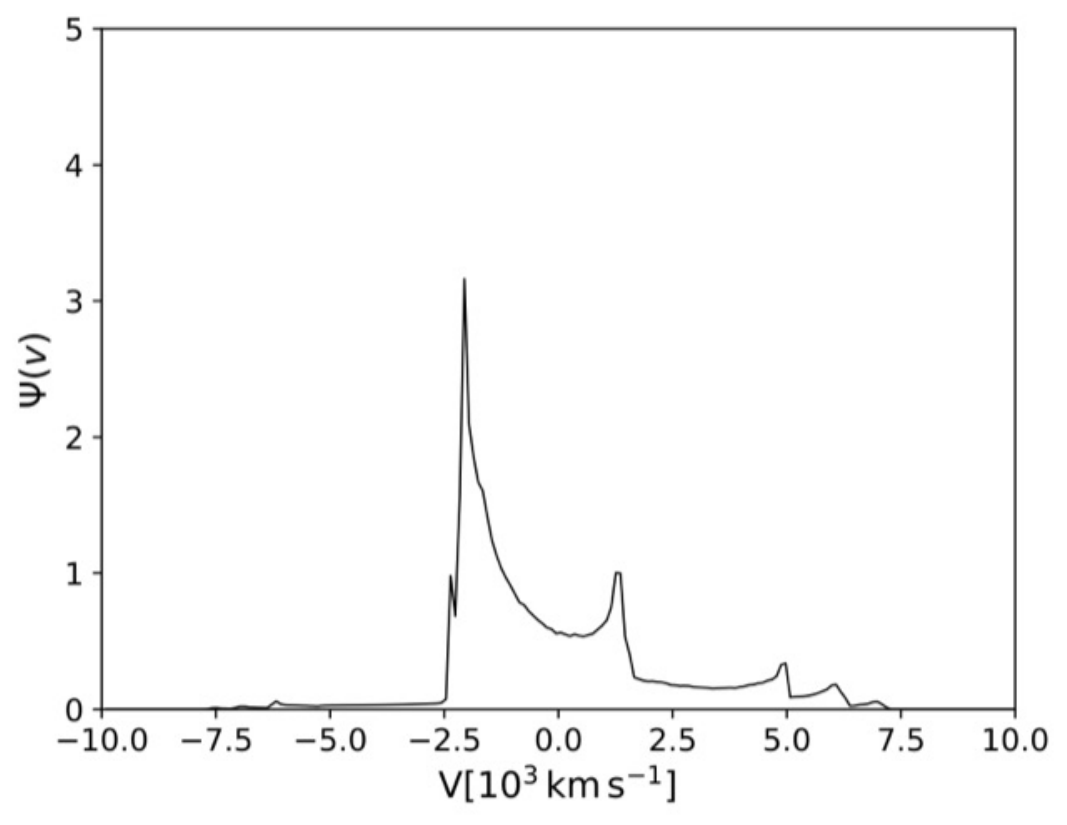}
    \caption{}
\end{subfigure}
\begin{subfigure}{6cm}
    \centering
    \includegraphics[width=5cm]{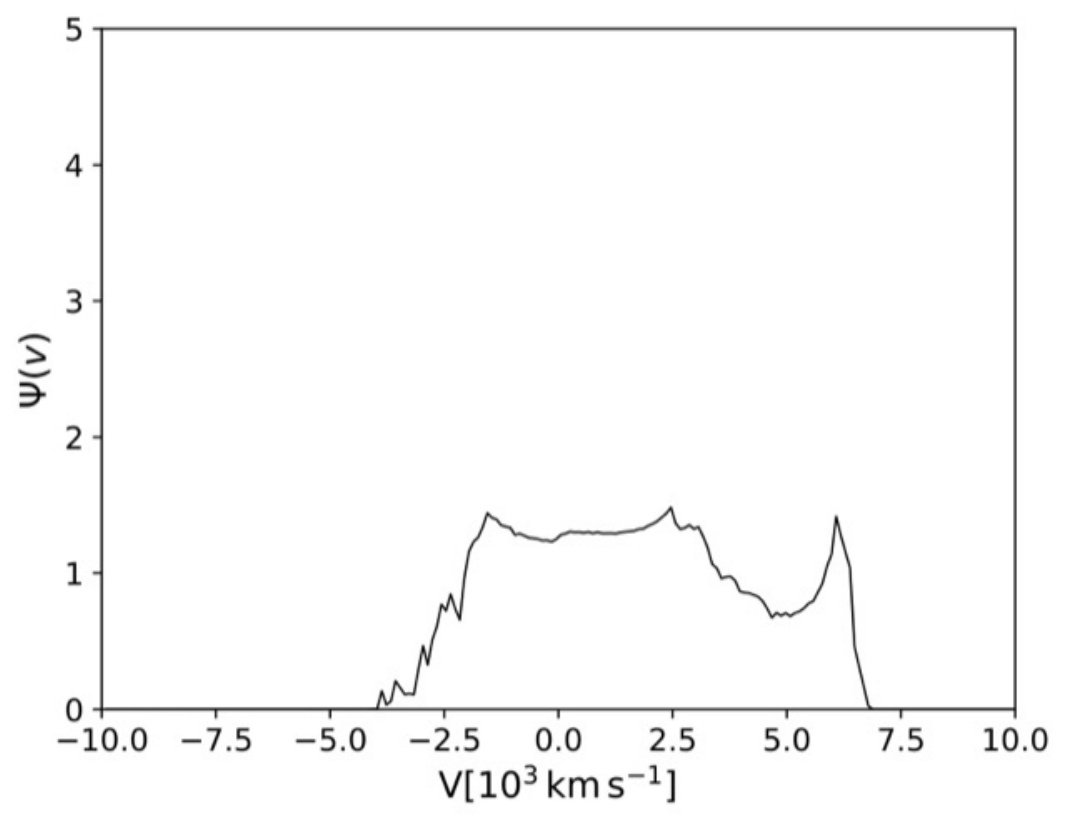}
    \caption{}
\end{subfigure}
\begin{subfigure}{6cm}
\centering
    \includegraphics[width=5cm]{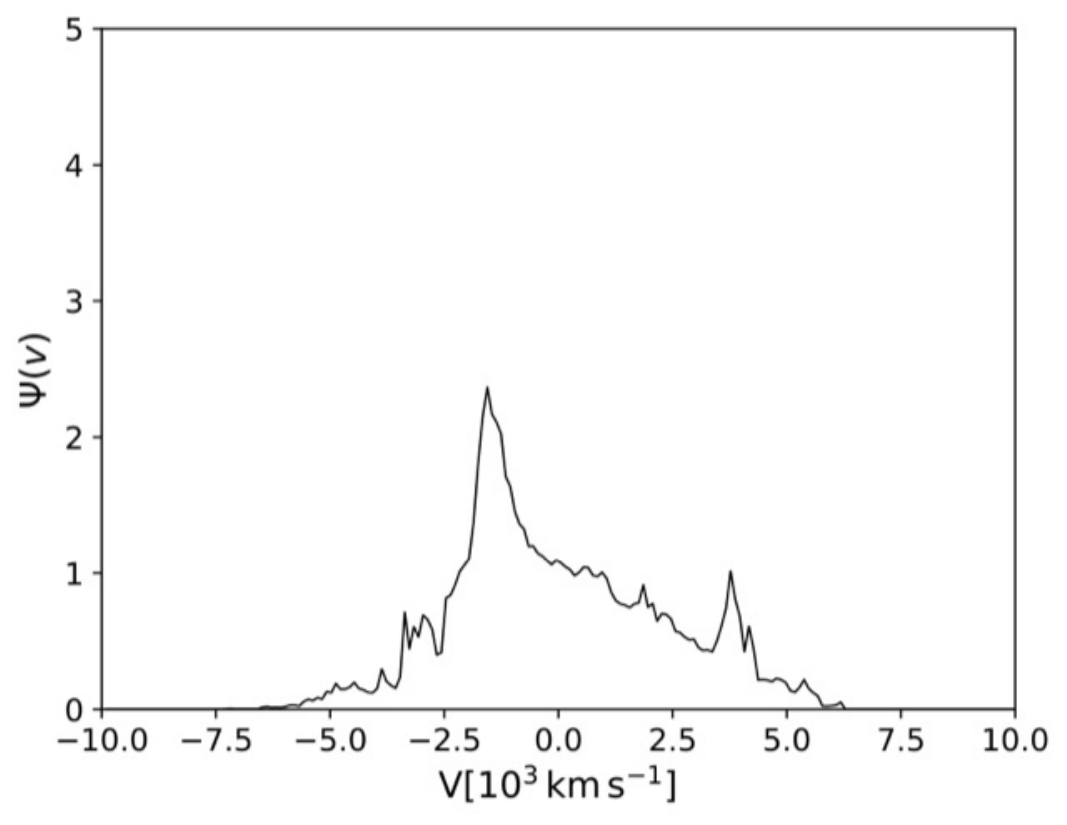}
    \caption{}
\end{subfigure}
\begin{subfigure}{6cm}
\centering
    \includegraphics[width=5cm]{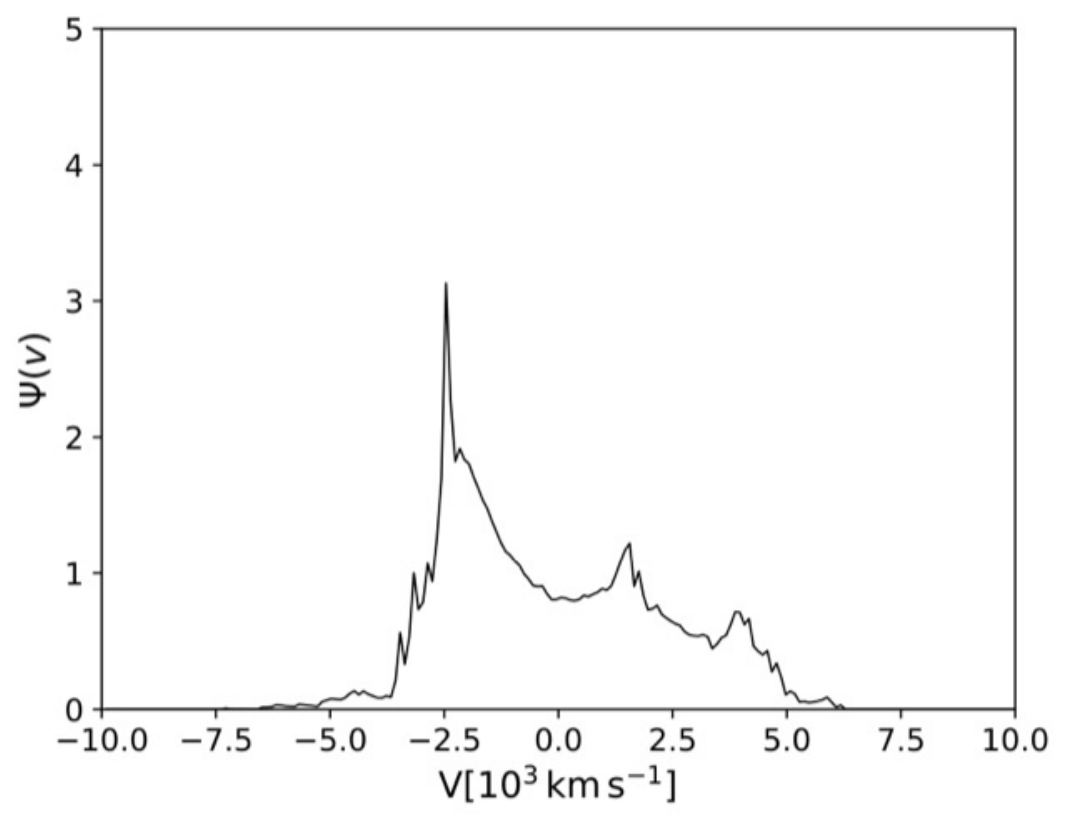}
    \caption{}
\end{subfigure}
\caption{Spectrum corresponding to theoretical 2DTF  maps obtained for different geometries of portions of cloud orbits in the binary system given  in Fig. \ref{fig:tf4}.}
\label{fig:ttf4}
\end{figure*}

Figures \ref{fig:ttf1}-\ref{fig:ttf4} display a series of velocity profiles obtained by integrating $\Psi(v,\tau)$
over the time delay \citep{1982ApJ...255..419B},

\begin{equation}
\Psi(v)=\int^{\tau_{max}}_{0} \Psi(v,\tau) \mathrm{d}\tau
,\end{equation}

\noindent where  $\Psi(v,\tau)\geq 0$ for   $\tau_\mathrm{max} \in (0, \infty) $. It is a tim- averaged line profile  that  is the convolution of $\Psi(v,\tau)$ with a time-averaged  continuum light curve because we do not have measured continuum variations (see Eq. \ref{bland}).  Thus, 
the line profiles  viewed here are time-averaged  representatives of the emission  line shapes that are  expected for our models presented in Figures \ref{fig:tf1}--\ref{fig:tf4}.

The 1D projections of 2DTF of single BLR models (see Fig \ref{fig:tf1}) are shown in Fig. \ref{fig:ttf1}. The   width and appearance  of the spectral line  is sensitive to the orientation of the cloud orbits. For fixed eccentricities and large $\Omega$, increasing $\omega$ clearly broadens  and  blurs  the typical double-peaked line profile of a Keplerian thin disc until one dominant horn is formed while the other peak is weakened.  Simultaneously increasing either $\Omega$ or $\omega$ and reducing to zero the other orientation angle leads to double-peak profiles. Randomisation of eccentricities and/or orientation of clouds does not affect the appearance of the spectra significantly (stability of elliptical orbits is not considered).  

The most interesting feature for these spectral line shapes is the asymmetrical double peak (with one peak more prominent than other). The features  in Fig. \ref{fig:ttf1} (c), (g)  are detected in the spectra of PKS1739+18C observed by \citet[][see their Fig. 8]{2010ApJS..187..416L}. The feature from 
Fig. \ref{fig:ttf1} (b) is detected in spectra of Pictor A that was also observed by  \citet[][see their Fig. 19 right column of panels]{2010ApJS..187..416L}.  Interestingly, \citet[][see their Fig. 2]{doi.org/10.1088/0067-0049/201/2/23} found a similar shape of H$\alpha$ and  H$\beta$ emission lines on { SDSS J001224.02-102226.2} in their systematic search for  close SMBBHs and rapidly recoiling black holes.  Moreover, the features shown in Fig. \ref{fig:ttf1} (h) are  very similar  in appearance to the  emission-line profiles of the disc-wind model present by \citet[][compare to their Fig. 4 panel (5a)]{2019ApJ...870...16N},   the Hubble space telescope observations of the H$\alpha$ line of NGC 3147 \citep[][see their  Fig. 2]{Bianchi19}, and the
   H$\beta$ emission line on  { SDSS J021259.60-003029.5}   \citep[][see their Fig. 2]{doi.org/10.1088/0067-0049/201/2/23}.
A similar feature  is observed in the changing look  NGC 3516 \citep[][see their Fig. 4]{2018ApJ...866..133D}. This example  underlines that caution must be taken in  interpreting disc-wind models that are solely inferred from spectral line shapes in the absence of detailed
kinematic and 2DTF modelling.

Next, we  present 1D dynamical models of emission lines  obtained from 2DTF of the binary SMBH presented in Fig. \ref{fig:tf2}.  The  plots in Fig. \ref{fig:ttf2} reveal the primary effect of the phase of the SMBBH system: a prominent increase in 
flux at the line center. Profiles in the initial and the last phase  of  the coplanar circular binary system (see  Fig  \ref{fig:ttf2}(a) and (c)) are broader than  
those obtained from its non-coplanar version   (see  Fig  \ref{fig:ttf2}(d) and (f)). Their orientation also depends on the phase. In the middle of the orbital period, the only difference is at the top of profile.
In the case of the elliptical binary system, the emission line shapes depend on the orbital phase of the binary, but 
 a small orientation angle of pericenter of the cloud orbits  $\omega$ almost flattens the top of emission lines (see Fig. \ref{fig:ttf2}(g)--(i)). Conversely, higher values of   $\omega$ tend to
 broaden the line shape in the initial orbital phase (Fig. \ref{fig:ttf2}(j)), and  prominent core with double peaks in the  middle ((Fig. \ref{fig:ttf2}(k)) and final orbital phase (Fig. \ref{fig:ttf2}(l)). The wings are also asymmetric: the right wings are stronger because both SMBHs contribute to the system.

The effects of randomisation of eccentricities and/or orientation angles of the cloud orbit on the emission line shapes  are shown in Fig. \ref{fig:ttf3} (we did not consider the stability of the elliptical orbits). They were obtained from 2DTF maps presented in Fig. \ref{fig:tf3}.  The first row (Fig. \ref{fig:ttf3}(a--c)) shows emission lines  obtained for the coplanar elliptical SMBBH  system and cloud orbits with random eccentricities and  higher values of $\omega$ (than in the case of Fig.\ref{fig:ttf2}(g--i)). The contribution of both SMBHs is clearly visible in the middle and final portion of orbital phase.   Increasing the inclination of the elliptical orbit of  the more massive SMBH and decreasing the angle of pericenter of the cloud orbits that have random eccentricities  blurs the contribution of the emission  of the less massive SMBH  (see Fig. \ref{fig:ttf3}(d--f)).
However, in the same non-coplanar settings of the SMBBH system, if we randomise the orientations of the clouds in both BLRs, but  fix  the eccentricities of the clouds of the more massive SMBH  to a higher value (0.5) than the value of the less massive  SMBH (0.1), the contribution of the less massive component is almost diminished and only a hint of its presence is visible as an asymmetry in the line profiles. 
On the other hand, a randomisation of the eccentricities and orientations of the clouds in both BLRs of  the non-coplanar case terminates the contribution of the smaller SMBH  around the middle (see Fig. \ref{fig:ttf3}(k)) and end of orbital phase (see Fig. \ref{fig:ttf3}(l)). A weak impression of   the companion  is visible at the beginning of the orbital period (see Fig. \ref{fig:ttf3}(j)).

The most interesting feature in Fig. \ref{fig:ttf2}(a), (c), (d), (f) and  Fig. \ref{fig:ttf3}(b--f) is an intermediate  peak. This  has been observed in the spectral lines of a few objects: 3C 390.3 by \citet[][see their Fig.1]{2011A&A...528A.130P}, \citet[][see their Fig. 3]{2014A&A...572A..66P}, NGC 4151  \citep[][see their Fig. 6]{2008A&A...486...99S}, and E1821+643 \citep[][see their Fig. 15]{2016ApJS..222...25S}. Additionally, the features shown in    Fig. \ref{fig:ttf3} (e-f) and (h-i)  are detected in the spectra of NGC 5548 \citep[][see their Fig. 3]{1987ApJ...312...79P}. Features in panels (k-l) are detected in  the spectra of the spectral line in Mrk 668 \citep[][see their Fig. 1 lower spectrum]{2016ApJ...817...42L}.

Likewise, broad-line features such as shown in Fig. \ref{fig:ttf2} (j) and Fig. \ref{fig:ttf3} (a) are detected 
in BLR disc-wind models. The lines that form in the vicinity of the disc-wind base  look broad and symmetric  \citep[see][]{2017PASA...34...42Y}. 
As already mentioned,   disc-wind models and binary black holes can produce similar  spectral line shapes, and a more thorough analysis is needed.

 Next, we considered  that only one year of the orbital period of the clouds of the non-coplanar  elliptical binary  SMBH system are  observed  in different SMBH orbital phases. As we show in  Fig \ref{fig:ttf4}, the emission line shapes vary remarkably.
 When the clouds are on circular trajectories in both BLRs, the right peak  is more prominent  at the beginning (see Fig. \ref{fig:ttf4}(a)) and middle of the orbital phase  (see Fig. \ref{fig:ttf3}(b)). However, elliptical BLRs will produce  a prominent left peak only in the middle and  at the end of the orbital phase motion  (see Fig. \ref{fig:ttf4}(e), (f)).
On the other hand, randomisation of the eccentricities in both BLRs broadens the emission lines  (see Fig. \ref{fig:ttf4}(g)-(i),  but we did not consider the stability of elliptical orbits). The features in panel (a)   are similar to the spectra of PKS 1346-11 \citep[][see their Fig. 1]{2003ApJ...599..886E}, and the H$\beta$ emission line on{  SDSS J022014.57-072859.1} \citep{doi.org/10.1088/0067-0049/201/2/23}. The feature in panel (b) is also similar to artificial spectra in \cite[][see their Fig. 3(b)]{2004PASP..116..465H}. The line shape in panel (c) looks like the H$\beta$ emission line on { SDSS J074007.28+410903.6} \citep{doi.org/10.1088/0067-0049/201/2/23}.
 The shape shown in panel (f) resembles the line features in   Mrk 668 \citep[][see their Fig. 1 lower spectrum]{2016ApJ...817...42L}. The spectral line in panel (e) is a mirror image of the H$\alpha$ line observed on { SDSS J020011.53-093126.2} \citep{doi.org/10.1088/0067-0049/201/2/23}.

\section{Conclusions}\label{sec:con}

Here,  by extending previous analysis in Paper I and II, we have presented a 3D geometrical model that self-consistently predicts the 2DTF  signatures of  an SMBBH system consisting  of two binary SMBHs on elliptical orbits with elliptical disc-like BLRs. 
We considered  a full  set of orbital  parameters of bot SMBHs in the system and clouds in both BLRs because our first goal was to understand the typical 
signatures of the BLR features in 2DTFs.
 We identified a
number of characteristic features that might help in assessing
the SMBBH system and in evaluating its
parameters. Our main findings are listed below.
\begin{enumerate}
 \item Simple and composite 2DTFs  of elliptical disc-like BLRs   have a deformed bell shape. The slope gradients and wing asymmetry of the deformed bells are predominantly controlled  by   the  orbital orientation of the clouds. In particular, the 2DTF could serve as an advanced diagnostic
tool to distinguish the BLR models on the basis of quantitative
measurements.

\item Both simple and  composite 2DTF  exhibit further differentiation  between  randomly oriented or randomly elongated {  clouds} trajectories  and those where  the orbital eccentricity or orientation of the cloud is fixed.
The randomisation of  the orbital parameters  of { the clouds} tends  to produce filaments in bells, which appeared to be  more  asymmetric and chaotic when the orbital orientation is random.   As we discussed in section 3.0,  in our model of randomised motion we did not
 consider cloud collisions, which should be taken  into
 account, and we postpone this consideration to a future study. An inclined SMBH orbit deforms the  2DTF bell size corresponding to its BLR.

\item In particular,  we found a simple  2DTF  (inferred from a single SMBH model) that we  calculated for    small angles of orientation and random eccentricities of the { clouds} orbits,  similar to that observed  in  Mrk 50.  We did not consider the stability of elliptical orbits during randomised motion.

\item An overall degeneracy of the 2DTF maps is observed for the hypothetical case when the cloud orbit is observed for only one year. The presence of two discs is clear when the observations cover  the first tenth of the  orbital periods of the clouds.
This means that the length of monitoring campaigns must be
long enough to sample the full range of time lags, which can help  to  constrain the parameters of  dynamical models.

 \item The simulations show that an intermediate peak in  the
broad-line profiles such as that  observed in NGC 4151, NGC 5548, and 3C 390.3  can
indeed be reproduced by our elliptical binary model.

 \item We found   a remarkable coincidence  between
the  line distortions produced for the disc-wind
 and elliptical BLR models.
A good distinction between these
BLR models would require long-term and  quality-cadenced spectrophotometric monitoring.
\end{enumerate}

\begin{acknowledgements}
We  appreciate and thank an anonymous Referee  for invaluable comments.  This work is supported by the Ministry of Education, 
Science and Technological development of Republic Serbia  through the  
\emph{Astrophysical Spectroscopy of Extragalactic Objects} project number 176001,\emph{National Key R\&D Program of China} (grant 2016YFA0400701) and grants NSFC--
11173023, -11133006, -11373024, -11233003 and -11473002, and by\emph{CAS Key Research Program of
Frontier Sciences}, QYZDJ--SSW--SLH007.   
\end{acknowledgements}

%

\begin{thebibliography}{}

\bibitem[Afanasiev et al. (2019)]{af19} 
Afanasiev, V. L., Popovi\'c, L. \v C., Shapovalova, A. I. 2019, MNRAS, 482, 4985 

\bibitem[Alexander(2011)]{http://www.aspbooks.org/publications/439/129.pdf} 
Alexander, T.  2011, in Astronomical Society of the Pacific Conference Series, Vol. 439, The Galactic Center: a Window to the Nuclear Environment of Disk Galaxies, ed. M. R. Morris, Q. D. Wang, \& F. Yuan, 129

\bibitem[Amaro-Seoane et al. (2010)]{doi.org/10.1111/j.1365-2966.2009.16104.x} 
Amaro-Seoane, P.,  Sesana, A.,  Hoffman, L.,  Benacquista, M. et al. 2010, MNRAS, 402, 2308


\bibitem[Arav et al. (1998)]{1998MNRAS.297..990A} 
Arav, N., Barlow, T. A., Laor, A., Sargent, W. L. W., Blandford, R. D. 1998, MNRAS, 297, 990


\bibitem[Bahcall et al. (1972)]{doi: 10.1086/151300} 
Bahcall, J. N., Kozlovsky, B.-Z., Salpeter, E. E. 1972, ApJ, 171, 467 

\bibitem[Begelman et al. (1980)]{1980Natur.287..307B}
Begelman, M. C., Blandford, R. D., Rees, M. J. 1980, Nature, 287, 307


\bibitem[Bentz \& Katz (2015)]{doi: 10.1086/679601} 
Bentz, M. C., Katz, S.   2015,   PASP, 127, 67


\bibitem[Bianchi et al. (2019)]{Bianchi19} 
Bianchi, S., Antonucci, R., Capetti, A., Chiaberge, M., et al. 2019, MNRAS in press




\bibitem[Blandford \& McKee (1982)]{1982ApJ...255..419B} 
Blandford, R.,  McKee, C. 1982, ApJ 255, 419


\bibitem[Bogdanovi{\' c} et al. (2009)]{2009ApJ...697..288B}
Bogdanovi{\' c}, T., Eracleous, M.,  Sigurdsson, S. 2009, ApJ, 697, 288


\bibitem[Bogdanovi{\' c} et al. (2007)]{2007ApJ...661L.147B} 
Bogdanovi{\' c}, T., Reynolds, C. S.,  Miller, M. C. 2007, ApJ, 661, L147


\bibitem[Bon et al. (2012)]{2012ApJ...759..118B}
Bon, E., Jovanovi{\'c}, P., Marziani, P., Shapovalova, A. I.,  Bon, N., Borka -Jovanovi{\'c}, V., Borka, D.,  Sulentic, J., Popovi{\'c}, L. {\v C}. 2012, ApJ, 759, id. 118, 9


\bibitem[Bowen et al. (2017)]{2017ApJ...838...42B}
Bowen, D. B., Campanelli, M., Krolik, J. H., Mewes, V., Noble S. C. 2017, ApJ, 838, 42

\bibitem[Bowen et al. (2018)]{2018ApJ...853L..17B}
Bowen, D. B., Mewes, V., Campanelli, M., Noble, S. C., Krolik, J. H., Zilh{\~a}o, M. 2018, ApJL, 853, L17

\bibitem[Bowen et al. (2019)]{2019ApJ...879...76B}
Bowen, D. B., Mewes, V., Noble, S. C., Avara, M., Campanelli, M., Krolik, J. H., 2019, ApJ,  879,  9 


\bibitem[Bradt (2008)]{ISBN-13: 978-1107677241} 
Bradt, H. 2008,  Astrophysics Processes: The Physics Of Astronomical Phenomena,Cambridge University Press  



\bibitem[Brouwer and Clemence (1961)]{BC61} 
Brouwer, D. and Clemence, G.M. 1961. Methods of Celestial Mechanics. Academic Press, New York.

\bibitem[Burke and Collins (2013)]{bu13} Burke, C., Collins, C. A. 2013, MNRAS, 434, 2856 

\bibitem[Capelo et al. (2015)]{ca15} 
Capelo, P. R., Volonteri, M., Dotti, M., Bellovary, J. M., Mayer, L., Governato, F. 2015, MNRAS, 447, 2123 


\bibitem[Courvoisier et al. (1996)]{C1996A&A...308L..17C}
Courvoisier, T. J.-L., Paltani, S., Walter, R. 1996, A\&A, 308, L17

\bibitem[Courvoisier and T{\"u}rler(2005)]{DOI: 10.1051/0004-6361:20040527}
Courvoisier, T. J. L., T{\"u}rler, M., 2005, A\&A, 444, 417


 
\bibitem[d’Ascoli et al. (2018)]{2018ApJ...865..140D}
d’Ascoli, S., Noble, S. C., Bowen, D. B.,Campanelli, M., Krolik, J. H., Mewes, V.  2018, ApJ, 865, 140

\bibitem[Deane et al. (2015)]{DOI:10.1038/nature13454}
Deane, R. P., Paragi,  Z., Jarvis, M. J., Coriat, M. et al. 2014, Nature, 511, 57

 
\bibitem[Decarli et al. (2013)]{2013MNRAS.433.1492D}
Decarli, R.,  Dotti, M., Fumagalli, M.,  Tsalmantza, P.,  Montuori, C.,  Lusso, E.,  Hogg, D. W., Prochaska, J. X. 2013, MNRAS, MNRAS, 433, 1492




\bibitem[Doan et al. (2019)]{Doan19}
Doan, A.,  Eracleous, M.,  Runnoe, J. C., Liu, J.,  Mathes, G.,  Flohic,  M. L. G., 2019, accepted by MNRAS



\bibitem[De Rosa et al. (2018)]{2018ApJ...866..133D}
De Rosa, G., Fausnaugh, M. M.,  Grier, C. J.,  Peterson, B. M., Denney, K. D., Horne, K., Bentz, M. C.; Ciroi, S. et al. 2018, ApJ, 866, id133 


\bibitem[D’Orazio et al. (2016)]{2016MNRAS.459.2379D}
D’Orazio, D. J., Haiman, Z., Duffell, P., MacFadyen, A., Farris, B. 2016, MNRAS, 459, 2379

\bibitem[Dotti et al. (2009)]{2009MNRAS.398L..73D} 
Dotti, M., Montuori, C., Decarli, R., Volonteri, M., Colpi, M., Haardt, F.  2009, MNRAS, 398, L73



\bibitem[Du et al. (2015)]{doi: 10.1088/0004-637X/806/1/22} 
Du, P., Hu, C., Lu, K.-X., Huang, Y.-K., Cheng, C., Qiu, J., et al. 2015, ApJ, 806, 22

\bibitem[Eracleous \& Halpern (1994)]{1994ApJS...90....1E}
Eracleous, M. \& Halpern, J. P.   1994, ApJS,  90,1

\bibitem[Eracleous \& Halpern (2003)] {2003ApJ...599..886E}
Eracleous, M., Halpern, J. P.  2003, ApJ, 599, 886 

\bibitem[Eracleous et al. (1997)] {1997ApJ...490..216E}
Eracleous, M., Halpern, J. P., Gilbert, A. M., Newman, J. A., Filippenko, A. V. 1997, ApJ, 490, 216 


\bibitem[Eracleous et al. (2012)] {doi.org/10.1088/0067-0049/201/2/23}
Eracleous,  M., Boroson, T. A.,  Halpern, J.P,  Liu, J.  2012, ApJSS, 20, 21pp


\bibitem[Eracleous et al. (1995)]{1995ApJ...438..610E}
Eracleous, M., Livio, M., Halpern, J. P., \& Storchi-Bergmann, T.  1995, ApJ, 438, 610

\bibitem[Farris et al. (2014)]{2014ApJ...783..134F}
Farris, B. D., Duffell, P., MacFadyen, A. I. ,  Haiman Z.  2014, ApJ, 783, 134

\bibitem[Farris et al. (2015a)]{2015MNRAS.447L..80F}
Farris, B. D., Duffell, P., MacFadyen, A. I., Haiman Z. 2015a, MNRAS, 447, L80

\bibitem[Farris et al. (2015b)]{2015MNRAS.446L..36F}
Farris, B. D., Duffell P., MacFadyen, A. I., Haiman, Z. 2015b MNRAS 446 L36


\bibitem[Gaskell (2006)]{1996LNP...471..165G}
Gaskell, M. C., 1996, Jets from Stars and Galactic Nuclei. Proceedings of a Workshop Held at Bad Honnef, Germany, 3 - 7 July 1995. Edited by Wolfgang Kundt. Springer-Verlag Berlin Heidelberg New York. Also Lecture Notes in Physics, volume 471, p.165


\bibitem[Gaskell (2010)]{2010Natur.463E...1G}
Gaskell, C. M. 2010, Nature, 463, 1

\bibitem[Gilbert et al. (1999)]{1999ASPC..175..189G}
Gilbert, A. M., Eracleous, M., Filippenko, A. V.,  Halpern, J. P. 1999, ASP
Conf. Ser. 175: Structure and Kinematics of Quasar Broad Line Regions, 189

 \bibitem[Graham et al. (2015)]{2015Natur.518...74G}
Graham, M. J., Djorgovski, S. G., Stern, D., Glikman, E. et al. 2015, Nature, 518, 74


\bibitem[Guo et al. (2019)]{2019MNRAS.482.3288G}
Guo, H., Liu, X., Shen, Y.,  Loeb, A., Monroe, T.,  Prochaska, J. X., 2019, MNRAS,  482, 3288


\bibitem[Hayasaki et al. (2008)]{2008ApJ...682.1134H}
Hayasaki, K., Mineshige, S., Ho, L. C., 2008, ApJ,  682,  1134




\bibitem[Heckman et al. (2009)]{2009ApJ...695..363H}
Heckman, T. M., Krolik, J. H., Moran, S. M., Schnittman, J.,  Gezari, S. 2009, ApJ, 695, 363


\bibitem[Horne et al. (2004)]{2004PASP..116..465H}
 Horne, K.,  Peterson, B. M., Collier, S. J., Netzer, H.  2004, PASP, 116, 465 
    
\bibitem[Jiang et al. (2012)]{ji12} Jiang, T., Hogg, D. W., Blanton, M. R.  2012, ApJ, 759, 140 


 \bibitem[Ju et al. (2013)]{J2013ApJ...777...44J}               
Ju, W., Greene, J. E., Rafikov, R. R., Bickerton, S. J., Badenes, C. 2013, ApJ,  777,   44, 16





\bibitem[Komossa et al. (2008)]{2008ApJ...678L..81K}
Komossa, S., Zhou, H.,  Lu, H.  2008, ApJ, 678, L81




\bibitem[Kova{\v c}evi{\' c} et al. (2018)]{2018MNRAS.475.2051K}
Kova{\v c}evi{\' c}, A.,  P{\'e}rez-Hern{\'a}ndez, E., Popovi{\'c}, L. {\v C}., Shapovalova, A. I., Kollatschny, W.,  Ili{\'c}, D.  2018, MNRAS, 475, p.2051

\bibitem[Kova{\v c}evi{\' c} et al. (2019)]{2019ApJ...871...32K}
Kova{\v c}evi{\'c}, A. B., Popovi{\'c}, L. {\v C}.,  Simi{\'c}, S., Ili{\'c},  D.  2019, ApJ,  871,  id. 32


\bibitem[Lewis et al. (2010)]{2010ApJS..187..416L}
Lewis, K. T., Eracleous, M., Storchi-Bergmann, T. 2010, ApJS, 187, 416



\bibitem[Li  et al. (2019)]{2016ApJ...822....4L}
Li, Y.-R., Wang, J.-M., Ho, L. C., Lu, K.-X., Qiu, J., Du, Pu et al. 2016, ApJ,  822,  id. 4, 21  





\bibitem[Li  et al. (2019)]{2019ApJS..241...33L}
Li, Y.-R., Wang, J.-M.,Zhang, Z.-X., Wang, K., et al.  2019, ApJSS, 241, 33



\bibitem[Liu  et al. (2016)]{2016ApJ...817...42L}
Liu, J., Eracleous, M., Halpern, J. P.  2016, ApJ, 817, id.c42


\bibitem[Liu  et al. (2018)]{2018ApJ...859L..12L}
Liu, T., Gezari, S., Miller, M. C.  2018,  ApJL, 859,  L12


\bibitem[Liu  et al. (2014)]{2014ApJ...789..140L}
Liu, X., Shen, Y., Bian, F., Loeb, A., Tremaine, S. 2014, ApJ,  789,  140, 22 


\bibitem[Mathews and Capriotti (1985)]{1985aagq.conf..185M}
Mathews, W. G.,  Capriotti, E. R., 1985, Structure and dynamics of the broad line region. In: Astrophysics of active galaxies and quasi-stellar objects. Mill Valley, CA, University Science Books, 1985



\bibitem[Mingarelli et al. (2017)]{2017NatAs...1..886M} 
Mingarelli, C. M. F, Lazio, T. J. W. ,  Sesana, A., Greene, J. E. et al.  2017, Nature Astronomy,  12, 886


\bibitem[Montuori et al. (2011)]{2011MNRAS.412...26M}
Montuori, C., Dotti, M., Colpi, M., Decarli, R.,  Haardt, F. 2011, MNRAS, 412, 26


\bibitem[Montuori et al. (2012)]{2012MNRAS.425.1633M}
Montuori, C., Dotti, M., Haardt, F., Colpi, M., Decarli, R. 2012, MNRAS, 425, 1633

 
\bibitem[Moody et al. (2019)]{2019ApJ...875...66M}
Moody, M. S. L., Shi, J.-M.,Stone, J. M. 2019, ApJ, 875, 66


\bibitem[Mu{\~n}oz and Lai (2016)]{2016ApJ...827...43M}
Mu{\~n}oz, D. J.,  Lai, D., 2016,  ApJ, 827, 43

\bibitem[Murray \& Dermott (1999)]{MD99}
Murray, C. D. Dermott, S. F. 1999, Solar system dynamics  (Cambridge: Cambridge University Press), ch.2

\bibitem[Netzer (2015)]{2015ARA&A..53..365N} 
Netzer, H.  2015, ARAA, 53, 365


\bibitem[Netzer \& Peterson (1997)]{1997ASSL..218...85N} 
Netzer, H.,  Peterson, B. M. 1997,  Astronomical Time Series, eds D. Maoz, A. Sternberg, and E. Leibowitz (Dordrecht: Kluwer Academic Publishers), 85



\bibitem[Nguyen (2019)]{2019ApJ...870...16N}
Nguyen, K., Bogdanovi{\'c}, T.,  Runnoe, J. C., Eracleous, M. et al. 2019, ApJ, 870, id.16

\bibitem[Pancoast et al. (2012)]{10.1088/0004-637X/754/1/49} 
Pancoast,  A.,   Brewer, B. J.,  Treu, T.,   Barth, A. J.,  . Bennert, V. N. et al. 2012, ApJ, 754,49



\bibitem[Peterson (1987)]{1987ApJ...312...79P}
Peterson, B. M. 1987, 312, 79


\bibitem[Peterson (1993)]{1993PASP..105..247P}
Peterson, B. M. 1993, PASP, 105, 247



\bibitem[Peterson (2006)]{https://doi.org/10.1007/3-540-34621-X_3}
Peterson, B. 2006 The Broad-Line Region in Active Galactic Nuclei. In: Alloin D., Johnson R., Lira P. (eds) Physics of Active Galactic Nuclei at all Scales. Lecture Notes in Physics, vol 693. Springer, Berlin, Heidelberg


\bibitem[Popovi{\'c} (2012)]{2012NewAR..56...74P}
Popovi{\'c}, L. {\v C}.  2012, New Astronomy Reviews,  56,  2-3,  74


\bibitem[Popovi{\'c} et al. (2004)]{2004A&A...423..909P}
Popovi{\'c}, L. {\v C}., Mediavilla, E., Bon, E., Ili{\'c}, D.  2004, A\&A, 423, 909


\bibitem[Popovi{\'c} et al. (2014)]{2014A&A...572A..66P}   
Popovi{\'c}, L. {\v C}., Shapovalova, A. I.,   Ili{\'c}, D., Burenkov, A.. et al.  2014, A\&A, 572, id. A66


\bibitem[Popovi{\'c} et al. (2011)]{2011A&A...528A.130P}   
Popovi{\'c}, L. {\v C}., Shapovalova, A. I.,   Ili{\'c}, D.,  Kova\v cevi\'c, A.,  Kollatschny, W. et al.  2011, A\&A, 528, id. A130


\bibitem[Runnoe et al. (2017)]{2017MNRAS.468.1683R}
Runnoe, J. C., Eracleous, M.,  Pennell, A.,  Mathes, G.,  Boroson, T.,  Sigurdsson, S.,  Bogdanovi{\' c}, T., Halpern, J. P., Liu, J.,  Brown, S. 2017, MNRAS, 468, 1683



\bibitem[Ryan et al. (2017)]{2017ApJ...835..199R}
Ryan, G.,  MacFadyen, A. 2017, ApJ, 835, 199


\bibitem[Scheafer et al. (2016)]{2016AJ....152..213S}
Schaefer, G. H., Hummel, C. A., Gies, D. R.,  Zavala, R. T. et al. 2016, ApJ, 152, id. 213

 \bibitem[Sergeev et al. (2000)]{2000A&A...356...41S}
Sergeev, S. G., Pronik, V. I., Sergeeva, E. A., 2000, A\&A,  356, 41

 
\bibitem[Shapovalova et al. (2019)]{2019MNRAS.485.4790S}
Shapovalova, A. I.; Popovi\'c, L.\v C., Afanasiev, V. L., Ili\' c, D., Kova\v cevi\'c, A., Burenkov, A. N., Chavushyan, V. H., Mar\v ceta-Mandi\' c, et al. \ 2019, MNRAS, 485, 4790


\bibitem[Shapovalova et al. (2016)]{2016ApJS..222...25S}
Shapovalova, A. I., Popovi{\'c}, L. {\v C}., Chavushyan, V. H., Burenkov, A. N., Ili{\'c}, D. et al. 2016, ApJS, 222, id. 25

\bibitem[Shapovalova et al. (2017)]{2017MNRAS.466.4759S}
Shapovalova, A. I., Popovi{\'c}, L. {\v C}., Chavushyan, V. H.,  Afanasiev, V. L., , Ili{\'c}, D.,  Kova\v cevi\'c, A., Burenkov, A. N., Kollatschny, W. et al. 2017, MNRAS, 466, 4759


\bibitem[Shapovalova et al. (2008)]{2008A&A...486...99S}
Shapovalova, A. I., Popovi{\'c}, L. {\v C}., Collin, S., Burenkov, A. N. et al. 2008, A\&A, 486, 99



\bibitem[Shen et al. (2013)]{2013ApJ...775...49S}
Shen, Y., Liu, X., Loeb, A., Tremaine, S.  2013, ApJ, 755,   23

\bibitem[Shen et al. (2016)]{doi: 10.3847/0004-637X/818/1/30}
Shen, Y., Horne, K., Grier, C. J., Peterson, B. M., Denney, K. D., Trump, J. R., et al. 2016, ApJ, 818, 30 


\bibitem[Simi{\'c} and Popovi{\'c}(2016)]{2016Ap&SS.361...59S}
Simi{\'c}, S.,  Popovi{\'c}, L. {\v C}. 2016, ApJSS, 361,  id.59, 10 pp.


\bibitem[Songsheng et al. (2019)]{Songsheng19}
Songsheng, Y.-Y. Wang, J.-M. et al. 2019, submitted.


\bibitem[Strateva (2003)]{2003AJ....126.1720S} 
Strateva I. V., Strauss, M. A., Hao, L., Schlegel, D. J., Hall P. B., Gunn, J. E.,
Li, L., Ivezi{\'c}, {\v Z}. et al. 2003, AJ, 126, 1720

\bibitem[Storchi-Bergmanni et al. (2003)]{2003ApJ...598..956S} 
Storchi-Bergmann, T., Nemmen da Silva, R., Eracleous, M., Halpern, J. P.  et al. 2003, ApJ, 598, 956

\bibitem[Storchi-Bergmanni et al. (2017)]{2017ApJ...835..236S}
Storchi-Bergmann, T., Schimoia, J. S., Peterson, B. M., Elvis, M. Denney, K. D., Eracleous, M., Nemmen, R. S. 2017, ApJ,  835,  id. 236, 13 pp.

\bibitem[Sturm et al. (2018)]{2018Natur.563..657G}
Sturm, E., Dexter, J., Pfuhl, O.,Stock, M. R., Davies, R. I., Lutz, D. et al.  2018,  Nature, 563,  7733, 657


\bibitem[Tang et al. (2018)]{2018MNRAS.476.2249T}
Tang, Y., Haiman, Z., MacFadyen, A. 2018, MNRAS, 476, 2249


\bibitem[Tsalmantza et al. (2011)]{2011ApJ...738...20T}
Tsalmantza, P., Decarli, R., Dotti, M., Hogg, D. W. 2011, ApJ,  738,   9



\bibitem[Wang et al. (2009)]{2009ApJ...705L..76W} 
Wang, J., Chen, Y., Hu, C., Mao, W., Zhang, S.,  Bian, W.  2009, ApJ, 705, L76


 \bibitem[Wang et al. (2017)]{2017ApJ...834..129W}
Wang, L.,  Greene, J. E., Ju, W.,  Rafikov, R. R., Ruan, J. J., Schneider, D. P.  2017,  ApJ, 834, 13 

\bibitem[Wang et al. (2018)]{2018ApJ...862..171W} 
Wang, J-M.,  Songsheng, Y.-Y.,  Rong, L.,   Yu, Z. 2018, ApJ, 862, id. 171



\bibitem[Wyithe \&  Loeb (2003)]{2003ApJ...595..614W} 
 Wyithe, J.S.B, Loeb, A.  2003, ApJ, 590, 691

\bibitem[Xu \& Komossa (2009)]{2009ApJ...705L..20X}
Xu, D.,  Komossa, S.  2009, ApJ,705, L20
2017PASA...34...42Y

\bibitem[Yong at al. (2016)]{2017PASA...34...42Y}
Yong, S. Y., Webster, R. L., King, A. L., Bate, N. F. et al.   2017,   PASA,  34, id.e042


\end{thebibliography}
%

\begin{appendix} 
\section{Geometric description of orbits}\label{app:geo}

Firstly, we  introduce the general framework we used  for any orbital motion description,   which is  standard
astrometric notation (Figure \ref{fig:cosis}(a)). If the reference coordinate system $XYZ$ with origin in object $ O$ is arbitrary chosen, let the object of mass M  be in  an elliptical orbit about   object  $O$ moving in anticlockwise direction,  so that its  orbital position  radius vector $OM$ has  a value  $\bf{ r}$ at time $t$.
If a set of axes $O\xi$ and $O\eta$ are taken in the plane of the orbit with $O\xi$ along the
major axis of the elliptical  orbit towards the pericenter $P$ and $O\eta$, perpendicular to the major axis, then the coordinates of $M$ relative to this set of axes are $\xi$ and $\eta$  given by


\begin{eqnarray} 
\vec{r}=\left\{\begin{array}{c}  \displaystyle \xi\\ 
\displaystyle
                               \eta\\
                               \displaystyle
                                   0 
             \end{array} 
             \right\rbrace=\left\{\begin{array}{c}
         \displaystyle r \cos f\\ \displaystyle  r\sin f\\ 0\end{array}\right\},
\end{eqnarray}             
and 
\begin{eqnarray}             
\vec{v}=\left\lbrace\begin{array}{c} \displaystyle  \dot{\xi} \\ \displaystyle  \dot{\eta}\\0  \end{array}\right\rbrace
       =\left\{\begin{array}{c}
       \displaystyle{\frac{-na\sin f}{\sqrt{1-e^{2}}}}\\
       \displaystyle{\frac{na(e+\cos f)}{\sqrt{1-e^2}}}\\
       0 \end{array}\right\}
\label{prve}
 ,\end{eqnarray}

where      $f= \angle POM $ is  a true anomaly defined as the angle
at the focus $O$ between the direction of perihelion and the position radius vector $OM$ of the body M,  $r=\frac{p}{1+e\cos f}$  is the norm of  the position radius vector $OM$, $a$ is the semimajor axis of the elliptical orbit, $e$ is the eccentricity of the elliptical orbit  that defines the amplitude of the oscillations in $\bf{r}$ at time $t$. Furthermore,  $T\text{ and } n$ are the orbital period and  the mean angular frequency (or the mean motion, defined as $n=2\pi/T$) of object M. 
Because 
integrating the velocity in an elliptical  orbit is difficult,  a convenient method uses the  concept of a reference circle centred  on the centre of the elliptical orbit with a radius equal to the semimajor axis, so that the vectors of position and velocity are given as follows:

\begin{eqnarray} 
\vec{r}=\left\{\begin{array}{c}
      \displaystyle       a(\cos E-e)\\ \displaystyle  a\sqrt{1-e^2}\sin E\\ 0\end{array}\right\},
\end{eqnarray}             

and 
\begin{eqnarray}             
\vec{v}=\left\lbrace \begin{array}{c}
       \displaystyle{\frac{-na^{2}}{r}\sin E}\\
       \displaystyle{\frac{na^{2}}{r} \sqrt{1-e^{2}}\cos E}\\
       0 \end{array}\right\rbrace
       \label{druge}
 ,\end{eqnarray}
where $E$ is the eccentric anomaly defined in Appendix \ref{app:appendecc}, an angular variable that also represents the phase of the oscillation in $\bf{r,}$ and $r=a(1-e\cos E)$ is the   norm of the position  radius vector at  a given time. 
 The point of closest approach (i.e. the orbital pericenter)  of body M to  O
is at $E = f=0$, so that  $E =f= \pi$ corresponds to the apocenter and $ p$ is  the semilatus rectum of the elliptical orbit of object M.


The representation of the orbit of object M  in 3D is also illustrated in Fig. \ref{fig:cosis}(a). The orbital  plane intersects  the  reference plane of the coordinate system on a line called the line
of nodes {N$\prime$}ON. If the direction in which the object moves on its orbit is anticlockwise,
N is referred to as the ascending node.  Then the longitude of the ascending
node $\Omega$ is given by  $\angle XON$ measured in the reference plane from $0$ to $2\pi$. The inclination $i$ is the angle between the reference and the orbital plane. The argument of pericenter $\omega=\angle NOP$ is  the angle from the body's ascending node to its pericenter P, measured in the direction of motion in the orbital plane.

The expression  of the orbital position and velocity in 3D is found by  transforming the  state vector from   the apsidal frame to 3D   by applying
\begin{eqnarray}
\vec{r}^{\prime\prime}&=&\mathbb{P}^{T} \vec{r}, \label{inplane0}\\
\vec{v}^{\prime\prime}&=&\mathbb{P}^{T} \vec{v}, \label{inplane}
 \end{eqnarray}

 where in terms of inclination ($i$), argument of pericenter 
 ($\omega$) and angle of ascending node  ($\Omega$) of a given orbit,

\begin{equation}
\mathbb{P} =   \resizebox{0.45\textwidth}{!}{ $\begin{Bmatrix} 
\cos \Omega \cos \omega - \sin \Omega \sin \omega \cos\, i  &  \sin \Omega \cos \omega +\cos \Omega \sin \omega \cos\, i & \sin \omega \sin i\\
-\cos \Omega \sin \omega - \sin \Omega \cos \omega \cos i  &  -\sin \Omega \sin \omega +\cos \Omega \cos \omega \cos\, i & \cos \omega \sin i\\
\sin \Omega \sin\, i  &  -\cos \Omega \sin\, i &  \cos i \\
\end{Bmatrix}$}
\label{matrica}
.\end{equation}
Because the direction cosine matrix $\mathbb{P}$ consists of three rotations that  are orthogonal, 
its  inverse  is just its transpose or  $\mathbb{P}^T$.  The orientation of the elliptic orbit in the coordinate system depends on the inclination $i$,  the angle of the ascending node   $\Omega,$ and the argument of pericenter $\omega$.
If the inclination is zero, the orbit is located in the $X-Y$ plane. As the eccentricity increases,  the pericenter  will be  in the $+X$ direction. If the inclination is increased, the ellipse will be rotating around the $X$-axis, so that +$Y$ is rotated toward +$Z$. An increase in $\Omega$ is the rotation of +$X$ toward +$Y$. A variation  of $\omega$  alters the orbit position only in the orbital plane.

 \subsubsection{Introducing time in the model}

 We note that state vectors  are functions of the astrometric angles. The position and velocity of an SMBH and clouds in their  orbital planes  at a given time depend on their  phases  in orbits, which are defined by their  true anomalies \citep[e.g.][]{MD99}. In the circular orbit  limit $e\rightarrow 0,$ the orbital phase is simply
$f=2\pi\Phi(t-t_{0}),$ where $\Phi$ is the Keplerian orbital frequency and $t_0$ is the pericenter passage (see Paper I).  
However, for eccentric orbits, the  non-linear relative angular positions between two bodies in motion about their centre of mass is related  to linearly advancing time through the Kepler equation.
When  the object passes through pericenter  at $t=t_0$, and its orbital period is T, the dimensionless variable
 $l=2\pi ({t-t_{0}})/{T}=n(t-t_{0})$ is referred to as the mean anomaly, where $n$ is the mean angular frequency (or mean motion), $n=2\pi/T$. We numerically solved the transcendental Kepler equation for the eccentric anomaly $E$:
 
\begin{equation}
l=E- e \sin E
\label{Keplereq}
 .\end{equation}
 Furthermore, the true ($f$) and eccentric ($E$)  anomaly are related by (see Appendix \ref{app:appendecc}, Eq. \ref{ef9})
  
 \begin{equation}
\tan \frac{f}{2}=\sqrt{\frac{1+e}{1-e}}\tan \frac{ E}{2}
 \label{connection}
 ,\end{equation}
 where $ f$ establishes the temporal variation of the phase of the object on its orbit. 
 Thus we can obtain the position and velocity of the object  at any given time $t$.
Clearly, the phase is related to the inverse of the tangent of the eccentric anomaly, and the latter is related in transcendental form as the Kepler equation to the mean motion $n=2\pi/T$ and thus to the frequency $n=2\pi \Phi={2\pi}/{T}$.

Knowing the average angular frequency and the eccentricity of the orbit of the object,
we solved the transcendental Eq. \ref{Keplereq} for $E$ at a given time $t$, where $l$ is calculated  as $l = 2\pi(t -t_{0})/T$. The eccentric anomaly was then substituted into Eq.  \ref{connection} to give the  phase $f$.  After this, the  eccentric anomaly was plugged into Eq. \ref{druge}  to obtain the position and velocity at any point along the orbit.  We assumed that time and phase are measured from the moment of passage through pericenter, then the constant $t_0$ can be set to zero. 
 
\subsection{Relative motion in the orbit}\label{app:relmotapp}

Taking into account the relative motion of the SMBH with respect to the barycenter, we derived the transformation of the vectors between the barycentric and the comoving frame attached to the SMBH.
We assumed that the position of the more massive SMBH in  the barycentric coordinate frame of the coplanar binary case is 
$\vec {r}_\mathrm{b1}$. The moving frame of reference has its origin at the  SMBH and is orbiting the barycenter of the binary. The $x$ -axis is directed along the outward radial $\vec{ r}_\mathrm{b1}$ to the SMBH.  Thus the unit vector $\vec{\hat{ i}}$ along the moving axis is

\begin{equation}
 \vec{\hat {i}}=\frac{\vec{ r}_\mathrm{b1}}{{r}_\mathrm{b1}}
.\end{equation}

The z-axis of the moving frame  is normal to the orbital plane of the SMBH, lying in the direction of the angular momentum vector of the SMBH. Its  unit vector   is given by

\begin{equation}
 \vec{\hat{k}}=\frac{\vec{ r}_\mathrm{b1}\times \vec{ v}_\mathrm{b1}}{|{\vec {r} _\mathrm{b1}\times \vec {r} _{b1}}|}
.\end{equation}

The y -axis is perpendicular to both $\vec{\hat{ i}}$ and $\vec{\hat{ k}}$, with the unit vector completing the right triad of the moving frame, that is,

\begin{equation}
 \vec{\hat{ j}}=\vec{\hat{ k}}\times \vec{\hat{ i}}
.\end{equation}

The transformation of any vector $\vec{s}_\mathrm{B}$  from the barycentric to the local SMBH  moving  frame is given by

\begin{equation}
 \vec{s}_{\bullet}=\mathbb{Q}  \vec{ s}_\mathrm B
,\end{equation}

where  the matrix $\mathbb{Q} $ consists of rows of the unit vector coordinates defined above: 

\begin{equation}
\mathbb{Q} = \begin{Bmatrix} 
\vec{\hat{ i}}\\
\vec{\hat{ j}}\\
\vec{\hat{ k}} \\
\end{Bmatrix}
\label{matricaQ}
,\end{equation}
and the inverse transformation from the local to the barycentric system is given by
 $ \vec{{ s}_\mathrm{B}}=\mathbb{Q}^{-1}  \vec{ s}_{\bullet} $.

In the non-coplanar case the local reference plane of the inclined SMBH does not coincide with the general reference plane.

\section{ Relation between the eccentric and the true anomaly}\label{app:appendecc}
\begin{figure}[ht!]
\includegraphics[trim=20 50 25  40,clip,width=0.45\textwidth]{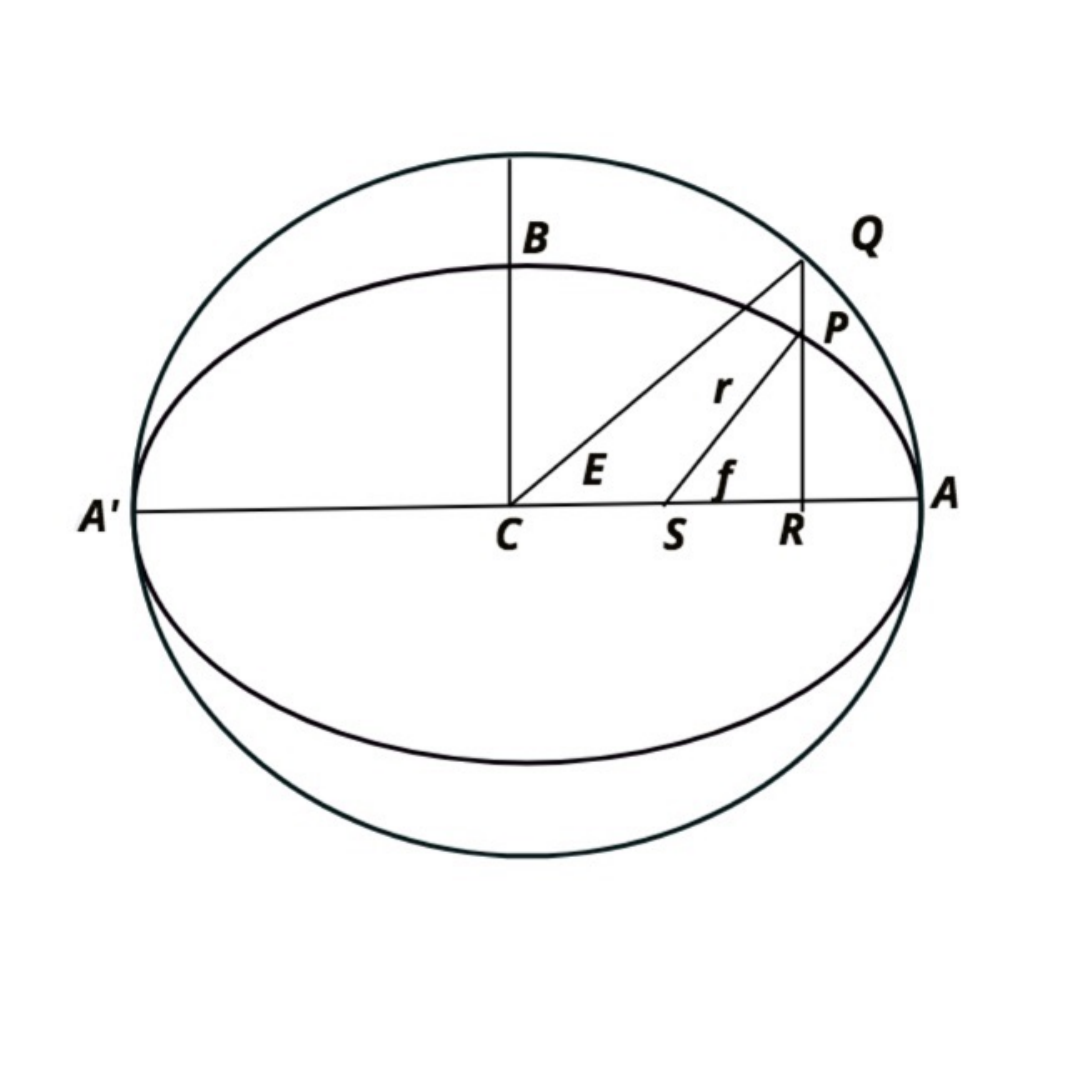}
\vspace{0.5pt}
\caption{Scheme for deriving the relation between the true and mean anomaly.  C is the common centre of the ellipse and the auxiliary circle. S is the focus of the ellipse. The semimajor  and semiminor axes are $CA=a$ and $CB=b,$ respectively. The orbital postion of the object is P, its perpendicular projection on the semimiajor axis is R, and the intersection of PR and the auxiliary circle is Q. $|SP|=r$ and $CS=ae$, where e is the eccentricity of ellipse. The angles $\angle QCA \text{ and } \angle PSA$ are the eccentric and the true anomaly, respectively. The radius of the auxiliary circle equals the orbital semimajor axis. \label{fig:eccf}}
\end{figure}

The relation between the eccentric (E) and the true anomaly (f)  defined in Eq. \ref{connection} is derived  by 
following geometrical consideration.
A circumcircle is described on AA$^{\prime}$ as  the diameter around the ellipse, as shown in Fig \ref{fig:eccf}.
PR is perpendicular to AA$^{\prime}$ and intersects the circumcircle in Q, the angle QCA is the eccentric anomaly E, and it is related to the true anomaly f.

From the geometry, it is easy to see that

\begin{eqnarray}
SR&=&CR-CS=a \cos E - ae,\\
SR&=&r \cos f
\label{ef1}
,\end{eqnarray}

and hence

\begin{eqnarray}
r \cos f&=&a (\cos E-e)
\label{ef2}
.\end{eqnarray}
According to the property of the ellipses and circumscribed circles,

\begin{eqnarray}
\frac{PR}{QR}&=&\frac{b}{a},\\
r \sin f&=&b \sin E=a\sqrt{1-e^2} \sin E
\label{ef3}
.\end{eqnarray}
Squaring and summing Eqs. \ref{ef2} and \ref{ef3}, we obtain
\begin{eqnarray}
r&=&a\left(1-e\cos E\right)
\label{ef4}
.\end{eqnarray}

Now, using trigonometry, 
\begin{eqnarray}
r\cos f&=&r\left(\cos^{2} \frac{f}{2}-\sin^{2}\frac{f}{2}\right)=r\left(1-2\sin^{2}\frac{f}{2}\right)
\label{ef5}
,\end{eqnarray}

so that 

\begin{eqnarray}
2r\sin^{2} \frac{f}{2}&=&r\left(1-\cos f\right)
\label{ef6}
.\end{eqnarray}

Using Eqs. \ref{ef2} and \ref{ef4}, we obtain
\begin{eqnarray}
2r\sin^{2} \frac{f}{2}&=&a\left(1+e\right)\left(1-\cos E\right)
\label{ef7}
.\end{eqnarray}
With similar reasoning, it is easy to obtain
\begin{eqnarray}
2r\cos^{2} \frac{f}{2}&=&a\left(1-e\right)\left(1+\cos E\right)
\label{ef8}
.\end{eqnarray}

Dividing Eq \ref{ef7} by Eq. \ref{ef8}, we finally obtain the relation between the eccentric and the true anomaly,

\begin{eqnarray}
\tan \frac{f}{2}&=&\sqrt{\frac{1+e}{1-e}}\tan \frac{ E}{2}
\label{ef9}
.\end{eqnarray}

The eccentric anomaly is an angular variable that represents the phase of the oscillation in radial position of a body on an elliptical orbit.

\section{ Expansions of vectorial equations for velocity-delay maps}\label{app:expandtauv}

Here we list expansions of the equations of the barycentric state vectors of $M_1$ (see Eq.\ref{baricentric}), the radial position of the  jth cloud in its BLR with respect to the barycenter (see Eq. \ref{rdisk}), the barycentric velocity of the cloud (Eq. \ref{barvel}), and the projection of the cloud barycentric vector of the relative position and velocity   (see Eqs. \ref{delay} and \ref{barvel1}) on the observed  line of sight ( $\vec{n}_{\mathrm {obs}}=({0},-\sin{i_0},-\cos{i_0})$). For  practical reasons, the subscript 1  is omitted for SMBH orbital elements, and  we use  the short notations

\begin{eqnarray}
 b=a\sqrt{1-e^2},\\
 r=a(1-e\cos E),\\
 h=\sqrt{G\frac{M_{1}+M_{2}}{a_\mathrm{B}}} \frac{M_{1}M_{2}}{M_{1}+M_{2}}b_\mathrm{B},\label{inangm}\\
 \vec{\omega}_{\bullet/\mathrm{B}}={h}\left \{
\begin{array}{c}
 \sin \Omega \sin i \\
- \cos \Omega \sin i\\
\cos i
\end{array}
 \right\}, \label{L} \\
 n=\frac{2\pi}{T}\\
 w_{11}= \cos \Omega \cos\omega - \sin \Omega \sin \omega \cos i, \label{bpr} \\ 
 w_{12}=\sin \Omega \cos\omega + \cos \Omega \sin \omega \cos i,\\
 w_{13}=\sin \omega \sin i, \\
 w_{21}=-\cos \Omega \sin\omega - \sin \Omega \cos \omega \cos i,\\
 w_{22}=-\sin \Omega \sin\omega + \cos \Omega \cos \omega \cos i,\\
 w_{23}=\cos \omega \sin i,\\
 \mathbb{Q}^{-1}=[q_{ij}], i,j=1,3  \label{bdr}  \
,\end{eqnarray}
where $h$ in Eq.\ref{inangm} is the norm of the total angular momentum of the barycentric system, which is written in terms of the semimajor $a_\mathrm{B}$ and semiminor $b_\mathrm{B}$ axis of the ellipse that is swept by the vector of the relative position of $M_2$ with respect to $M_1$ \citep[see][]{ISBN-13: 978-1107677241}.    We note that in Eq. \ref{bpr} - \ref{bdr}   numbers 1 and 2  have descriptive meaning and are not the vector components of a certain  SMBH. 
Their forms  are invariant under coordinate system transformations. 
Likewise, we can introduce short notations for  the parameters related to the cloud orbits in a moving frame, 
\begin{eqnarray}
 b_\mathrm{c}=a_\mathrm{c}\sqrt{1-e^2_\mathrm{c}},\\
 r_\mathrm{c}=a_\mathrm{c}(1-e_\mathrm{c}\cos E_\mathrm{c}),\\
 n_\mathrm{c}=\frac{2\pi}{T_\mathrm{c}}\\
 w^{\mathrm c}_{11}= \cos \Omega_{c} \cos\omega_{c} - \sin \Omega_{c} \sin \omega_{c} \cos i_{c}, \label{bprc} \\ 
 w^{c}_{12}=\sin \Omega_\mathrm{c} \cos\omega_\mathrm{c} + \cos \Omega_\mathrm{c} \sin \omega_\mathrm{c} \cos i_\mathrm{c},\\
 w^\mathrm{c}_{13}=\sin \omega_\mathrm{c} \sin i_\mathrm{c}, \\
 w^\mathrm{c}_{21}=-\cos \Omega_\mathrm{c} \sin\omega_\mathrm{c} - \sin \Omega_\mathrm{c} \cos \omega_\mathrm{c} \cos i_\mathrm{c},\\
 w^\mathrm{c}_{22}=-\sin \Omega_\mathrm{c} \sin\omega_\mathrm{c} + \cos \Omega_\mathrm{c} \cos \omega_\mathrm{c} \cos i_\mathrm{c},\\
 w^\mathrm{c}_{23}=\cos \omega_\mathrm{c} \sin i_\mathrm{c},\\
 \mathbb{Q}^{-1}=[q_{ij}], i,j=1,3  \label{bdrc}  \
.\end{eqnarray}

 The  transformation  $\mathbb{Q}^{-1} $  is an inverse of  the transformation defined  by Eq. \ref{matricaQ}, and allows us to transform the vector from the comoving to the barycentric frame. We note that  Eqs. \ref{bpr}-\ref{bdr} and \ref{bprc}-\ref{bdrc}   could be reduced to a simpler form for 
 $\Omega=0$ and $\Omega_\mathrm{c}=0$.

 For a given set of elements, the mean motion $n$ and the time, the barycentric coordinates,  and the velocity components of $M_{1}$ can be computed as follows:

\begin{eqnarray}
\vec{r}_{\mathrm b}&=&r \frac{M_2}{{M}_{1}+M_{2}}\left \{
\begin{array}{l}
\displaystyle
a w_{11}\cos E+b w_{21}\sin E - a w_{11} \\
\displaystyle
a w_{12} \cos E+b w_{22}\sin E-a e w_{11}\\
\displaystyle
a w_{13}\cos E+b w_{23}\sin E-a e w_{13}
\end{array}
 \right\}
\end{eqnarray}

\begin{eqnarray}
\vec{v}_{\mathrm b}&=&\frac{{M_2}}{M_{1}+M_{2}}\frac{na}{r}\left \{
\begin{array}{l}
\displaystyle
b w_{21}\cos E-a w_{11}\sin E  \\
\displaystyle
b w_{22} \cos E-a w_{12}\sin E\\
\displaystyle
b  w_{23} \cos E-a w_{13}\sin E
\end{array}
 \right\}
.\end{eqnarray}

Similarly, the vector of the cloud  position and velocity in a 3D SMBH  comoving frame is given as

\begin{eqnarray}
\vec{\varrho}_{\bullet}&=&r_\mathrm{c}\left \{
\begin{array}{l}
\displaystyle
a_\mathrm{c} w^\mathrm{c}_{11}\cos E_\mathrm{c}+b_\mathrm{c} w^\mathrm{c}_{21}\sin E_\mathrm{c} - a_\mathrm{c} w^\mathrm{c}_{11} \\
\displaystyle
a_\mathrm{c}w^\mathrm{c}_{12} \cos E_\mathrm{c}+b_\mathrm{c}w^\mathrm{c}_{22}\sin E_\mathrm{c}-a_\mathrm{c} e_\mathrm{c} w^\mathrm{c}_{12}\\
\displaystyle
a_\mathrm{c}w^\mathrm{c}_{13}\cos E_\mathrm{c}+b_\mathrm{c}w^\mathrm{c}_{23}\sin E_\mathrm{c}-a\mathrm{c}e_\mathrm{c} w^\mathrm{c}_{13}
\end{array}
 \right\}
  \label{comorho}
\end{eqnarray}

\begin{eqnarray}
\dot{\vec{\varrho}_{\bullet}}&=&\frac{n_\mathrm{c}a_\mathrm{c}}{r_\mathrm{c}}\left \{
\begin{array}{l}
\displaystyle
b_\mathrm{c} w^\mathrm{c}_{21}\cos E_\mathrm{c}-a_\mathrm{c}w^\mathrm{c}_{11}\sin E_\mathrm{c}  \\
\displaystyle
b_\mathrm{c}w^\mathrm{c}_{22} \cos E_\mathrm{c}-a_\mathrm{c} w^\mathrm{c}_{12}\sin E_\mathrm{c}\\
\displaystyle
b_\mathrm{c}w^\mathrm{c}_{23}\cos E_\mathrm{c}-a_\mathrm{c} w^\mathrm{c}_{13}\sin E_\mathrm{c}
\end{array}
 \right\}
 \label{comovel}
.\end{eqnarray}

Thus the relative position of the cloud in the barycentric  frame  is calculated by  the transformation  $\mathbb{Q}^{-1} $ , which is the inverse of  the transformation defined  by Eq. \ref{matricaQ} at each time instant:
\begin{equation}
[\vec{\varrho}]_\mathrm{B}=\mathbb{Q}^{-1}[\vec{\varrho}]_{\bullet}
\label{qrhotrans}
 \end{equation}
The characterisation of the motion of the SMBH coordinate frame is given by specifying the
angular velocity vector of the SMBH. Because the SMBH coordinate frame is also a body frame
of the SMBH, the angular velocity of the SMBH is also the angular velocity of the
frame. The barycentric  angular velocity of the SMBH  is given by Eq. \ref{inangm}.
The velocity of the cloud in the barycentric frame as time-derivative is  calculated through the basic kinematic equation or transport theorem (see  Eq. \ref{barvel}).
Thus,  the elements of the projections of $[\vec{\varrho}]_\mathrm{B}$ on the observer's line of sight  are given by
\begin{minipage}[h]{0.95\textwidth}
\begin{equation}
\begin{split}
{[\vec{\varrho}]_B}\cdot \vec{n}_{\mathrm obs}&=- {\sin}\, {i_0}
 \bigg(q_{21} {r_c} ({a_c} w^\mathrm{c}_{11}  {\cos}\,
   {E_c}-{a_\mathrm{c} w^\mathrm{c}_{11}}+{b_c}{w^\mathrm{c}_{21}}  {\sin}\,{E_c})+{q_{22}} {r_c} (-{a_c} {e_c}
   {w^\mathrm{c}_{12}}+{a_c} {w^\mathrm{c}_{12}}  {\cos}\, {E_c}+{b_c}
   {w^\mathrm{c}_{22}}  {\sin}\, {E_c})+ \\
  & \quad {q_{23}} {r_c} (-{a_c} 
   {e_c} {w^\mathrm{c}_{13}}+{a_c} {w^\mathrm{c}_{13}}  {\cos}\,
   {E\_c}+{b_c} {w^\mathrm{c}_{23}}  {\sin}\, {E_c})\bigg)- {\cos}\, {i_0}
  \bigg ({q_{31}} {r_c} ({a_c} {w^\mathrm{c}_{11}}  {\cos}\,
   {E_c}-{a_\mathrm{c}w^\mathrm{c}_{11}}+{b_c}{w^\mathrm{c}_{21}}  {\sin}\,
   {E_c})+  \\
   &\quad  {q_{32}} {r_c} (-{a_c} {e_c}
   {w^\mathrm{c}_{12}}+{a_c} {w^\mathrm{c}_{12}}  {\cos}\, {E_c}+{b_c}
   {w^\mathrm{c}_{22}}\,  {\sin}\, {E_c})+{q_{33}} {r_c} (-{a_c}
   {e_c} {w^\mathrm{c}_{13}}+{a_c} {w^\mathrm{c}_{13}}  {\cos}\,
   {E_c}+{b_c} {w^\mathrm{c}_{23}}  {\sin}\, {E_c})\bigg)
   \end{split}
\end{equation}
 \end{minipage}
 
 Likewise, the elements of the projections  $\vec{V}$ on the observer's line of sight is obtained as follows:

\begin{minipage}[h]{0.95\textwidth}
\begin{equation}
\begin{split}
V_{z}&=\vec{V} \cdot \vec{n}_{\mathrm obs}= 
 \frac{1}{r_c}\Bigg(-a_c \bigg( {\cos}\, {E_c} {b_c }{n_c} \left( {\cos}\,{i_0}\, {w^\mathrm{c}_{23}} + {w^\mathrm{c}_{22}} \, {\sin}\,{i_0}\right) + 
                h {r^{2}_c} \big( {\cos}\,{
                     E_c} {w^\mathrm{c}_{11}} ( {\cos}\,{i_0}  {\cos}\,{\Omega}\,  {s}\,{i} + 
                       {\cos}\,{i}  {\sin}\,{i_0}) +\\
                      &\quad  ( {\cos}\,{E_c} - {e_c})  {\sin}\,{
                     i} ( {\cos}\,{i_0} {w^\mathrm{c}_{12}} - {w^\mathrm{c}_{13}}  {\sin}\,{i_0})  {\sin}\,{\Omega}\big)\bigg) + 
             \bigg( {a^{2}_c} {n_c} ( {\cos}\,{i_0} {w^\mathrm{c}_{13}} + {w^\mathrm{c}_{12}}  {\sin}\,{i_0})  {\sin}\,{E_c} + \\
             &\quad     h {r^{2}_c} ({a w^\mathrm{c}_{11}} ( {\cos}\,{i_0}  {\cos}\,{\Omega} \, {\sin}\,{i} + 
                    {\cos}\,{i_0}  {\sin}\,{
                     i_0})\bigg) -  \bigg({b_\mathrm{c}} {w^\mathrm{c}_{21}} ( {\cos}\,{i_0}  {\cos}\,{\Omega}\,  {\sin}\,{i} + 
                       {\cos}\,{i_0}  {\sin}\,{i_0}) + \\
             &\quad      {b_\mathrm{c}}  {\sin}\,{i}\, ( {\cos}\,{i_0}\, {w^\mathrm{c}_{22}} - {w^\mathrm{c}_{23}}  {\sin}\,{i_0})\,  {\sin}\,{\Omega}\bigg)\,  {\sin}\,{  E_c}\Bigg )\end{split}
\end{equation}
\end{minipage}
\end{appendix}

\end{document}